\documentclass[11pt,english]{article}
\usepackage[T1]{fontenc}
\usepackage[latin9]{inputenc}
\usepackage{calc}
\usepackage{mathtools}
\usepackage{amsmath}
\usepackage{amssymb}
\usepackage{stackrel}
\usepackage{graphicx}
\usepackage[numbers]{natbib}

\makeatletter

\providecommand{\tabularnewline}{\\}

\pdfoutput=1
\usepackage[T1]{fontenc}
\usepackage[latin9]{inputenc}
\usepackage{color}
\usepackage{float}
\usepackage{graphicx}
\usepackage{graphics}
\usepackage{esint}
\usepackage{enumerate}

\makeatletter

\providecommand{\tabularnewline}{\\}

\usepackage{latexsym}
\usepackage{epsfig}
\usepackage{graphicx}
\usepackage[colorlinks,linkcolor=blue]{hyperref}
\allowdisplaybreaks 

\usepackage{babel}
\usepackage{tcolorbox}

\usepackage{tikz}

\makeatother

\usepackage{babel}
\begin{document}
{}~ \hfill\vbox{\hbox{CTP-SCU/2025023}}\break
\vskip 2.5cm
\centerline{\Large \bf Toward a worldsheet theory of entanglement entropy}

\vspace*{11.0ex}
\centerline{\large  Houwen Wu and Shuxuan Ying}
\vspace*{7.0ex}
\vspace*{3.0ex}

\centerline{\large \it College of Physics}
\centerline{\large \it Sichuan University}
\centerline{\large \it Chengdu, 610065, China} \vspace*{1.0ex}

\vspace*{3.0ex}
\centerline{\large \it Department of Physics}
\centerline{\large \it Chongqing University}
\centerline{\large \it Chongqing, 401331, China} \vspace*{1.0ex}
\vspace*{3.0ex}

\centerline{iverwu@scu.edu.cn, ysxuan@cqu.edu.cn}
\vspace*{8.0ex}
\centerline{\bf Abstract} \bigskip \smallskip
We propose a new action for entanglement entropy in the framework of the AdS$_{3}$/CFT$_{2}$ correspondence. This action is constructed directly from the entanglement entropy of the CFT$_{2}$, and we show that the Einstein equations of AdS$_{3}$ gravity can be derived from it. In the near-coincidence limit, using Riemann normal coordinates, the action reduces to a string worldsheet action in a curved background that naturally includes the symmetric spacetime metric, an antisymmetric Kalb-Ramond field, and a dilaton. The Kalb-Ramond field gives rise to a string charge density, from which we demonstrate that bit threads can be exactly reproduced. This correspondence provides a clear physical interpretation of bit threads. Exploiting this correspondence, we establish explicit relations between the emergent string worldsheet and the Ryu-Takayanagi (RT) surface, providing new insights into entanglement entropy. In particular, entanglement entropy can be computed from open string charge, while Bekenstein-Hawking entropy arises from closed string charge through open-closed string duality. These results suggest a unified picture in which the Susskind-Uglum conjecture, open-closed string duality, and the ER=EPR proposal emerge as equivalent manifestations of the same underlying principle. Finally, we propose a quantization of the RT surface, pointing to a possible connection with loop quantum gravity that refines Wall's conjecture.

\vfill 
\eject
\baselineskip=16pt
\vspace*{10.0ex}
\tableofcontents

\section{Introduction}

The black hole information paradox remains a central problem in modern
theoretical physics. A significant development in this context was
the work of Ryu and Takayanagi, who established a relationship between
the entanglement entropy (specifically, the von Neumann entropy) in
two-dimensional conformal field theory (CFT$_{2}$) and the area of
extremal surfaces---known as Ryu-Takayanagi (RT) surfaces---in the
bulk of AdS$_{3}$ \cite{Ryu:2006bv,Ryu:2006ef,Hubeny:2007xt}. Building
on this result, one can compute the von Neumann entropy of gravitational
systems via the quantum extremal surface prescription \cite{Faulkner:2013ana,Engelhardt:2014gca}.
This approach has been successfully applied to evaporating black holes
and their associated Hawking radiation, yielding results consistent
with the Page curve \cite{Penington:2019npb,Almheiri:2019psf}. Nevertheless,
the precise quantum state corresponding to the entanglement entropy
of Hawking radiation remains unclear, posing a key unresolved aspect
of the information paradox \cite{Almheiri:2020cfm}.

\begin{figure}[h]
\begin{centering}
\includegraphics[scale=0.4]{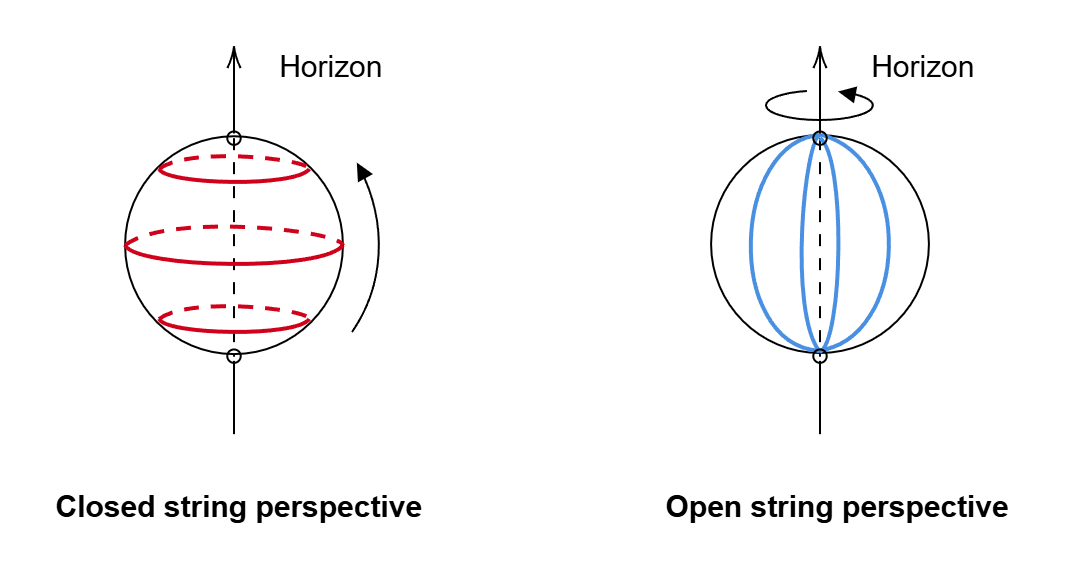}
\par\end{centering}
\caption{\label{fig:SU}The left panel of the figure illustrates the closed
string perspective in the Susskind and Uglum's setup. The blue circle
represents a closed string whose initial and final states interact
with the event horizon. In contrast, the right panel depicts the open
string configuration, indicated by the red lines, with endpoints fixed
on the horizon. In this case, the target space naturally factorizes
into two Hilbert spaces corresponding to the open strings inside and
outside the event horizon: $\mathcal{H}\subseteq\mathcal{H}_{out}\otimes\mathcal{H}_{in}$.
This factorization provides a statistical interpretation of entanglement
entropy, as discussed in \cite{Ahmadain:2022eso}.}
\end{figure}

Susskind and Uglum proposed that black hole entropy could be derived
from the closed string worldsheet, where the string sphere is punctured
twice by the black hole horizon, as illustrated in the left panel
of figure (\ref{fig:SU}) \cite{Susskind:1994sm}. This picture can
also be interpreted from the open string perspective, in which the
string endpoints are fixed on the horizon. In this framework, a slice
of the closed string (the sphere) intersecting the horizon appears
as an open string to a Rindler observer outside the event horizon
(right panel of figure (\ref{fig:SU})), offering a statistical interpretation
of entanglement entropy. Susskind and Uglum's conjecture thus suggests
that both black hole entropy and entanglement entropy can be computed
in string theory, potentially allowing the identification of their
underlying quantum states. Recently, Ahmadain and Wall provided a
proof of the closed string version of this conjecture using the sphere
diagram in off-shell closed string theory, reproducing the expected
result $\mathrm{Area}/4G_{N}$ \cite{Ahmadain:2022tew,Ahmadain:2022eso,Ahmadain:2024hdp}.
However, establishing the open string version remains challenging,
as the replica trick introduces a conical singularity that breaks
the conformal symmetry of the string worldsheet.

If Susskind and Uglum\textquoteright s conjecture is correct, there
must exist a correspondence between the string worldsheet and the
RT surface. Such a correspondence could plausibly serve as a guiding
principle for developing the open string perspective of the conjecture
in the framework of bosonic string theory, ultimately enabling the
computation of entanglement entropy using the string worldsheet. With
this framework in place, the black hole information paradox can be
revisited from a new perspective, potentially revealing the microscopic
quantum state of Hawking radiation. In previous works, qualitative
evidence for a relationship between the string worldsheet and the
RT surface has been found in both open and closed string field theories
\cite{Wang:2021aog,Jiang:2024noe}. Both entanglement entropy and
string field theory are fundamentally grounded in hyperbolic geometry.
Furthermore, key dynamical features---such as phase transitions in
the entanglement wedge cross section (EWCS) \cite{Takayanagi:2017knl}
or reflected surface \cite{Dutta:2019gen}, and the Batalin--Vilkovisky
(BV) master equations in open or closed string field theory \cite{Sen:1993kb,Sen:1994kx}---appear
to be intrinsically linked. These observations suggest that the correspondence
is not a mere coincidence due to shared geometric structures but reflects
a deeper, underlying connection. Nevertheless, a rigorous quantitative
proof of this correspondence remains elusive. A similar observation
was recently made in \cite{Bao:2025plr}: as expected, the RT surface
for multiboundary black holes---which in our framework corresponds
to the string vertices discussed earlier---can be derived directly
from the dual CFT$_{2}$.

Moreover, there are several related approaches to computing entanglement
entropy using string theory. The first method involves evaluating
the entropy in string theory via orbifolds \cite{Dabholkar:1994ai,Dabholkar:2022mxo,Lowe:1994ah,He:2014gva,Witten:2018xfj},
where the orbifold fixed point corresponds to the tip of a cone with
opening angle $\beta=2\pi/N$. However, the most successful approaches
to date \cite{Lowe:1994ah,He:2014gva,Witten:2018xfj} face difficulties
in addressing the timing of tachyon condensation, as required for
consistency with the Susskind--Uglum conjecture. On the other hand,
as emphasized in \cite{Ahmadain:2022eso}, another approach \cite{Balasubramanian:2018axm,Naseer:2020lwr}
suffers from a conceptual issue. In this approach, the \textquotedblleft position\textquotedblright{}
of a string is defined solely by the location of its center of mass.
However, due to string vibrations, different parts of the string can
extend outside the assigned region, making the center-of-mass prescription
inconsistent with the actual spatial support of the string.

In this paper, \emph{rather than pursuing a bottom-up strategy that
attempts to collect and synthesize fragmented evidence from various
theoretical frameworks, we adopt a top-down approach}. Specifically,
we aim to reformulate the theory of entanglement entropy directly
in the framework of AdS$_{3}$/CFT$_{2}$ correspondence. In other
words, we directly propose an action for entanglement entropy and
demonstrate that it reduces to the string worldsheet action in curved
spacetime. This approach yields an explicit and quantitative relation
between the string worldsheet and the RT surface. To approach this
aim, we begin by observing that the entanglement entropy in CFT$_{2}$
corresponds to the geodesic length $L\left(X,X^{\prime}\right)$ in
AdS$_{3}$:

\begin{equation}
S_{\mathrm{vN}}\left(A:B\right)=\frac{L\left(X,X^{\prime}\right)}{4G_{N}^{\left(3\right)}},
\end{equation}

\noindent where $G_{N}^{\left(3\right)}$ is a three-dimensional Newton\textquoteright s
constant. From this geodesic length, we can extract Synge's world
function \cite{Synge:1960}, which can be expressed in terms of the
entanglement entropy as

\begin{equation}
\Omega\left(X,X^{\prime}\right)=\frac{1}{2}L\left(X,X^{\prime}\right)^{2}=8\left[G_{N}^{\left(3\right)}S_{\mathrm{vN}}\left(A:B\right)\right]^{2}.
\end{equation}

\noindent We define the partial derivatives of $\Omega$ with respect
to $X$ and $X^{\prime}$ as $\Omega_{\mu}\coloneqq\partial\Omega/\partial X^{\mu}$
and $\Omega_{\mu^{\prime}}\coloneqq\partial\Omega/\partial X^{\mu^{\prime}}$,
respectively. The covariant derivative of $\Omega_{\mu}$ with respect
to $X$ is denoted as $\Omega_{\mu\nu}\coloneqq\nabla_{\nu}\Omega_{\mu}$,
while the mixed derivative with respect to $X^{\prime}$ is $\Omega_{\mu\nu^{\prime}}\coloneqq\partial_{\nu^{\prime}}\Omega_{\mu}$.
An important concept in this context is the coincidence limit, in
which $X^{\prime}\rightarrow X$, denoted by $\left[\cdots\right]$.
Under this limit, the spacetime metric derived from the geodesic length
is given by

\begin{equation}
g_{\mu\nu}\left(X\right)=-\left[\Omega_{\mu\nu^{\prime}}\right]=-\underset{X^{\prime}\rightarrow X}{\lim}\partial_{\mu}\partial_{\nu^{\prime}}\Omega\left(X,X^{\prime}\right),\qquad g_{\mu\nu}\left(X\right)=\left[\Omega_{\mu\nu}\right]=\underset{X^{\prime}\rightarrow X}{\lim}\nabla_{\nu}\partial_{\mu}\Omega\left(X,X^{\prime}\right).
\end{equation}

\noindent To retain more geometric information, we can consider the
\textbf{near-coincidence} limit $X\rightarrow X^{\prime}$ \cite{Poisson:2011nh},
yielding the expansion:

\begin{equation}
\Omega_{\mu^{\prime}\nu^{\prime}}\left(X,X^{\prime}\right)=g_{\mu^{\prime}\nu^{\prime}}\left(X^{\prime}\right)-\frac{1}{3}R_{\mu^{\prime}\lambda^{\prime}\nu^{\prime}\kappa^{\prime}}\left(X\right)\Omega^{\lambda^{\prime}}\left(X,X^{\prime}\right)\Omega^{\kappa^{\prime}}\left(X,X^{\prime}\right)+\ldots.
\end{equation}

\noindent This expression closely parallels the expansion of the closed
string worldsheet action in Riemann normal coordinates (RNC).

Here, a key insight emerges. In string theory, gravity is known to
arise from CFT. When a string propagates in a curved background, the
requirement that the worldsheet theory remain conformally invariant
at the quantum level imposes constraints on the background geometry---namely,
the string-theoretic version of the Einstein equations \cite{Callan:1985ia}.
This observation raises a compelling question: can we apply the same
logic to entanglement entropy using the expansion of $\Omega_{\mu^{\prime}\nu^{\prime}}$?
More precisely, can we impose conformal symmetry on this expansion
and derive Einstein equations? This question is well-motivated, especially
since $\Omega_{\mu^{\prime}\nu^{\prime}}$, which encodes geometric
data such as the geodesic separation between two points in spacetime,
is entirely constructed from CFT correlators. If $\Omega_{\mu^{\prime}\nu^{\prime}}$
arises from CFT, then requiring quantum-level conformal invariance
may indeed constrain the emergent geometry---potentially leading
to gravitational dynamics from entanglement itself.

Fortunately, a natural scalar quantity is already available: the Synge's
world function $\Omega\left(X,X^{\prime}\right)$, derived from entanglement
entropy, which can serve as the foundation for constructing an action.
This function satisfies the identity

\begin{equation}
2\Omega=g_{\mu\nu}\left(X\right)\Omega^{\mu}\Omega^{\nu}=\Omega_{\mu\nu}\Omega^{\mu}\Omega^{\nu},
\end{equation}

\noindent suggesting that the dynamics of entanglement may be governed
by an action analogous to that of a point particle or a string worldsheet.
The Synge's world function, defined as half the squared geodesic distance
between two bulk points $X$ and $X^{\prime}$, is computable from
the CFT$_{2}$ and naturally depends on the CFT$_{2}$ coordinates
$\left(\tau,\sigma\right)$. While a single geodesic corresponds to
the minimal RT surface anchored on a boundary interval, varying the
entangling region continuously generates a family of geodesics that
collectively sweep out a two-dimensional surface in the bulk. This
motivates parameterizing the non-local action as

\begin{equation}
S_{E}=\frac{1}{\ell^{2}}\int d^{2}\sigma^{\prime}\Omega_{\mu^{\prime}\nu^{\prime}}\partial_{\alpha^{\prime}}\Omega^{\mu^{\prime}}\partial^{\alpha^{\prime}}\Omega^{\nu^{\prime}},\qquad\alpha^{\prime}=\left(\tau^{\prime},\sigma^{\prime}\right).\label{eq:1 action}
\end{equation}

\noindent where $\ell$ denotes a charactestic length scale. We use
the prime notation here because, by expanding the action in the near-coincidence
limit $X\rightarrow X^{\prime}$\footnote{Since $X$ and $X^{\prime}$ are interchangeable, one may also drop the prime in the action (\ref{eq:1 action}) and take the limit $X^{\prime}\rightarrow X$.}
and employing RNC $\mathbb{X}$, we obtain

\begin{equation}
S_{E}=\int d^{2}\sigma\left(g_{\mathrm{ab}}\partial_{\alpha}\mathbb{X}^{\mathrm{a}}\partial^{\alpha}\mathbb{X}^{\mathrm{b}}-\frac{\ell^{2}}{3}R_{\mathrm{acbd}}\mathbb{X}^{\mathrm{c}}\mathbb{X}^{\mathrm{d}}\partial_{\alpha}\mathbb{X}^{\mathrm{a}}\partial^{\alpha}\mathbb{X}^{\mathrm{b}}+\ldots\right),
\end{equation}

\noindent which is precisely the string worldsheet action in a curved
background. Thus, in the near-coincidence limit, the expansion of
$\Omega\left(X,X^{\prime}\right)$ yields, at leading order, a local
quadratic action equivalent to the Polyakov kinetic term for the small
separation field $\mathbb{X}^{\mathrm{a}}$. Locally, the effective
dynamics of geodesic fluctuations is therefore identical to a string
worldsheet theory, with curvature corrections provided by higher-order
terms. The worldsheet boundary conditions then reduce---under the
usual Dirichlet/Neumann split and static gauge---to the target-space
geodesic equation governing the embedding of the boundary curve. In
this sense, the geodesic is recovered as the image of a specific worldsheet
boundary curve, while the worldsheet itself emerges as the two-dimensional
surface parametrizing the entire family of geodesics.

However, this action is incomplete. Its corresponding $\beta$-function
of $g_{\mathrm{ab}}$ does not yield an Einstein's equations that
admits AdS$_{3}$ as a solution, which contradicts the foundational
assumption that $\Omega$ originates from the geodesic length in AdS$_{3}$.
To resolve this inconsistency, the theory must be supplemented by
an antisymmetric tensor field $B_{\mathrm{ab}}$ (the Kalb--Ramond
field) and a constant dilaton $\phi$, in addition to the symmetric
metric $g_{\mathrm{ab}}$. The inclusion of $B_{\mathrm{ab}}$ is
essential for capturing the full gravitational dynamics of the background,
especially those consistent with string theory and AdS$_{3}$ geometry.
Given that entanglement entropy is associated with the spacetime metric
$g_{\mathrm{ab}}$, it is natural to expect that there exists a corresponding
entanglement-related quantity sourced by the Kalb--Ramond field $B_{\mathrm{ab}}$.
To identify this quantity, let us recall the physical role of the
Kalb--Ramond field in string theory. It describes a macroscopic string
carrying Kalb--Ramond charge, propagating in the background and interacting
with background fields \cite{Dabholkar:1990yf,Sen:1992yt,Duff:1994an}.
The corresponding equation of motion is given by:

\begin{equation}
\frac{\partial H^{\mathrm{abc}}}{\partial x^{\mathrm{c}}}=\kappa^{2}j^{\mathrm{ab}}.
\end{equation}

\noindent where the Kalb--Ramond charge density vector is given by
$\overrightarrow{j}^{\mathrm{0}}\equiv j^{\mathrm{0a}}$. This current
is divergenceless. Moreover, since we are working in the AdS$_{3}$/CFT$_{2}$
correspondence, the bulk gravity is weak, i.e. $\kappa^{2}=8\pi G_{N}^{\left(3\right)}\rightarrow0$.
To ensure that these string sources do not backreact on the AdS$_{3}$
metric at leading order, we must include a constant dilaton $\phi=\phi_{0}$
in the action. This yields a weak string coupling $g_{s}=\exp\left(\phi_{0}\right)\ll1$.
Consequently, all three massless sectors of the closed string---the
spacetime metric $g_{\mathrm{ab}}$, the Kalb--Ramond field $B_{\mathrm{ab}}$,
and the dilaton $\phi$---are indispensable for establishing the
correspondence between the string worldsheet theory and entanglement
entropy. On the other hand, in the context of holographic entanglement,
a natural candidate for such a divergenceless vector already exists:
the bit threads, which represent the flow of entanglement entropy
across the RT surface \cite{Freedman:2016zud}. This suggests the
possibility of a direct identification between the string charge density
$\overrightarrow{j}^{0}$ and the bit-thread vector $v$. To test
this proposal, we first highlight three key properties of $\overrightarrow{j}^{0}$:
\begin{enumerate}
\item $\overrightarrow{j}^{0}$ arises from string sources carrying Kalb--Ramond
charge and interacting with background fields. A convenient description
employs the static gauge, in which the current flows along a single
spacetime direction.
\item $\overrightarrow{j}^{0}$ is localized along a line in spacetime,
being proportional to a delta function. To make contact with bit threads,
which are continuous vector fields, we must consider an ensemble of
multiple parallel open strings.
\item At the string endpoints, $\overrightarrow{j}^{0}$ reduces to the
unit normal vector $\overrightarrow{n}$ on the D-brane. This D-brane
plays the role of the RT surface in the bit-thread formulation.
\end{enumerate}
These three features motivate us to place multiple parallel open strings
on a time slice of AdS$_{3}$. The main challenges are (i) how to
impose the static gauge on this time slice, and (ii) how to arrange
the corresponding D-brane along the geodesic representing the RT surface.
Both issues are resolved by considering the BTZ black brane geometry,
which can be obtained by introducing multiple string sources in the
limits $\kappa^{2}=8\pi G_{N}^{\left(3\right)}\rightarrow0$ and $g_{s}\ll1$:

\begin{equation}
dS_{\pm}^{2}=\frac{l_{\mathrm{AdS}}^{2}}{z_{\pm}^{2}}\left(-\left(1-\left(z_{\pm}/b\right)^{2}\right)dt^{2}+\frac{dx_{\pm}^{2}}{1-\left(z_{\pm}/b\right)^{2}}+dz_{\pm}^{2}\right).
\end{equation}

\begin{figure}[h]
\begin{centering}
\includegraphics[scale=0.35]{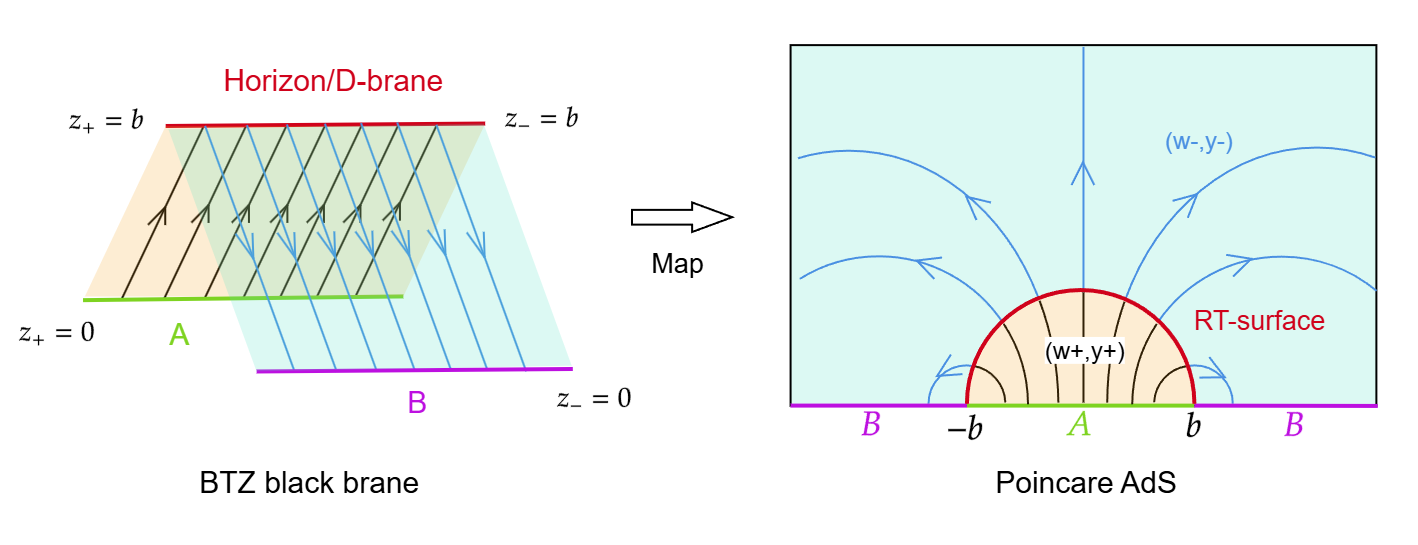}
\par\end{centering}
\caption{\label{fig:intromap}The left-hand panel illustrates two sets of parallel
open strings stretched along the $z_{+}$- and $z_{-}$-directions
in two copies of the BTZ black brane geometry. The red line at $z_{+}=z_{-}=b$
represents the planar horizon, or equivalently the D-brane, since
the open strings terminate on it. When the two sets of parallel open
strings uniformly cover the horizon, the number and locations of strings
on both sides coincide, allowing them to be smoothly connected. After
performing the appropriate coordinate transformation, we obtain the
charge density vector in the familiar Poincaré patch of a time slice
of AdS$_{3}$, as shown in the right-hand panel. In this picture,
the red D-brane naturally extends along the RT surface, making the
correspondence with bit threads manifest. In both panels, the arrows
indicate the direction of the charge density vector.}
\end{figure}

\noindent In this background, we introduce two sets of parallel open
strings: set 1 stretches along the $z_{+}$-direction from $z_{+}=b$
to $0$, while set 2 stretches along the $z_{-}$-direction from $z_{-}=0$
to $b$, as illustrated in figure (\ref{fig:intromap}). 
 The string endpoints (D-branes) naturally sit at $z_{+}=z_{-}=b$,
corresponding to the planar horizon of the BTZ black brane. Because
the parallel strings carry Kalb--Ramond charge, when they uniformly
cover the horizon the two sets of strings can be smoothly connected,
ensuring that the charge density vector $\overrightarrow{j}^{0}$
remains divergenceless despite the presence of the horizon. By performing
an appropriate coordinate transformation \cite{Casini:2011kv,Espindola:2018ozt,Caggioli:2024uza},
these two sets of open strings map onto two regions of the Poincaré
patch of a time slice of AdS$_{3}$: $dS_{\pm}^{2}=\frac{l_{\mathrm{AdS}}^{2}}{w_{\pm}^{2}}\left(dy_{\pm}^{2}+dw_{\pm}^{2}\right)$.
Under this mapping, the D-brane precisely aligns with the geodesic
that defines the RT surface. In other words, this $D$-brane thus
plays the role of the entangling surface, or more precisely, the entangling
brane ($E$-brane) \cite{Donnelly:2016jet}. From this construction,
we can obtain the vector field from string theory

\begin{equation}
v=\frac{l_{\mathrm{AdS}}}{2G_{N}^{\left(3\right)}}\overrightarrow{j}^{0}=\frac{1}{4G_{N}^{\left(3\right)}}\frac{1}{l_{\mathrm{AdS}}}\left(\frac{2bw}{\sqrt{\left(b^{2}-y^{2}-w^{2}\right)^{2}+4b^{2}w^{2}}}\right)^{2}\left(\frac{b^{2}-y^{2}+w^{2}}{2b},\frac{yw}{b}\right),
\end{equation}

\noindent which coincides precisely with the bit threads. This provides
evidence for a possible correspondence between the string worldsheet
and the RT surface: the Kalb--Ramond charge density vector$\overrightarrow{j}^{0}$
maps directly to the bit-thread vector $v$. Furthermore, the D-brane
extends along a geodesic in the AdS$_{3}$ time slice, corresponding
to the RT surface of the entangling region.

This correspondence provides a new framework to calculate and understand
the origin of entanglement entropy. The entanglement entropy can be
computed by counting the number $\mathcal{N}$ of open strings---each
carrying Kalb--Ramond charge---that intersect the entangling surface:

\begin{equation}
S_{\mathrm{vN}}=\frac{L_{\gamma_{A}}}{4G_{N}^{\left(3\right)}}=\frac{l_{\mathrm{AdS}}}{4G_{N}^{\left(3\right)}}\cdot2\mathcal{N}=\frac{c}{3}\ln\left(\frac{2b}{\epsilon}\right),
\end{equation}

\noindent where $2b$ denotes the length of the interval corresponding
to the entangling region $A$, as illustrated in (\ref{fig:intromap}).
This interpretation provides a microscopic, string-theoretic origin
for entanglement entropy, where each charged open string contributes
a discrete unit of information. By open--closed string duality, the
open string charge transforms into a closed string charge. Since the
corresponding closed string winds around the horizon and is non-contractible,
its total charge is non-vanishing and is given by

\begin{equation}
Q=2\pi r,
\end{equation}

\noindent where $r$ is the horizon radius. Consequently, the Bekenstein--Hawking
entropy follows as

\begin{equation}
S_{\mathrm{BH}}=\frac{Q}{4G_{N}^{\left(3\right)}}=\frac{2\pi r}{4G_{N}^{\left(3\right)}},
\end{equation}

\noindent which exactly reproduces the BTZ black hole entropy. This
result verifies the Susskind--Uglum conjecture within the AdS$_{3}$/CFT$_{2}$
framework: open strings contribute to entanglement entropy, while
closed strings contribute to black hole entropy. Furthermore, it demonstrates
that the two-punctured spheres of the original Susskind--Uglum construction
are replaced by hyperbolic cylinders in AdS$_{3}$. The open-- and
closed--string descriptions extend naturally to this cylinder geometry.
In this realization, the entangling surface or horizon, which punctures
the two-sphere in the original setup, is identified with the waist
of the hyperbolic cylinder in AdS$_{3}$, corresponding to the throat
of the wormhole.

Building on this new perspective, we obtain a string-theoretic interpretation
of ER = EPR. In this picture, the closed string winds around the wormhole
horizon, with its winding charge directly accounting for the Bekenstein--Hawking
entropy. As the entanglement entropy between subsystems $A$ and $B$
decreases, the bulk horizon shrinks. When the horizon radius falls
below the string length scale, the winding closed string develops
tachyonic modes. This triggers closed string tachyon condensation,
reducing the total winding charge $Q$ to zero. The vanishing of $Q$
implies that the closed string winding around the compactified dimension
becomes contractible, signaling that the originally connected spacetime
with a finite compact dimension splits into two disconnected components.

All of these observations suggest that open--closed string duality,
ER = EPR, and the Susskind--Uglum conjecture are not merely analogous
but represent manifestations of a deeper equivalence. In all three
frameworks, the transition between dual descriptions---whether in
string theory, spacetime geometry, or entanglement entropy---reflects
the same underlying principle.

Finally, instead of considering multiple parallel open strings carrying
Kalb--Ramond charge---each contributing to the entanglement entropy---we
now restrict to a finite number of strings. In this case, the entanglement
entropy becomes discretized according to the number of strings, providing
strong evidence that the RT surface itself should be quantized. In
the continuum limit, $\mathcal{N}\rightarrow\infty$ with inter-string
spacing $d_{s}\rightarrow0$, the system effectively approaches a
continuous description, and the result reduces to the well-known CFT$_{2}$
entanglement entropy. In our picture, an oriented open string intersected
by the RT surface is divided into two segments, and the entanglement
entropy naturally measures the quantum correlations between these
two parts. This structure closely parallels loop quantum gravity (LQG),
where the entangling surface cuts an oriented Wilson line state $\gamma$
into two parts, and the entanglement entropy between them is quantized,
yielding discrete values. This analogy suggests a potential bridge
between string theory and LQG, realized through the discrete nature
of entanglement entropy. More concretely, by performing a Schmidt
decomposition of both the open string charge density and the LQG Wilson
line, one can split each system into two parts and then recombine
halves to form a new configuration. Since both descriptions share
the same quantized wormhole horizon, and because oriented open strings
can attach to oriented Wilson lines at the same points on the horizon,
the two theories must coincide on the wormhole throat. This observation
resonates with Wall\textquoteright s conjecture, suggesting that the
wormhole throat could provide a possible framework where string theory
and LQG exhibit a deep connection through entanglement between the
string CFT and the LQG CFT \cite{Wall:2023myf}.

This paper is organized as follows. In Section 2, we outline the motivation,
showing how Einstein equations can emerge from CFT$_{2}$ through
string theory and arguing that entanglement entropy possesses a closely
related structure. Section 3 constructs an action directly from the
entanglement entropy of CFT$_{2}$. In the near-coincidence limit
and Riemann normal coordinates, this effective action reduces to the
string worldsheet action in curved spacetime. To ensure that the emergent
gravity is AdS$_{3}$, we incorporate the Kalb--Ramond field and
a constant dilaton into the action. This inclusion introduces the
Kalb--Ramond charge density vector and suggests a new method to compute
entanglement entropy using the antisymmetric field. In Section 4,
we show that bit threads can be exactly reproduced from multiple parallel
strings carrying Kalb--Ramond charge, thereby establishing an explicit
correspondence between the string worldsheet and the RT surface. Section
5 explores new results derived from this worldsheet formulation of
entanglement entropy: the standard entanglement entropy is obtained
from open string charge; through open--closed string duality, the
closed string charge yields the Bekenstein--Hawking entropy of the
BTZ black hole; the ER=EPR proposal, along with Susskind and Uglum\textquoteright s
conjecture, can be unified within a single framework; and the entanglement
entropy, together with the dual RT surface, becomes quantized. Finally,
Section 6 is devoted to conclusions and further discussion.

\section{Motivation}

It is important to highlight an approach demonstrating that gravity
can emerge from a conformal field theory---namely, string theory.
When strings propagate in a curved background, requiring the worldsheet
theory to remain conformally invariant after quantization imposes
additional dynamical equations on the background metric. These equations
serve as the string version of Einstein equations. In the following,
we review how gravity emerges from the string worldsheet \cite{Tong:2009np}
and explore its connection to entanglement entropy, which motivates
our investigation of a new action for the entanglement entropy/RT
surface.

Let us recall the Polyakov action for string theory in conformal gauge:

\begin{equation}
S=\frac{1}{4\pi\alpha^{\prime}}\int d^{2}\sigma g_{\mu\nu}\left(X\right)\partial_{\alpha}X^{\mu}\partial^{\alpha}X^{\nu}.
\end{equation}

\noindent where $g_{\mu\nu}\left(X\right)$ represents the target-space
metric. To determine its dynamics, we first compute the $\beta$-function
associated with the coupling constant $g_{\mu\nu}\left(X\right)$.
If the $\beta$-function vanishes, the quantum version of the Polyakov
action remains conformally invariant. To perform this calculation,
we introduce Riemann normal coordinates around a spacetime point $\bar{X}^{\mu}$,
expanding the coordinates as

\begin{equation}
X^{\mu}\left(\tau,\sigma\right)=\bar{X}^{\mu}+\sqrt{\alpha^{\prime}}\mathbb{X}^{\mu}\left(\tau,\sigma\right),
\end{equation}
where $\mathbb{X}$ is dimensionless. Under this expansion, the metric
takes the form

\begin{equation}
g_{\mu\nu}\left(X\right)=g_{\mu\nu}\left(\bar{X}\right)-\frac{\alpha^{\prime}}{3}R_{\mu\lambda\nu\kappa}\left(\bar{X}\right)\mathbb{X}^{\lambda}\mathbb{X}^{\kappa}+\ldots
\end{equation}

\noindent Substituting this into the Polyakov action yields

\begin{equation}
S=\frac{1}{4\pi}\int d^{2}\sigma\left(g_{\mu\nu}\partial\mathbb{X}^{\mu}\partial\mathbb{X}^{\nu}-\frac{\alpha^{\prime}}{3}R_{\mu\lambda\nu\kappa}\mathbb{X}^{\lambda}\mathbb{X}^{\kappa}\partial\mathbb{X}^{\mu}\partial\mathbb{X}^{\nu}+\ldots\right).\label{eq:polyakov}
\end{equation}
This formulation reveals an interacting quantum field theory, where
divergences arise from one-loop diagrams. To examine this, we recall
the propagator for the scalar field:

\begin{equation}
\left\langle \mathbb{X}^{\lambda}\left(\sigma\right)\mathbb{X}^{\kappa}\left(\sigma^{\prime}\right)\right\rangle =-\frac{1}{2}\delta^{\lambda\kappa}\ln\left|\sigma-\sigma^{\prime}\right|^{2}.
\end{equation}

\noindent At short lengths $\sigma\rightarrow\sigma^{\prime}$, this
propagator exhibits a divergence, which can be regularized using dimensional
regularization with $d=2+\epsilon$, leading to

\begin{equation}
\underset{\sigma\rightarrow\sigma^{\prime}}{\lim}\left\langle \mathbb{X}^{\lambda}\left(\sigma\right)\mathbb{X}^{\kappa}\left(\sigma^{\prime}\right)\right\rangle \rightarrow\frac{\delta^{\lambda\kappa}}{\epsilon}.
\end{equation}

\noindent In this limit, the term 

\begin{equation}
-\alpha^{\prime}R_{\mu\lambda\nu\kappa}\mathbb{X}^{\lambda}\mathbb{X}^{\kappa}\partial\mathbb{X}^{\mu}\partial\mathbb{X}^{\nu}\rightarrow-\frac{\alpha^{\prime}}{\epsilon}R_{\mu\nu}\partial\mathbb{X}^{\mu}\partial\mathbb{X}^{\nu}.
\end{equation}

\noindent This divergence can be canceled by adding the counterterm

\begin{equation}
\frac{\alpha^{\prime}}{\epsilon}R_{\mu\nu}\partial\mathbb{X}^{\mu}\partial\mathbb{X}^{\nu}.
\end{equation}

\noindent which is equivalent to the renormalization

\begin{equation}
\mathbb{X}^{\mu}\rightarrow\mathbb{X}^{\mu}+\frac{\alpha^{\prime}}{\epsilon}R_{\:\nu}^{\mu}\mathbb{X}^{\nu},
\end{equation}

\noindent or, equivalently,

\begin{equation}
g_{\mu\nu}\rightarrow g_{\mu\nu}+\frac{\alpha^{\prime}}{\epsilon}R_{\mu\nu},
\end{equation}

\noindent Therefore, the renormalization procedure introduces a UV
cutoff, implying that the physical quantities of the theory after
quantization generally depend on the energy scale $\epsilon$. This
dependence breaks conformal symmetry. To preserve conformal symmetry
at the quantum level, we must require that the coupling constants---such
as the background metric $G_{\mu\nu}$ ---remain independent of $\epsilon$.
In other words, we demand that the corresponding $\beta$-function
vanishes:

\begin{equation}
\beta\left(g\right)=\alpha^{\prime}R_{\mu\nu}=0.
\end{equation}

\noindent which is precisely the vacuum Einstein equation. 

On the other hand, let us examine how a similar structure arises in
the context of entanglement entropy. To proceed, we introduce Synge's
world function $\Omega$ \cite{Synge:1960}. The good review of Synge's
world function can be found in ref. \cite{Poisson:2011nh}. Consider
a geodesic $z^{\mu}\left(\tau\right)$ that connects two points in
spacetime: the base point $X^{\prime}$ and the field point $X$.
The affine parameter $\tau$ runs from $\tau_{0}$ to $\tau_{1}$,
such that:

\begin{equation}
z\left(\tau_{0}\right)\coloneqq X^{\prime},\qquad z\left(\tau_{1}\right)\coloneqq X.
\end{equation}

\begin{figure}[h]
\begin{centering}
\includegraphics[scale=0.3]{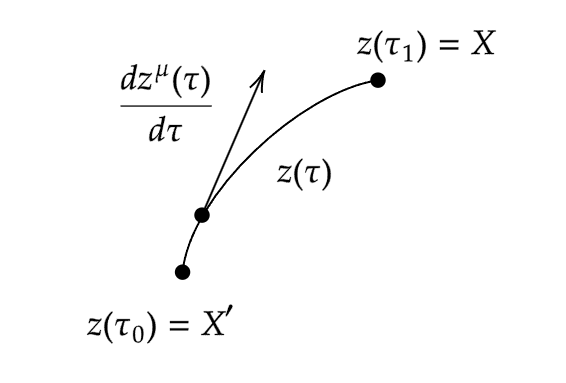}
\par\end{centering}
\caption{\label{fig:synge}This figure illustrates a geodesic parameterized
by $z\left(\tau\right)$, connecting the base point $X^{\prime}$
to the field point $X$. The vector $dz^{\mu}/d\tau$ represents the
tangent vector along the geodesic trajectory.}
\end{figure}

\noindent The tangent vector to the geodesic is $dz^{\mu}\left(\tau\right)/d\tau$,
as illustrated in figure (\ref{fig:synge}). 

\noindent Then, Synge\textquoteright s world function $\Omega\left(X,X^{\prime}\right)$
is a scalar function of these two points, defined as:

\begin{equation}
\Omega\left(X,X^{\prime}\right)=\frac{1}{2}\left(\tau_{1}-\tau_{0}\right)\int_{\tau_{0}}^{\tau_{1}}g_{\mu\nu}\left(z\right)\frac{dz^{\mu}}{d\tau}\frac{dz^{\nu}}{d\tau}d\tau.
\end{equation}

\noindent If we denote the geodesic length between two spacetime points
$X$ and $X^{\prime}$ as $L\left(X,X^{\prime}\right)$, Synge's world
function can be rewritten in terms of the geodesic length as

\begin{equation}
\Omega\left(X,X^{\prime}\right)=\frac{1}{2}L\left(X,X^{\prime}\right)^{2}.
\end{equation}

\noindent The partial derivatives of $\Omega$ with respect to $X$
and $X^{\prime}$ are denoted as $\Omega_{\mu}\coloneqq\partial\Omega/\partial X^{\mu}$
and $\Omega_{\mu^{\prime}}\coloneqq\partial\Omega/\partial X^{\mu^{\prime}}$,
respectively. Furthermore, the covariant derivative of $\Omega_{\mu}$
with respect to $X$ is given by $\Omega_{\mu\nu}\coloneqq\nabla_{\nu}\Omega_{\mu}$,
while its derivative with respect to $X^{\prime}$ is $\Omega_{\mu\nu^{\prime}}\coloneqq\partial_{\nu^{\prime}}\Omega_{\mu}$.
With these definitions, we introduce an important limit: the coincidence
limit, which refers to the limit $X^{\prime}\rightarrow X$, denoted
as $\left[\cdots\right]$. Under this limit, the spacetime metric
can be expressed as

\begin{equation}
g_{\mu\nu}\left(X\right)=-\left[\Omega_{\mu\nu^{\prime}}\right]=-\underset{X^{\prime}\rightarrow X}{\lim}\partial_{\mu}\partial_{\nu^{\prime}}\Omega\left(X,X^{\prime}\right),\qquad g_{\mu\nu}\left(X\right)=\left[\Omega_{\mu\nu}\right]=\underset{X^{\prime}\rightarrow X}{\lim}\nabla_{\nu}\partial_{\mu}\Omega\left(X,X^{\prime}\right).
\end{equation}

\noindent However, taking this limit eliminates valuable information.
Remarkably, it is possible to extract spacetime information without
imposing the coincidence limit. Instead, we can expand this quantity
in the \textbf{near-coincidence} regime $X\rightarrow X^{\prime}$:

\begin{equation}
\Omega_{\mu^{\prime}\nu^{\prime}}\left(X,X^{\prime}\right)=g_{\mu^{\prime}\nu^{\prime}}\left(X^{\prime}\right)-\frac{1}{3}R_{\mu^{\prime}\lambda^{\prime}\nu^{\prime}\kappa^{\prime}}\left(X^{\prime}\right)\Omega^{\lambda^{\prime}}\left(X,X^{\prime}\right)\Omega^{\kappa^{\prime}}\left(X,X^{\prime}\right)+\ldots.\label{eq:expansion}
\end{equation}

\noindent This expansion closely resembles a procedure in string theory.
As previously reviewed, requiring the worldsheet theory to be invariant
under conformal transformations leads to the Einstein equations. Here,
we pose a similar question: Can we impose conformal symmetry to obtain
the Einstein equations? This is a reasonable question, especially
since the quantity $\Omega_{\mu^{\prime}\nu^{\prime}}$ is entirely
derived from CFT data. If $\Omega_{\mu^{\prime}\nu^{\prime}}$, encoding
geometric information such as the geodesic separation between two
spacetime points, emerges from CFT correlators, then requiring conformal
invariance at the quantum level may constrain the background geometry.
The challenge lies in how to impose conformal symmetry on a known
quantity such as the geodesic length. Fortunately, such a possibility
exists. The geodesic length in AdS$_{3}$ can be related to the entanglement
entropy of a CFT$_{2}$ via the RT formula, which is inherently conformally
invariant:

\begin{equation}
S_{\mathrm{vN}}\left(A:B\right)=\frac{L\left(X,X^{\prime}\right)}{4G_{N}}.\label{eq:RT}
\end{equation}

\noindent At this step, the AdS$_{3}$/CFT$_{2}$ correspondence enters
the story. It is worth noting that this requirement also ensures that
the theory remains conformally invariant after quantization. The key
point is that the geodesic length is a classical quantity, while entanglement
entropy arises from the quantum theory. The RT formula serves as a
bridge between these two regimes, relating a classical geometric object
in the bulk to a quantum information-theoretic quantity in the boundary
theory. Therefore, maintaining conformal invariance across both sides
is essential for the consistency of the correspondence. Based on this
argument, if we assume that there exists an action for entanglement
entropy expressed in terms of $\Omega_{\mu^{\prime}\nu^{\prime}}\left(X^{\prime}\right)$,
we can substitute the entanglement entropy expression (\ref{eq:RT})
into the expansion (\ref{eq:expansion}), which leads to the following
form of the $\beta$-function for the spacetime metric $g$:

\begin{equation}
\beta\left(g\right)=\ell^{2}R_{\mu\nu}=\;?
\end{equation}

\noindent However, the precise form of this expression cannot be determined
at this stage, as the full action remains unspecified.

In conclusion, whether in string theory or entanglement entropy, the
same Einstein equations may be emerged through a similar procedure:
\noindent \begin{center}
\begin{tabular}{|c|c|c|}
\hline 
 & $\begin{array}{c}
\\
\\
\end{array}$\textbf{String theory}$\begin{array}{c}
\\
\\
\end{array}$ & $\begin{array}{c}
\\
\\
\end{array}$\textbf{Entanglement entropy}$\begin{array}{c}
\\
\\
\end{array}$\tabularnewline
\hline 
1st: Expansion & $\begin{array}{c}
\\
g_{\mu\nu}=g_{\mu\nu}-\frac{\alpha^{\prime}}{3}R_{\mu\lambda\nu\kappa}\mathbb{\mathbb{X}}^{\lambda}\mathbb{\mathbb{X}}^{\kappa}+\ldots\\
\\
\end{array}$ & $\Omega_{\mu^{\prime}\nu^{\prime}}=g_{\mu^{\prime}\nu^{\prime}}-\frac{1}{3}R_{\mu^{\prime}\lambda^{\prime}\nu^{\prime}\kappa^{\prime}}\Omega^{\lambda^{\prime}}\Omega^{\kappa^{\prime}}+\ldots$\tabularnewline
\hline 
2nd: Impose symmetry & $\begin{array}{c}
\\
\beta\left(g\right)=0\\
\\
\end{array}$ & $L\left(X,X^{\prime}\right)=4G_{N}S_{\mathrm{vN}}$\tabularnewline
\hline 
3rd: Dynamics for $g_{\mu\nu}$ & $\begin{array}{c}
\\
R_{\mu\nu}=0\\
\\
\end{array}$ & $R_{\mu\nu}=\;?$\tabularnewline
\hline 
\end{tabular}
\par\end{center}

\vspace*{3.0ex}

Since, from the perspective of entanglement entropy, we impose conformal
symmetry by requiring the geodesic length to correspond to the entanglement
entropy---arising from the quantization of the CFT---it is equivalent
to selecting the relevant $\beta$-function. This suggests that we
can replace the RT prescription with a $\beta$-function constraint.
If this replacement is made, then entanglement entropy effectively
becomes a manifestation of string theory, allowing us to formulate
its action in direct analogy to the Polyakov action. 

In the remainder of this paper, we will demonstrate that the geodesic
length---or, equivalently, the symmetric spacetime metric $g_{\mu\nu}$---is
insufficient to construct a complete and consistent theory. Consequently,
it is not possible to derive the full gravitational field equations
in the bulk of AdS$_{3}$ using only $g_{\mu\nu}$. The resolution,
as in string theory, is to introduce the antisymmetric Kalb--Ramond
field, which plays a crucial role in ensuring consistency.

\section{The worldsheet action for entanglement entropy}

In this section, we aim to construct an action for the entanglement
entropy that fulfills the following criteria:
\begin{itemize}
\item The action must be formulated in terms of the entanglement entropy
in CFT$_{2}$, and therefore it should depend explicitly on the coordinates
of the CFT$_{2}$.
\item The Einstein equations governing AdS$_{3}$ gravity should be derived
from this action.
\end{itemize}
Before proposing the explicit form of the action, we begin by reviewing
recent developments in the study of entanglement entropy for disjoint
entangling regions. This will provide crucial insight into how entanglement
entropy facilitates the mapping between CFT$_{2}$ coordinates and
the corresponding AdS$_{3}$ bulk geometry.

\begin{figure}[h]
\begin{centering}
\includegraphics[scale=0.35]{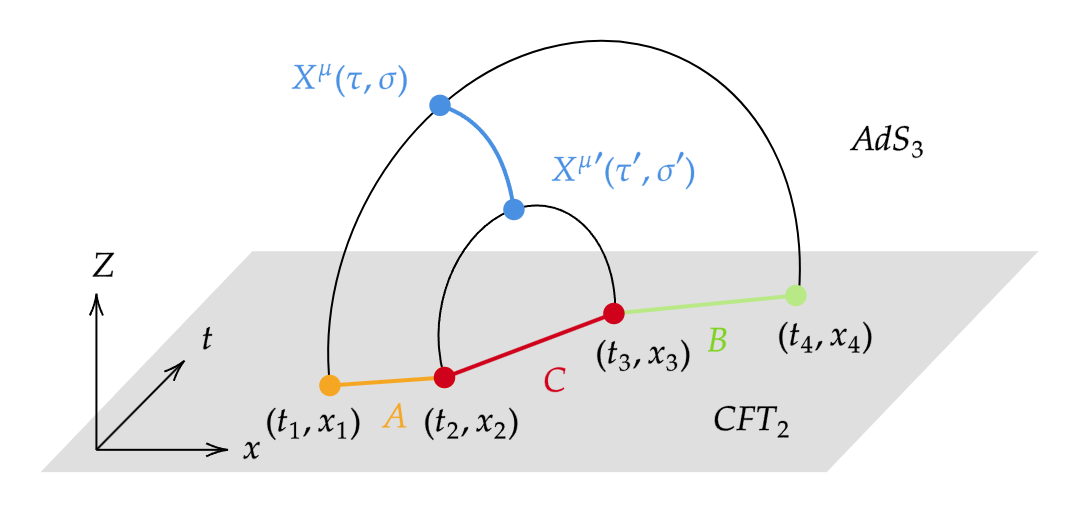}
\par\end{centering}
\caption{\label{fig:geodesic}This figure illustrates the blue geodesic in
AdS$_{3}$, known as the entanglement wedge cross section (EWCS),
whose length is interpreted as the bulk dual of the entanglement entropy
between two disconnected entangling regions $A$ and $B$. This result
implies that the four boundary coordinates $\left(t_{i},x_{i}\right)$
in the CFT$_{2}$ can be conformally mapped to two coordinates $\left(\tau_{i},\sigma_{i}\right)$
in a transformed CFT, which in turn correspond to two bulk points
in AdS$_{3}$.}
\end{figure}

Let us consider two disjoint spacelike intervals, $A=\left[x_{1},x_{2}\right]$
and $B=\left[x_{3},x_{4}\right]$, separated by the region $C=\left[x_{2},x_{3}\right]$.
By performing a boost, these intervals transform into $A=\left[\xi_{1},\xi_{2}\right]=\left[\left(x_{1}+it_{1}\right),\left(x_{2}+it_{2}\right)\right]$
and $B=\left[\xi_{3},\xi_{4}\right]=\left[\left(x_{3}+it_{3}\right),\left(x_{4}+it_{4}\right)\right]$,
as illustrated in figure (\ref{fig:geodesic}). The entanglement entropy
between regions $A$ and $B$ can be computed either through field-theoretic
method \cite{Cardy:2016fqc,Jiang:2024ijx,Jiang:2025tqu} or via the
holographic approach \cite{Basu:2022nds,Wen:2022jxr}:

\begin{equation}
S_{\mathrm{vN}}\left(A:B\right)=\frac{c}{12}\log\left(\frac{1+\sqrt{\eta}}{1-\sqrt{\eta}}\right)+\frac{c}{12}\log\left(\frac{1+\sqrt{\bar{\eta}}}{1-\sqrt{\bar{\eta}}}\right),\label{eq:EE in cross ratio}
\end{equation}

\noindent where the cross-ratio is defined by

\noindent 
\begin{equation}
\eta=\frac{\xi_{21}\xi_{43}}{\xi_{31}\xi_{42}},\qquad\xi_{ij}\equiv\xi_{i}-\xi_{j},\qquad\xi_{i}=x_{i}+it_{i}.
\end{equation}

\noindent Note $\left(t_{i},x_{i}\right)$ are ordinary coordinate
of CFT$_{2}$. We may regroup these coordinates into effective bulk
coordinates $\left(T,X,Z\right)$. The specific regrouping transformation
is given in \cite{Jiang:2025tqu,Basu:2022nds,Wen:2022jxr}, while
the general transformation was derived in \cite{Jiang:2024hjz} and
reads

\begin{eqnarray}
X+iT & = & \xi_{1}+\frac{\left|\xi_{13}\right|\left|\xi_{14}\right|}{\left|\xi_{12}\right|\left|\xi_{13}\right|+\left|\xi_{34}\right|\left|\xi_{24}\right|}\left|\xi_{21}\right|,\nonumber \\
X^{\prime}+iT^{\prime} & = & \xi_{2}+\frac{\left|\xi_{23}\right|\left|\xi_{24}\right|}{\left|\xi_{13}\right|\left|\xi_{34}\right|+\left|\xi_{12}\right|\left|\xi_{24}\right|}\left|\xi_{21}\right|,\nonumber \\
Z & = & \frac{\left|\xi_{14}\right|\sqrt{\left|\xi_{12}\right|\left|\xi_{13}\right|\left|\xi_{34}\right|\left|\xi_{24}\right|}}{\left|\xi_{12}\right|\left|\xi_{13}\right|+\left|\xi_{34}\right|\left|\xi_{24}\right|},\nonumber \\
Z^{\prime} & = & \frac{\left|\xi_{23}\right|\sqrt{\left|\xi_{12}\right|\left|\xi_{13}\right|\left|\xi_{34}\right|\left|\xi_{24}\right|}}{\left|\xi_{13}\right|\left|\xi_{34}\right|+\left|\xi_{12}\right|\left|\xi_{24}\right|}.
\end{eqnarray}

\noindent Then, the cross-ratio becomes

\begin{equation}
\eta=\left|\frac{\xi_{21}\xi_{43}}{\xi_{31}\xi_{42}}\right|=\frac{\triangle Z^{2}+\triangle X^{2}-\triangle T^{2}}{\left(Z+Z^{\prime}\right)^{2}+\triangle X^{2}-\triangle T^{2}}.
\end{equation}

\noindent The entanglement entropy then takes the familiar form:

\begin{equation}
S_{\mathrm{vN}}\left(A:B\right)=\frac{c}{6}\cosh^{-1}\left(1+\frac{\triangle Z^{2}+\triangle X^{2}-\triangle T^{2}}{2ZZ^{\prime}}\right).
\end{equation}

\noindent This result implies that in CFT$_{2}$, the four-points
$\left(t_{1},x_{1}\right)$, $\left(t_{2},x_{2}\right)$, $\left(t_{3},x_{3}\right)$,
and $\left(t_{4},x_{4}\right)$ can be effectively regrouped into
two new points,

\begin{equation}
\left(\tau\left(t_{i},x_{i}\right),\sigma\left(t_{i},x_{i}\right)\right),\qquad\mathrm{and}\qquad\left(\tau^{\prime}\left(t_{i},x_{i}\right),\sigma^{\prime}\left(t_{i},x_{i}\right)\right),\qquad i=1,\ldots,4.
\end{equation}

\begin{figure}[h]
\begin{centering}
\includegraphics[scale=0.3]{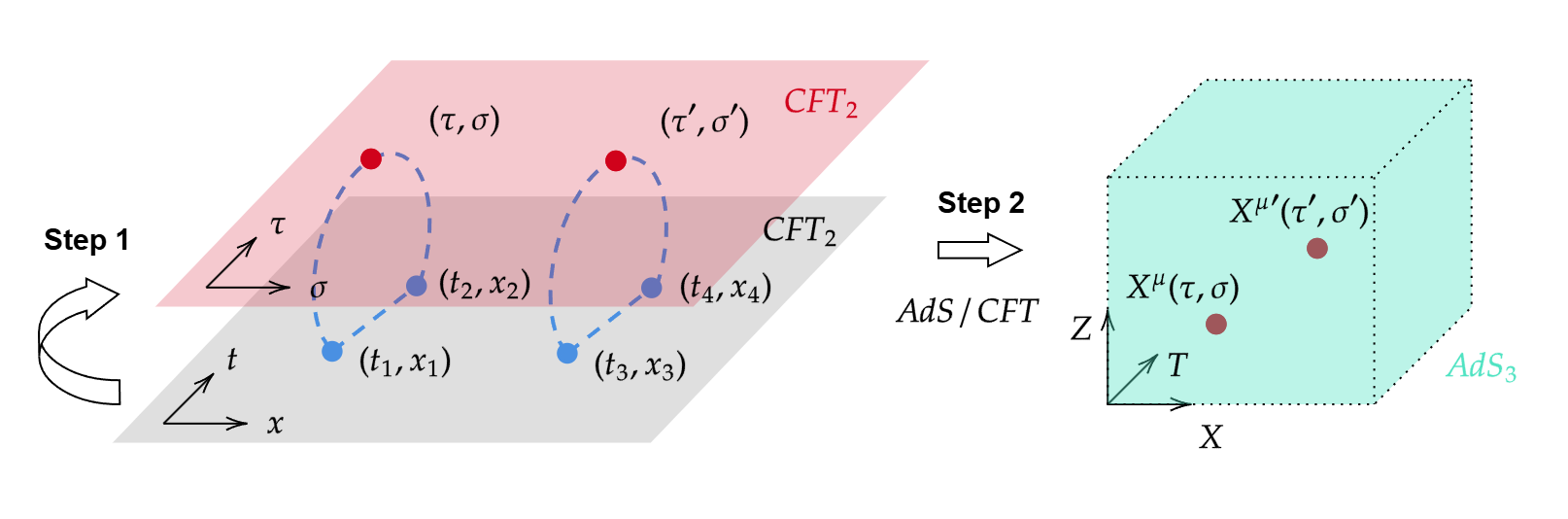}
\par\end{centering}
\caption{\label{fig:map}\textbf{Step 1}: The coordinates of four points $\left(t_{i},x_{i}\right)$
in CFT$_{2}$ can be reorganized into two effective points $\left(\tau_{i},\sigma_{i}\right)$
through a conformal transformation. \textbf{Step 2}: These two points
$\left(\tau_{i},\sigma_{i}\right)$ are then related to bulk coordinates
in AdS$_{3}$ via the AdS$_{3}$/CFT$_{2}$ correspondence. }
\end{figure}

\noindent Through the AdS$_{3}$/CFT$_{2}$ correspondence---specifically
the RT formula---\textbf{these CFT$_{2}$ coordinates $\left(\tau,\sigma\right)$
can be mapped to the spacetime coordinates $X^{\mu}$ of AdS$_{3}$
bulk}, corresponding to the two boundary points of the geodesic as
illustrated in figure (\ref{fig:map}). These boundary points in the
bulk are then parameterized by

\begin{equation}
X^{\mu}\left(\tau,\sigma\right)\qquad\mathrm{and}\qquad X^{\mu\prime}\left(\tau^{\prime},\sigma^{\prime}\right),
\end{equation}

\noindent where $\mu=0,1,2$. Beyond the example discussed here, a
well-known instance of this correspondence arises in the study of
conformal blocks involving internal twist operators in CFT$_{2}$
\cite{Hijano:2015qja,Hijano:2015zsa,Hirai:2018jwy}. In such cases,
the two bulk points---interpreted as endpoints of a bulk-to-bulk
propagator in AdS$_{3}$ ---emerge naturally from the global four-point
conformal blocks of the boundary theory.

This observation bears a close resemblance to the string worldsheet,
where the CFT$_{2}$ is embedded into the target space $X^{\mu}$.
The key difference, however, is that when computing entanglement entropy
for two entangling regions, the coordinates $\left(\tau,\sigma\right)$
and $\left(\tau^{\prime},\sigma^{\prime}\right)$ are fixed. Consequently,
the dual spacetime points $X^{\mu}$ and $X^{\mu\prime}$ in AdS are
uniquely determined. This means that the geodesic connecting these
two points can be described using only a single affine parameter,
rather than requiring a full two-dimensional worldsheet description.
In this sense, it is not a genuine worldsheet. This is consistent
with the general understanding that a geodesic is a fixed geometric
feature of the background, unlike a string that propagates dynamically
in spacetime.

\begin{figure}[h]
\begin{centering}
\includegraphics[scale=0.25]{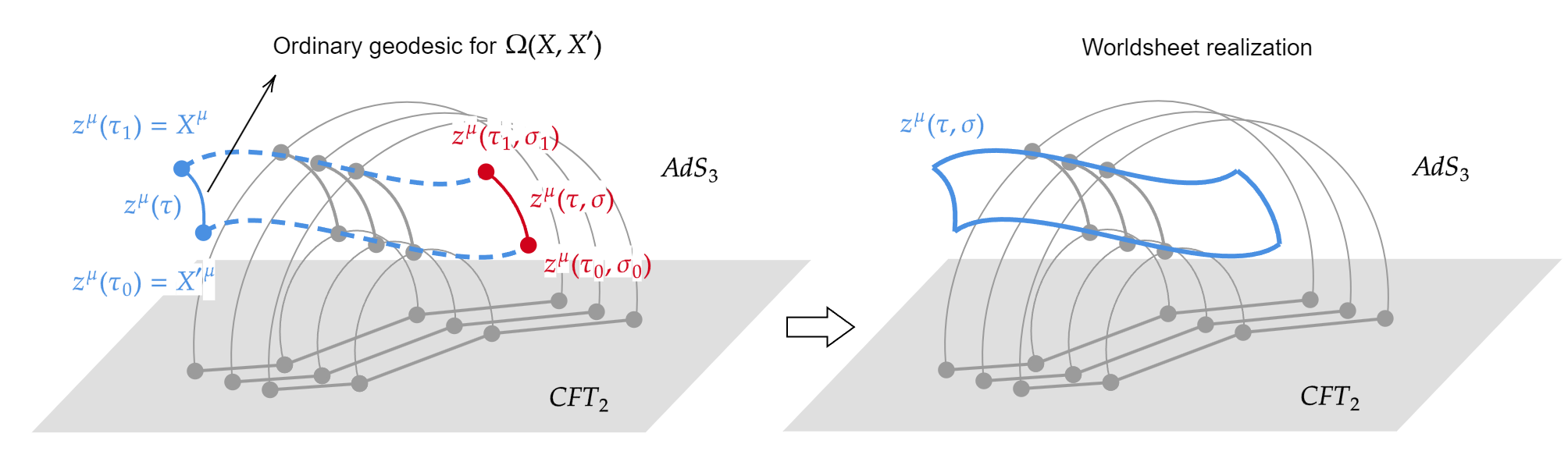}
\par\end{centering}
\caption{\label{fig:move_geodesic}This figure illustrates how a geodesic effectively
sweeps out a two-dimensional surface in AdS$_{3}$. In the traditional
holographic prescription, the entanglement entropy of CFT$_{2}$ is
dual to the geodesic length in AdS$_{3}$, with the geodesic endpoints
anchored at the boundaries of the chosen entangling region in the
CFT. As the entangling region is deformed---through translation,
rotation, stretching, or shrinking---the corresponding bulk geodesic
is continuously deformed as well. Collectively, this family of geodesics
sweeps out a two-dimensional surface in the bulk, which is naturally
analogous to a string worldsheet.}
\end{figure}

The situation changes if the entangling regions are varied in time.
In this case, both the entanglement entropy and its dual geodesic
length evolve accordingly. As a result, the geodesic effectively sweeps
through AdS$_{3}$, as illustrated in figure (\ref{fig:move_geodesic}).
The evolving trajectory generates a two-dimensional surface that can
be described by an embedding function $X^{\mu}\left(\tau,\sigma\right)$.
In other words, By promoting the two bulk endpoints $X^{\mu}$ and
$X^{\mu\prime}$ to fields $X^{\mu}\left(\tau,\sigma\right)$, $X^{\mu\prime}\left(\tau^{\prime},\sigma^{\prime}\right)$
we obtain a two-parameter family of geodesics. The union (envelope)
of this family defines an emergent two-dimensional surface in the
bulk. Motivated by the analogy with string theory, one can then construct
an action for this surface---analogous to the Polyakov action---that
captures the dynamics of entanglement entropy as the entangling regions
change.

\vspace*{5.0ex}

Based on these considerations, we now derive key insights for constructing
the action, drawing inspiration from how actions are formulated in
string theory:
\begin{enumerate}
\item Integrating the minimal area $\int dA$. 
\item Introducing prefactors such as $\ell^{2}$ to ensure the action is
dimensionless.
\item Parameterizing the action with appropriate variables, e.g., $\left(\tau,\sigma\right)$
for strings.
\item Building the action entirely from quantities associated with entanglement
entropy.
\end{enumerate}
In our setup, a natural area term arises:

\begin{equation}
2\Omega=g_{\mu\nu}\left(X\right)\Omega^{\mu}\Omega^{\nu}=\Omega_{\mu\nu}\Omega^{\mu}\Omega^{\nu},
\end{equation}

\noindent which is determined by the entanglement entropy (\ref{eq:EE in cross ratio})
of CFT$_{2}$. By introducing two parameters $\left(\tau,\sigma\right)$,
the single worldline can be promoted to a two-dimensional extended
object (worldsheet). This does not require redefining $\Omega$; instead,
the derivatives of $\Omega$ with respect to the two coordinates can
be reinterpreted as spanning tangent directions on the effective worldsheet.
This structure suggests an action closely analogous to that of the
string worldsheet:

\noindent\fbox{\begin{minipage}[t]{1\columnwidth - 2\fboxsep - 2\fboxrule}%
\begin{equation}
S_{E}=\frac{1}{\ell^{2}}\int d^{2}\sigma^{\prime}\Omega_{\mu^{\prime}\nu^{\prime}}\left(X,X^{\prime}\right)\partial_{\alpha^{\prime}}\Omega^{\mu^{\prime}}\left(X,X^{\prime}\right)\partial^{\alpha^{\prime}}\Omega^{\nu^{\prime}}\left(X,X^{\prime}\right),\label{eq:E theory action}
\end{equation}
\end{minipage}}

\medskip{}

\noindent where $\partial_{\alpha^{\prime}}=\left(\partial_{\tau^{\prime}},\partial_{\sigma^{\prime}}\right)$
acts only on the base point $X^{\prime}\left(\tau^{\prime},\sigma^{\prime}\right)$.
Both $X$ and $X^{\prime}$ are boundary-determined bulk points fixed
by the entangling regions in the CFT. Since the two endpoints $X$
and $X^{\prime}$ of any geodesic are interchangeable, there exists
a corresponding set of derivatives $\partial_{\alpha}=\left(\partial_{\tau},\partial_{\sigma}\right)$
acting on the field point $X\left(\tau,\sigma\right)$. As a result,
we obtain a non-local action constructed from the entanglement entropy,
which can describe the physics at either of the spacetime coordinates
$X$ or $X^{\prime}$. To verify the consistency of this action, one
must check whether it reproduces the entanglement entropy or geodesic
length. Before this verification, however, let us examine its physical
meaning. Here, instead of treating $X^{\mu}\left(\tau,\sigma\right)$
as the embedding of a true string, the embedding here is constructed
out of geodesics in AdS that are dual to entangling intervals in the
CFT$_{2}$. The Synge's world function $\Omega\left(X,X^{\prime}\right)$
(half the squared geodesic distance) replaces the role of $X^{\mu}$
as the basic dynamical variable. Consequently, the action is built
directly from entanglement entropy data (through $\Omega$) rather
than fundamental fields. The coordinates $\left(\tau,\sigma\right)$
parametrize the varying entangling regions. As these regions evolve,
their dual geodesics sweep out a two-dimensional surface in AdS. This
surface is not a fundamental string worldsheet but an effective entanglement
worldsheet: a bookkeeping device that tracks the dynamics of entanglement
entropy. The action (\ref{eq:E theory action}) therefore encodes
the dynamics of geodesic lengths (equivalently entanglement entropies)
as the entangling regions change. The quadratic form in derivatives
of $\Omega^{\mu^{\prime}}$ resembles the Polyakov action, ensuring
that the dynamics is governed by variations of geodesic distance. 

To gain intuition, we localize the action to a single spacetime point
by introducing a suitable parametrization and expanding around the
\textbf{near-coincidence limit} $X\rightarrow X^{\prime}$ \footnote{Note that we may also perform the expansion in the opposite limit
$X^{\prime}\rightarrow X$, as given in (\ref{eq:expansion}). In
this case, the action is modified to

\[
S_{E}=\frac{1}{\ell^{2}}\int d^{2}\sigma\Omega_{\mu\nu}\left(X,X^{\prime}\right)\partial_{\alpha}\Omega^{\mu}\left(X,X^{\prime}\right)\partial^{\alpha}\Omega^{\nu}\left(X,X^{\prime}\right).
\]

\noindent This action is equivalent to (\ref{eq:E theory action}),
since the two endpoints are interchangeable.}: 

\begin{equation}
\Omega_{\mu^{\prime}\nu^{\prime}}\left(X,X^{\prime}\right)=g_{\mu^{\prime}\nu^{\prime}}\left(X^{\prime}\right)-\frac{1}{3}R_{\mu^{\prime}\lambda^{\prime}\nu^{\prime}\kappa^{\prime}}\left(X^{\prime}\right)\Omega^{\lambda^{\prime}}\left(X,X^{\prime}\right)\Omega^{\kappa^{\prime}}\left(X,X^{\prime}\right)+\ldots.
\end{equation}

\noindent Substituting this expansion, the action (\ref{eq:E theory action})
becomes

\begin{equation}
S_{E}=\frac{1}{\ell^{2}}\int d^{2}\sigma^{\prime}\left(g_{\mu^{\prime}\nu^{\prime}}\partial_{\alpha^{\prime}}\Omega^{\mu^{\prime}}\partial^{\alpha^{\prime}}\Omega^{\nu^{\prime}}-\frac{1}{3}R_{\mu^{\prime}\lambda^{\prime}\nu^{\prime}\kappa^{\prime}}\Omega^{\lambda^{\prime}}\Omega^{\kappa^{\prime}}\partial_{\alpha^{\prime}}\Omega^{\mu^{\prime}}\partial^{\alpha^{\prime}}\Omega^{\nu^{\prime}}+\ldots\right),\label{eq:near coincidence}
\end{equation}

\noindent where $\ell$ can be interpreted as the minimal length scale
in the expansion. Next, we introduce a tetrad $e_{\mathrm{a}}^{\mu^{\prime}}\left(X^{\prime}\right)$
at a fixed base point $X^{\prime}$, allowing a neighboring point
to be parameterized as

\begin{equation}
\ell\mathbb{X}^{\mathrm{a}}=-e_{\mu^{\prime}}^{\mathrm{a}}\left(X^{\prime}\right)\Omega^{\mu^{\prime}}\left(X,X^{\prime}\right),\label{eq:RNCe}
\end{equation}

\noindent where the dual tetrad is defined as $e_{\mu^{\prime}}^{\mathrm{a}}=\eta^{\mathrm{ab}}g_{\mu^{\prime}\nu^{\prime}}e_{\mathrm{b}}^{\nu^{\prime}}$
and $\mathbb{X}$ is dimensionless. These coordinates $\mathbb{X}^{\mathrm{a}}$
correspond to the Riemann normal coordinates (RNC). In this framework,
the identity for $\Omega$ can be rewritten as

\begin{equation}
2\Omega\left(X,X^{\prime}\right)=g_{\mu^{\prime}\nu^{\prime}}\Omega^{\mu^{\prime}}\Omega^{\nu^{\prime}}=g_{\mathrm{ab}}e_{\mu^{\prime}}^{\mathrm{a}}e_{\nu^{\prime}}^{\mathrm{b}}\Omega^{\mu^{\prime}}\Omega^{\nu^{\prime}}=\ell^{2}g_{\mathrm{ab}}\mathbb{X}^{\mathrm{a}}\mathbb{X}^{\mathrm{b}},
\end{equation}

\noindent where $g_{\mathrm{ab}}\mathbb{X}^{\mathrm{a}}\mathbb{X}^{\mathrm{b}}$
denotes the squared geodesic distance between the base point $X^{\prime}$
and the field point $X$. A key feature of RNC is that they shift
the base point $X^{\prime}$ to the origin, so that $\mathbb{X}^{\mathrm{a}}=0$.
On the other hand, since the Synge's world function $\Omega\left(X,X^{\prime}\right)$
depends on both the base point $X^{\prime}$ and the field point $X$,
and since these points can be parameterized by the boundary CFT$_{2}$
coordinate $\left(\tau,\sigma;\tau^{\prime},\sigma^{\prime}\right)$,
we can write $\Omega\left(X,X^{\prime}\right)=\Omega\left(X\left(\tau,\sigma\right),X^{\prime}\left(\tau^{\prime},\sigma^{\prime}\right)\right)$.
Consequently, the RNC coordinates $\mathbb{X}^{\mathrm{a}}$ can also
be expressed in terms of the CFT$_{2}$ coordinate $\left(\tau,\sigma;\tau^{\prime},\sigma^{\prime}\right)$
as

\begin{equation}
\ell\mathbb{X}^{\mathrm{a}}\left(\tau,\sigma;\tau^{\prime},\sigma^{\prime}\right)=-e_{\mu^{\prime}}^{\mathrm{a}}\left(X^{\prime}\left(\tau^{\prime},\sigma^{\prime}\right)\right)\Omega^{\mu^{\prime}}\left(X\left(\tau,\sigma\right),X^{\prime}\left(\tau^{\prime},\sigma^{\prime}\right)\right),
\end{equation}

\noindent In other words, in RNC, we have

\begin{equation}
2\Omega\left(X,X^{\prime}\right)=\ell^{2}g_{\mathrm{ab}}\mathbb{X}^{\mathrm{a}}\left(\tau,\sigma;\tau^{\prime},\sigma^{\prime}\right)\mathbb{X}^{\mathrm{b}}\left(\tau,\sigma;\tau^{\prime},\sigma^{\prime}\right).
\end{equation}

\noindent Based on these results, the near-coincidence limit of the
action (\ref{eq:near coincidence}) can be rewritten in the RNC by
using (\ref{eq:RNCe}), which takes the form

\begin{equation}
S_{E}=\int d^{2}\sigma^{\prime}\left(g_{\mathrm{ab}}\partial_{\alpha^{\prime}}\mathbb{X}^{\mathrm{a}}\partial^{\alpha^{\prime}}\mathbb{X}^{\mathrm{b}}-\frac{\ell^{2}}{3}R_{\mathrm{acbd}}\mathbb{X}^{\mathrm{c}}\mathbb{X}^{\mathrm{d}}\partial_{\alpha^{\prime}}\mathbb{X}^{\mathrm{a}}\partial^{\alpha^{\prime}}\mathbb{X}^{\mathrm{b}}+\ldots\right),
\end{equation}

\noindent where $R_{\mathrm{acbd}}\coloneqq R_{\alpha^{\prime}\gamma^{\prime}\beta^{\prime}\delta^{\prime}}e_{\mathrm{a}}^{\alpha^{\prime}}e_{\mathrm{c}}^{\gamma^{\prime}}e_{\mathrm{b}}^{\beta^{\prime}}e_{\mathrm{d}}^{\delta^{\prime}}$.
At the beginning of our construction, the two arguments of Synge\textquoteright s
world function were kept distinct, $X=X\left(\tau,\sigma\right)$
and $X^{\prime}=X^{\prime}\left(\tau^{\prime},\sigma^{\prime}\right)$.
because the primed coordinates $\left(\tau^{\prime},\sigma^{\prime}\right)$
serve as the integration variables in the action. In the near-coincidence
limit and in RNC, we identify the two worldsheet points, $\left(\tau,\sigma\right)=\left(\tau^{\prime},\sigma^{\prime}\right)$.
After this identification, the primed coordinates no longer carry
any dynamical meaning and become dummy integration variables. They
can therefore be relabeled without affecting the integral. Following
standard practice in bitensor expansions, we rename $\left(\tau^{\prime},\sigma^{\prime}\right)$
as $\left(\tau,\sigma\right)$ and drop all primes on tensors and
derivatives. This procedure does not lose any information; it simply
reflects that the coincidence-limit expansion is performed about a
single worldsheet point. We thus obtain the final form of the near-coincidence
action,

\begin{equation}
S_{E}=\int d^{2}\sigma\left(g_{\mathrm{ab}}\partial_{\alpha}\mathbb{X}^{\mathrm{a}}\partial^{\alpha}\mathbb{X}^{\mathrm{b}}-\frac{\ell^{2}}{3}R_{\mathrm{acbd}}\mathbb{X}^{\mathrm{c}}\mathbb{X}^{\mathrm{d}}\partial_{\alpha}\mathbb{X}^{\mathrm{a}}\partial^{\alpha}\mathbb{X}^{\mathrm{b}}+\ldots\right),\label{eq:local action}
\end{equation}

\noindent This coincides with the form of the Polyakov action in a
curved background (\ref{eq:polyakov}) if we identify $\ell=\sqrt{\alpha^{\prime}}$.
Therefore, in the near-coincidence limit, the expansion yields a local
quadratic action reminiscent of the Polyakov kinetic term for the
small separation field $\hat{X}^{\mathrm{a}}$. Thus, locally the
effective dynamics of geodesic fluctuations is identical to a string
worldsheet theory (with curvature corrections given by higher terms
in the expansion). A subtle question arises: does the geodesic itself
belong to the string worldsheet, given that the action takes a worldsheet-like
form? The answer is nuanced. Although the geodesic evolves like a
worldsheet, the geodesic itself is not literally a string but sweeps
out an effective worldsheet when followed over time. This suggests
the geodesic may be interpreted as analogous to a D1-brane trajectory
(an \textquoteleft entangling brane\textquoteright ), with open strings
attached to it. In this interpretation, the action effectively describes
the dynamics of this brane-like object, rather than a fundamental
string. We stress that this identification is conjectural: it is not
derived from conventional string theory, but rather suggested by the
structural similarities between geodesic dynamics, entanglement entropy,
and string worldsheet theory. In this view, the action describes the
dynamics of this effective brane, rather than a fundamental string.
This reinterpretation not only clarifies the geometric picture but
also provides a more precise route for deriving entanglement entropy---or
equivalently, the geodesic equation---from the action.

To understand this D-brane configuration, we attempt to reproduce
the geodesic equation from the action (\ref{eq:E theory action}).
We begin by varying the action with respect to the base point $X^{\rho^{\prime}}$,
such that $X^{\rho^{\prime}}\rightarrow X^{\rho^{\prime}}+\delta X^{\rho^{\prime}}$,
while keeping $X^{\prime}$ fixed. The variation of $\Omega_{\mu^{\prime}\nu^{\prime}}$
yields $\delta\Omega_{\mu^{\prime}\nu^{\prime}}=\partial_{\rho^{\prime}}\Omega_{\mu^{\prime}\nu^{\prime}}\delta X^{\rho^{\prime}}$,
and that of $\partial_{\alpha^{\prime}}\Omega^{\mu^{\prime}}$ gives
$\delta\Omega_{\mu^{\prime}\nu^{\prime}}=\partial_{\rho^{\prime}}\Omega_{\mu^{\prime}\nu^{\prime}}\delta X^{\rho^{\prime}}$.
Substituting these variations, the action becomes

\begin{eqnarray}
\delta S_{E} & = & \frac{1}{\ell^{2}}\int d^{2}\sigma^{\prime}\left[\partial_{\rho^{\prime}}\Omega_{\mu^{\prime}\nu^{\prime}}\partial_{\alpha^{\prime}}\Omega^{\mu^{\prime}}\partial^{\alpha^{\prime}}\Omega^{\nu^{\prime}}-2\partial_{\alpha^{\prime}}\left(\Omega_{\mu^{\prime}\nu^{\prime}}\partial^{\alpha^{\prime}}\Omega^{\nu^{\prime}}\right)\partial_{\rho^{\prime}}\Omega^{\mu^{\prime}}\right]\delta X^{\rho^{\prime}}\nonumber \\
 &  & +\frac{1}{\ell^{2}}\int_{\partial\Sigma}d\tau^{\prime}\left[2n_{\alpha^{\prime}}\Omega_{\mu^{\prime}\nu^{\prime}}\partial_{\rho^{\prime}}\Omega^{\mu^{\prime}}\partial^{\alpha^{\prime}}\Omega^{\nu^{\prime}}\delta X^{\rho^{\prime}}\right]_{\sigma=\mathrm{bdy}},
\end{eqnarray}

\noindent where $n_{\alpha}$ is normal to the boundary $\partial\Sigma$.
Requiring the action to be stationary under arbitrary $\delta X^{\rho}$
yields the EOM and boundary condition:

\begin{eqnarray}
\mathrm{EOM:} &  & \partial_{\rho^{\prime}}\Omega_{\mu^{\prime}\nu^{\prime}}\partial_{\alpha^{\prime}}\Omega^{\mu^{\prime}}\partial^{\alpha^{\prime}}\Omega^{\nu^{\prime}}-2\partial_{\alpha^{\prime}}\left(\Omega_{\mu^{\prime}\nu^{\prime}}\partial^{\alpha^{\prime}}\Omega^{\nu^{\prime}}\right)\partial_{\rho^{\prime}}\Omega^{\mu^{\prime}}=0,\nonumber \\
\mathrm{B.C.:} &  & \left.n_{\alpha^{\prime}}\Omega_{\mu^{\prime}\nu^{\prime}}\partial_{\rho^{\prime}}\Omega^{\mu^{\prime}}\partial^{\alpha^{\prime}}\Omega^{\nu^{\prime}}\delta X^{\rho^{\prime}}\right|_{\partial\Sigma}=0,\label{eq:EOM ori}
\end{eqnarray}

\noindent In the coincidence limit, these reduce to

\begin{equation}
\left[\Omega_{\mu^{\prime}\nu^{\prime}}\right]=g_{\mu\nu},\qquad\left[\Omega_{\mu^{\prime}}\right]=0,\qquad\left[\partial_{\alpha^{\prime}}\Omega^{\mu^{\prime}}\right]=\partial_{\alpha}X^{\mu},
\end{equation}

\noindent which leads to the standard string-theoretic form:

\begin{eqnarray}
\mathrm{EOM:} &  & \partial_{\alpha}\partial^{\alpha}X^{\lambda}+\Gamma_{\mu\nu}^{\lambda}\partial_{\alpha}X^{\mu}\partial^{\alpha}X^{\nu}=0,\nonumber \\
\mathrm{B.C.:} &  & \left.g_{\mu\nu}n_{\alpha}\partial^{\alpha}X^{\mu}\delta X^{\nu}\right|_{\partial\Sigma}=0.\label{eq:EOM limi}
\end{eqnarray}

\begin{figure}[h]
\begin{centering}
\includegraphics[scale=0.4]{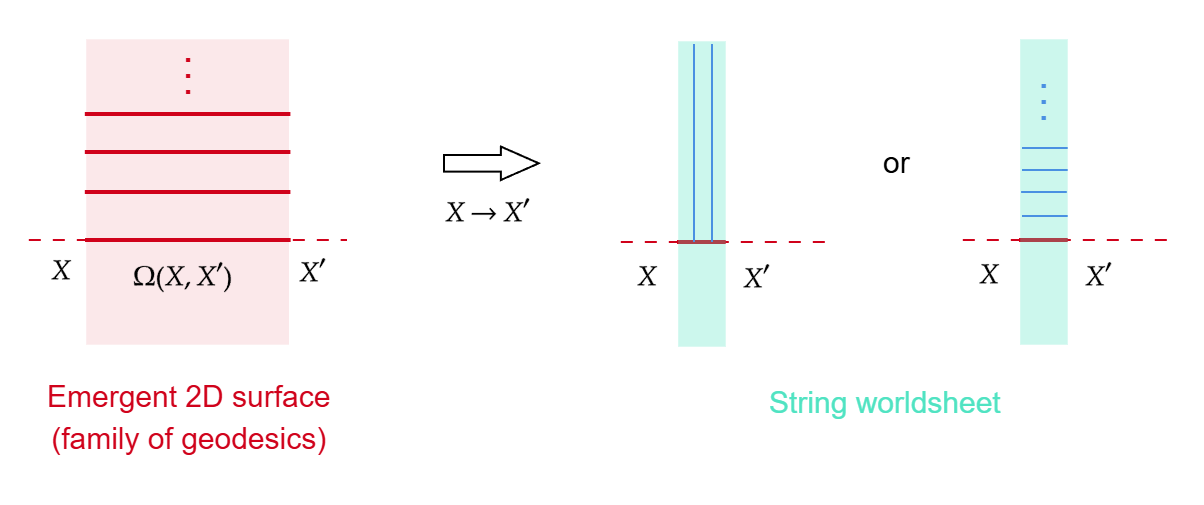}
\par\end{centering}
\caption{\label{fig:local}The left panel shows how a family of geodesics between
$X$ and $X^{\prime}$ (red lines) sweeps out an effective two-dimensional
surface. The length of each geodesic corresponds to the bulk dual
of a CFT entanglement entropy. In the near-coincidence limit $X\rightarrow X^{\prime}$,
this surface acquires the structure of a string worldsheet, with the
Polyakov action emerging naturally. The right panel illustrates this
worldsheet. Depending on the gauge choice, the open string (blue lines)
can fluctuate along horizontal or vertical directions, while its endpoints
obey the geodesic equation for a point-like particle.}
\end{figure}

\noindent To obtain the endpoint geodesic equation, we fix the static
gauge so that the boundary is parameterized by $\tau$. We then decompose
the indices $\mu$ into tangent directions $a$ (along the D-brane/RT
surface) and normal directions $i$ (transverse), or vice versa. For
the normal directions we impose Dirichlet boundary conditions, $\left.\delta X^{i}\right|_{\partial\Sigma}=0$,
while for the tangential directions we impose Neumann boundary conditions,
$\left.g_{\mu\nu}n_{\alpha}\partial^{\alpha}X^{a}\right|_{\partial\Sigma}=0$.
Applying these conditions, the EOM at the boundary becomes

\begin{equation}
\ddot{X}^{\mathrm{a}}+\Gamma_{\textrm{bc}}^{\mathrm{a}}\dot{X}^{\mathrm{b}}\dot{X}^{\mathrm{c}}=0.\label{eq:geodesic EOM bounadry}
\end{equation}

\noindent which is precisely the geodesic equation for the endpoints
along the D-brane. This result can be interpreted as shown in figure
(\ref{fig:local}). In our setup, the spacetime point $X$, representing
a geodesic endpoint, is parameterized by two-dimensional coordinates
$\left(\tau,\sigma\right)$. Consequently, the geodesic sweeps out
a two-dimensional surface whose dynamics are governed by the non-local
EOM and boundary conditions (\ref{eq:EOM ori}). Taking the near-coincidence
limit $X\rightarrow X^{\prime}$ reduces these to the local EOM and
boundary conditions (\ref{eq:EOM limi}), in full agreement with string
theory. Under the appropriate boundary conditions, the boundary $\partial\Sigma$
obeys the geodesic EOM (\ref{eq:geodesic EOM bounadry}). Thus, the
proposed action (\ref{eq:E theory action}) indeed contains the expected
geodesic dynamics.

Nevertheless, this geodesic equation is general and applies to arbitrary
geodesics. At this stage, we cannot restrict the geodesics to reside
exclusively in AdS$_{3}$. Moreover, it remains unclear how the gravitational
field equations for the AdS$_{3}$ gravity emerge from this action.
These issues contradict our initial premise that $\Omega$, and hence
the action, originates from entanglement entropy in the AdS$_{3}$/CFT$_{2}$
framework. In other words, the action constructed thus far is incomplete.

To obtain the complete action, let us recall how Einstein equations
emerge from string worldsheet theory. Since our proposed action reduces
to the Polyakov action in the near-coincidence limit, the same logic
applies here. In conventional string theory, preservation of conformal
symmetry at the quantum level requires the vanishing of the $\beta$-function
for $g_{\mathrm{ab}}$. For instance, when the Polyakov action includes
only the spacetime metric, the corresponding $\beta$-function is

\begin{equation}
\beta\left(g\right)=\ell^{2}R_{\mathrm{ab}}.
\end{equation}

\noindent In our case, if we demand that $\Omega$ corresponds to
the squared geodesic length in the bulk of AdS$_{3}$, then the metric
$g_{\mathrm{ab}}$ must be the AdS$_{3}$ metric. Substituting this
into the $\beta$-function gives

\begin{equation}
\beta\left(g\right)=\ell^{2}R_{\mathrm{ab}}=-2\frac{\ell^{2}}{l_{\mathrm{AdS}}^{2}}g_{\mathrm{ab}}\neq0.
\end{equation}

\noindent This $\beta$-function is manifestly non-vanishing, which
implies that it is impossible to derive the gravitational field equations
of AdS$_{3}$ only from the RT surface and its corresponding $\Omega$.
To achieve a vanishing $\beta$-function when evaluated on the AdS$_{3}$
background, the only resolution is to include an antisymmetric field
from the outset. This leads us to modify the action as follows: 

\noindent\fbox{\begin{minipage}[t]{1\columnwidth - 2\fboxsep - 2\fboxrule}%
\begin{equation}
S_{E}=\frac{1}{\ell^{2}}\int d^{2}\sigma^{\prime}\sqrt{\gamma}\left(\gamma^{\alpha^{\prime}\beta^{\prime}}\Omega_{\mu^{\prime}\nu^{\prime}}\left(X,X^{\prime}\right)+i\epsilon^{\alpha^{\prime}\beta^{\prime}}\mathcal{B}_{\mu^{\prime}\nu^{\prime}}\left(X,X^{\prime}\right)\right)\partial_{\alpha^{\prime}}\Omega^{\mu^{\prime}}\left(X,X^{\prime}\right)\partial_{\beta^{\prime}}\Omega^{\nu^{\prime}}\left(X,X^{\prime}\right).\label{eq:E theory action-1}
\end{equation}
\end{minipage}}

\medskip{}

\noindent where $\mathcal{B}_{\mu^{\prime}\nu^{\prime}}\left(X,X^{\prime}\right)$
becomes

\begin{eqnarray}
\mathcal{B}_{\mu^{\prime}\nu^{\prime}}\left(X^{\prime}\right) & = & B_{\mu^{\prime}\nu^{\prime}}\left(X^{\prime}\right)+\nabla_{\lambda^{\prime}}B_{\mu^{\prime}\nu^{\prime}}\Omega^{\lambda^{\prime}}\left(X,X^{\prime}\right)+\frac{1}{2}\left(\nabla_{\alpha^{\prime}}\nabla_{\beta^{\prime}}B_{\mu^{\prime}\nu^{\prime}}\left(X^{\prime}\right)-\frac{1}{3}R_{\;\alpha^{\prime}\mu^{\prime}\beta^{\prime}}^{\lambda^{\prime}}B_{\lambda^{\prime}\nu^{\prime}}\right.\nonumber \\
 &  & \left.-\frac{1}{3}R_{\;\alpha^{\prime}\nu^{\prime}\beta^{\prime}}^{\lambda^{\prime}}B_{\lambda^{\prime}\mu^{\prime}}\right)\Omega^{\alpha^{\prime}}\left(X,X^{\prime}\right)\Omega^{\beta^{\prime}}\left(X,X^{\prime}\right)+\ldots,
\end{eqnarray}

\noindent in the \textbf{near-coincidence limit}. Moreover, in addition
to the inclusion of the Kalb--Ramond field, it is also necessary
to incorporate the dilaton field $\phi$ and the cosmological constant
term $4/k$, as is standard in non-critical string theory. The role
of the dilaton will be explained in Section 4. All massless sectors
of the closed string thus become indispensable in establishing the
correspondence between the string worldsheet theory and entanglement
entropy. Employing RNC and using the relation $B_{\mathrm{ab}}\coloneqq B_{\alpha^{\prime}\beta^{\prime}}e_{\mathrm{a}}^{\alpha^{\prime}}e_{\mathrm{b}}^{\beta^{\prime}}$,
the vanishing of the $\beta$-functions for $g_{\mathrm{ab}}$, $B_{\mathrm{ab}}$
and $\phi$ leads to the following gravitational field equations: 

\begin{eqnarray}
R_{\mathrm{ab}}+2\nabla_{\mathrm{a}}\nabla_{\mathrm{b}}\phi-\frac{1}{4}H_{\mathrm{acd}}H_{\mathrm{b}}^{\:\mathrm{cd}} & = & 0,\nonumber \\
\nabla^{\mathrm{a}}\left(e^{-2\phi}H_{\mathrm{abc}}\right) & = & 0,\nonumber \\
4\nabla^{2}\phi-4\left(\nabla\phi\right)^{2}+R-\frac{1}{12}H^{2}+\frac{4}{k} & = & 0.
\end{eqnarray}

\noindent This system now admits a solution that is fully consistent
with our entanglement entropy setup, with $g_{\mathrm{ab}}$ given
by the AdS$_{3}$ metric, the Kalb--Ramond field strength $H_{\mathrm{abc}}=\frac{2}{l_{AdS}}\epsilon_{\mathrm{abc}}$,
a constant dilaton $\phi$ and $k=l_{AdS}^{2}$ \cite{Horowitz:1993jc}.
Under this solution, the connection $\Gamma_{\mathrm{bc}}^{\mathrm{a}}$
introduced in equation (\ref{eq:geodesic EOM bounadry}) is modified
to include the contribution from the Kalb--Ramond field: $\Gamma_{\mathrm{bc}}^{\mathrm{a}}=\Gamma_{\mathrm{bc}}^{\mathrm{a}}-\frac{1}{2}H_{\mathrm{bc}}^{\mathrm{a}}$.
With this modification, the geodesic equation reproduces the motion
of a point-like particle in AdS$_{3}$. 

In summary, we argue that our proposed action provides a valid description
of the CFT$_{2}$ entanglement entropy for the following reasons:
\begin{enumerate}
\item The action is fully constructed from the entanglement entropy itself,
$\Omega\left(X,X^{\prime}\right)\sim S_{\mathrm{vN}}\left(A:B\right)^{2}$.
\item The equations of motion and boundary conditions derived from this
action naturally include the geodesic equation in AdS$_{3}$.
\item By imposing conformal symmetry, the action yields the Einstein equations
of AdS$_{3}$ gravity in the near-coincidence limit.
\end{enumerate}
Morover,
\begin{itemize}
\item This result suggests a framework for understanding the relation between
Klein--Gordon equation + renormalization group (RG) equations and
general relativity (GR). At the level of the effective worldsheet
theory, the dynamics are governed by a QFT whose equations of motion
reduce to the two-dimensional Klein--Gordon equation for bosonic
fields. Expanding around the near-coincidence limit and employing
RNC introduces interaction terms associated with background fields,
which in turn manifest as RG equations. The requirement of conformal
symmetry then enforces the emergence of the Einstein equations in
AdS$_{3}$. 
\end{itemize}
Finally, it is worth recalling previous developments of string theory
on AdS$_{3}$ in the context of the AdS/CFT correspondence. The subject
originates from the duality between type IIB superstring theory on
AdS$_{3}$$\times$S$^{3}$$\times$T$^{4}$ and a CFT in the moduli
space of the symmetric orbifold of T$^{4}$ \cite{Maldacena:1997re,Seiberg:1999xz}.
This duality was subsequently refined to identify the tensionless
limit of strings with minimal pure NS--NS flux with the symmetric
orbifold Sym$^{N}$(T$^{4}$) \cite{Gaberdiel:2018rqv,Eberhardt:2018ouy,Eberhardt:2019ywk}
, and later extended beyond the tensionless regime through deformations
by exactly marginal operators \cite{Eberhardt:2021vsx,Balthazar:2021xeh,Martinec:2022ofs,Yu:2024kxr,Yu:2025qnw}.
On the worldsheet, strings in AdS$_{3}$ are described by the $SL\left(2,\mathbb{R}\right)$
WZW model \cite{Kutasov:1999xu,Giveon:1998ns,deBoer:1998gyt}, where
the inclusion of spectrally flowed representations of the affine algebra
is essential to account for the physics of long strings \cite{Maldacena:2000hw,Maldacena:2000kv,Maldacena:2001km}.

In this work, however, our focus is not on the quantization of strings
in AdS$_{3}$ or on the construction of their precise CFT duals. Instead,
our analysis originates entirely from the study of entanglement entropy,
from which a Polyakov-like action emerges locally as a consequence
of the underlying entanglement dynamics, as discussed above.

In the following sections, we will present further evidence demonstrating
that this new action and framework provide novel methods and lead
to new results for the study of entanglement entropy.

\section{Kalb-Ramond charge density as bit threads}

If our derivation of the worldsheet action for entanglement entropy
is correct, there must exist an additional, previously unrecognized
antisymmetric field in the entanglement theory, which leads to the
consistent gravitational field equations for the AdS$_{3}$. Just
as the RT surface gives rise to a symmetric field---the spacetime
metric $g_{\mathrm{ab}}$---there should exist a bulk quantity on
equal footing that sources an antisymmetric field $B_{\mathrm{ab}}$:

\begin{equation}
g_{\mathrm{ab}}\iff B_{\mathrm{ab}}.
\end{equation}

\noindent To identify this quantity, let us recall the physical meaning
of the Kalb--Ramond field $B_{\mathrm{ab}}$. In string theory, the
worldsheet term $B_{\mathrm{ab}}\partial_{\alpha}X^{\mathrm{a}}\partial_{\beta}X^{\mathrm{b}}$
describes the coupling of the string to the antisymmetric Kalb--Ramond
field. This is the natural generalization of the electric coupling
between a point-like particle and the electromagnetic field. In this
sense, the string carries an electric Kalb--Ramond charge. To see
this, recall the definition of electric current in electromagnetism:

\begin{equation}
\frac{\partial F^{\mathrm{ab}}}{\partial x^{\mathrm{b}}}=j^{\mathrm{a}},
\end{equation}

\noindent where $j^{0}$ denotes the electric charge density. Analogously,
the field strength $H^{\mathrm{abc}}$ of the Kalb--Ramond field
satisfies:

\begin{equation}
\frac{\partial H^{\mathrm{abc}}}{\partial x^{\mathrm{c}}}=\kappa^{2}j^{\mathrm{ab}},
\end{equation}

\noindent The Kalb--Ramond charge density vector is given by $\overrightarrow{j}^{0}\equiv j^{\mathrm{0a}}$.
This vector possesses two important properties that guide us in identifying
its entanglement entropy analogue:
\begin{enumerate}
\item The charge density vector is tangent to the string.
\item The string charge density is a divergenceless vector $\nabla\cdot\overrightarrow{j}^{0}=0$.
\end{enumerate}
In the context of entanglement entropy, there exists a closely analogous
quantity: a divergenceless vector field defined on the RT surface,
encoding information about entanglement---namely, the bit threads
\cite{Freedman:2016zud}. These are flow lines in AdS spacetime that
begin and end on the boundary and thread through the bulk, providing
a dual description of entanglement entropy. Previous studies exploring
the connection between bit threads and Einstein\textquoteright s equations
can be found in \cite{Agon:2020mvu,Agon:2021tia}. For example, consider
the flux through region $A$ which is defined as

\begin{equation}
\int_{A}v\equiv\int_{A}\sqrt{h}n_{\mu}v^{\mu}=\int_{\gamma_{A}}v,
\end{equation}

\noindent where $h$ is the induced metric on $A$, and $n_{\mu}$
is the unit normal to $A$. The equality follows from the fact that
$A$ and $\gamma_{A}$ are homologous; thus, the net flux entering
the bulk volume between them must vanish. Moreover, the flux is bounded
by the area of the RT surface:

\begin{equation}
\int_{A}v=\int_{\gamma_{A}}v\leq C\,\mathrm{area}\left(\gamma_{A}\right).
\end{equation}

\noindent The entanglement entropy is given by

\begin{equation}
S_{\mathrm{vN}}=\frac{\mathrm{area}\left(\gamma_{A}\right)}{4G_{N}}=\underset{v}{\mathrm{max}}\int_{A}v,
\end{equation}

\noindent with $C=1/4G_{N}^{\left(3\right)}$. In short, in an oriented
and bounded Riemannian manifold with a positive constant $C=1/4G_{N}^{\left(3\right)}$,
the bit thread configuration is described by a divergenceless, bounded
vector field $v$. This vector field $v$, referred to as a flow,
satisfies the following conditions:
\begin{enumerate}
\item The field $v$ is perpendicular to the RT surface. If $v\left(A\right)$
denotes the set of flow lines sourced from a boundary region $A$,
then its flux equals the area of the RT surface $\gamma_{A}$ homologous
to $A$. 
\item The field is divergenceless and bounded: $\nabla\cdot v=0$, $\left|v\right|\leq1/4G_{N}^{\left(3\right)}$.
\end{enumerate}
Therefore, it is straightforward to verify the equivalence between
$\overrightarrow{j}^{0}$ and $v$, as both vector fields are defined
on the same time slice of AdS$_{3}$ and obey identical divergenceless
conditions: $\nabla\cdot\overrightarrow{j}^{0}=0$ and $\nabla\cdot v=0$.
If these two vectors coincide, it provides a natural identification
between the worldsheet theory and entanglement entropy: one may relate
the string charge density vector $\overrightarrow{j}^{0}$ to the
bit thread vector field $v$. On the other hand, in this configuration,
the Kalb--Ramond charge density vector $\overrightarrow{j}^{0}$
lies along the string in the direction of increasing $\sigma$, reflecting
the string\textquoteright s orientation. Moreover, it is important
to note that the reparameterization invariance of the string worldsheet
is broken in the presence of the Kalb--Ramond field. The reason is
that the corresponding term $B_{\mathrm{ab}}\partial_{\alpha}X^{\mathrm{a}}\partial_{\beta}X^{\mathrm{b}}$
in the worldsheet action changes sign under a reversal of the worldsheet
coordinates $\left(\tau,\sigma\right)$. As a result, strings become
oriented when the Kalb--Ramond field is included, and the charge
density vector $\overrightarrow{j}^{0}$ is aligned with the $\sigma$-direction,
rather than the $\tau$-direction. This orientation plays a crucial
role in identifying the dual of the charge density vector in the entanglement
entropy picture.

On the other hand, beyond establishing the relation between $\overrightarrow{j}^{0}$
and $v$, identifying the counterpart of the RT surface on the string
worldsheet is more subtle. This is because there is no natural prescription
to extract geodesics on the worldsheet embedded in AdS$_{3}$. However,
an important object in string theory that has not yet been incorporated
into our discussion---the D-brane---offers a consistent resolution.
Together with the previously proposed correspondence between $\overrightarrow{j}^{0}$
and the bit thread vector $v$, we identify three guiding constraints
that link the worldsheet geometry to the RT surface: 
\begin{enumerate}
\item The geodesic on the worldsheet must coincide with the RT surface.
This is because the function $\Omega\left(X,X^{\prime}\right)$ in
the action originates from the entanglement entropy, which is itself
determined by the RT surface.
\item Bit threads are orthogonal to the RT surface. Consequently, the string
charge density vector $\overrightarrow{j}^{0}$ (or equivalently,
the orientation of open strings) must also be orthogonal to the corresponding
geodesic on the worldsheet.
\item The string charge density vector $\overrightarrow{j}^{0}$ must be
divergenceless and cannot terminate at any point in spacetime. This
implies that strings must either be closed or extend infinitely; in
particular, open strings cannot end on a boundary unless a mechanism
exists to maintain current continuity.
\end{enumerate}
These conditions suggest a natural geometric configuration: while
it may be difficult to isolate geodesics on the worldsheet, we can
instead fix the boundary of the worldsheet to follow the geodesic
in AdS. In this picture, the boundary of the worldsheet---the D-brane---is
identified with the RT surface. Furthermore, to satisfy the divergenceless
condition, open strings cannot simply end on the D-brane. A consistent
solution is to consider pairs of open strings attached to opposite
sides of the D-brane. When the number of open strings reaches a maximal
density and all strings are orthogonal to the D-brane, strings from
the two sides can smoothly connect across the D-brane, ensuring the
continuity of $\overrightarrow{j}^{0}$ and the absence of any current
sink or source. 

In the following subsections, we will demonstrate how to approach
this set-up and explicitly derive the bit thread vector field $v$
from the string charge density $\overrightarrow{j}^{0}$.

\subsection{Set-up}

\begin{figure}[h]
\begin{centering}
\includegraphics[scale=0.33]{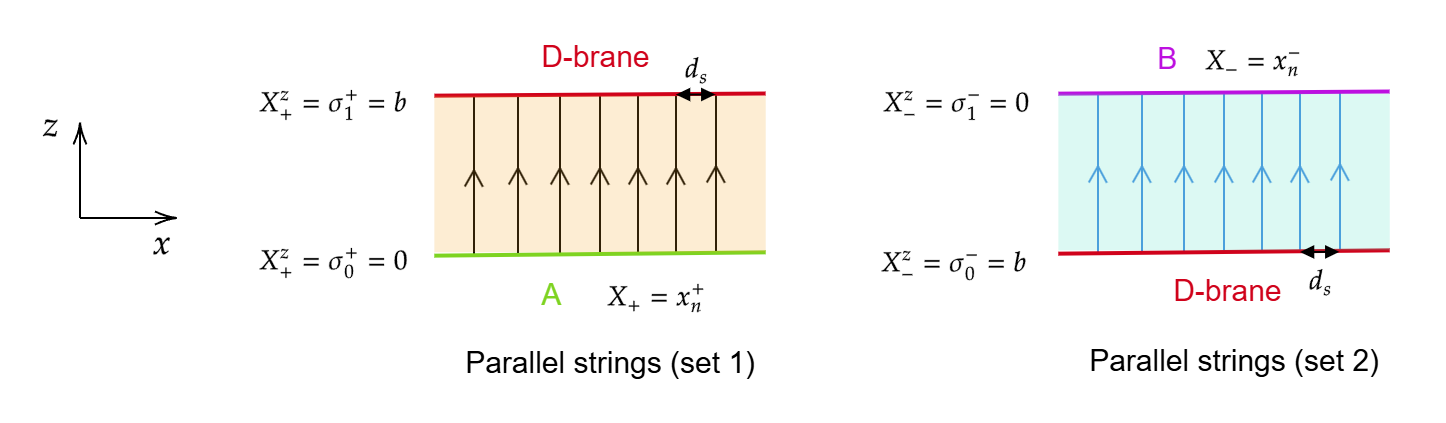}
\par\end{centering}
\caption{\label{fig:2world}This figure depicts two sets of parallel open strings
in static gauge, stretched along the $z$-direction with an inter-string
separation of $d_{s}$. The green, purple, and red solid lines represent
D-branes, with the red D-brane serving as the central focus of this
discussion, corresponding to the RT surface. The open string endpoints
satisfy Dirichlet boundary conditions in the $z$-direction. These
two sets of parallel strings can be embedded in a time slice of BTZ
black brane and smoothly joined along the red D-brane. Crucially,
the strings asymptotically approach the green and purple D-branes
at the AdS boundary, ensuring that the string charge density vector
remains divergenceless throughout the configuration. In the following
discussion, we will map these two sets of parallel strings onto two
regions of the Poincaré patch of a time slice of AdS$_{3}$, as illustrated
in figure (\ref{fig:Dbrane2in1}). This mapping will make the underlying
physics more transparent.}
\end{figure}

To make this configuration explicit, we consider two sets of stationary,
parallel open strings stretched along the $z_{+}$- and $z_{-}$-directions
in three-dimensional flat spacetime $\left(t_{\pm},z_{\pm},x_{\pm}\right)$,
as illustrated in figure (\ref{fig:2world}). The first set of parallel
strings, parameterized by $\left(\tau^{+},\sigma^{+}\right)$, and
the second set, parameterized by $\left(\tau^{-},\sigma^{-}\right)$,
are embedded in two-dimensional target spaces:

\begin{equation}
X_{-}^{t}\left(\tau^{-},\sigma^{-}\right),X_{-}^{z}\left(\tau^{-},\sigma^{-}\right),X_{-}^{x}\left(\tau^{-},\sigma^{-}\right),\qquad X_{+}^{t}\left(\tau^{+},\sigma^{+}\right),X_{+}^{z}\left(\tau^{+},\sigma^{+}\right),X_{+}^{z}\left(\tau^{+},\sigma^{+}\right),
\end{equation}

\noindent respectively. We adopt the static gauge by setting

\begin{equation}
X_{\pm}^{t}\left(\tau^{\pm},\sigma^{\pm}\right)=\tau^{\pm},\qquad X_{\pm}^{z}\left(\tau^{\pm},\sigma^{\pm}\right)=\sigma^{\pm},\qquad X_{\pm}^{x}\left(\tau^{\pm},\sigma^{\pm}\right)=x_{n}^{\pm},
\end{equation}

\noindent where the constants $x_{n}^{\pm}$ specify the endpoints
of the strings at the $n$-th position along the $x$-direction. The
separation between two adjacent strings is denoted by $d_{s}$. At
the string endpoints, we impose Dirichlet boundary conditions: $\left.\partial_{\tau^{\pm}}X_{\pm}^{z}\right|_{0,b}=0$.

We now embed the two worldsheets into a time slice of BTZ black brane,
equipped with the metric

\begin{equation}
dS_{\pm}^{2}=\frac{l_{\mathrm{AdS}}^{2}}{z_{\pm}^{2}}\left(-\left(1-\left(z_{\pm}/b\right)^{2}\right)dt^{2}+\frac{dx_{\pm}^{2}}{1-\left(z_{\pm}/b\right)^{2}}+dz_{\pm}^{2}\right),\label{eq:BTZ brane}
\end{equation}

\noindent where $t_{+}=t_{-}=t$, $x_{\pm}\in\mathbb{R}$ and $z_{\pm}\in\left(0,b\right]$.
In this coordinate system, the planar horizon---which we identify
with the RT surface---is located at $z_{\pm}=b$. This is the junction
of the two strings, and continuity across this junction implies $x_{+}=x_{-}=x$
along the sewing line. From the boundary conditions, we see that open
strings on either side of the D-brane can be smoothly joined. Specifically,
at $z_{\pm}=b$, the boundary conditions

\begin{equation}
\left.\partial_{\tau}X_{+}^{z}\right|_{b}=\left.\partial_{\tau}X_{-}^{z}\right|_{b}=0,\qquad\left.\partial_{\sigma^{+}}X_{+}^{z}\right|_{b}=\left.\partial_{\sigma^{-}}X_{-}^{z}\right|_{b}=0,\qquad X_{+}^{x}=X_{-}^{x}=x_{n},\label{eq:bc}
\end{equation}

\noindent ensure that both the temporal and spatial derivatives of
the string embedding functions match at the junction. To study whether
the string charge density vectors $\overrightarrow{j}_{\pm}^{0}$
from the two open strings can be connected at the D-brane, we recall
the general expression for the string current $j^{\mu\nu}$ for a
single string:

\begin{equation}
j^{\mu\nu}\left(x\right)=\int d\sigma^{2}\delta^{D}\left(x-X\left(\tau,\sigma\right)\right)\left(\partial_{\tau}X^{\mu}\partial_{\sigma}X^{\nu}-\partial_{\tau}X^{\nu}\partial_{\sigma}X^{\mu}\right),
\end{equation}

\noindent This expression is defined at spacetime points $x$ that
lie on the worldsheet. In the static gauge, where $X^{0}=\tau$, the
spatial components simplify to

\begin{equation}
\overrightarrow{j}^{0}\left(x\right)=\int d\sigma\delta\left(\overrightarrow{x}-\overrightarrow{X}\left(\tau,\sigma\right)\right)\partial_{\sigma}\overrightarrow{X}\left(\tau,\sigma\right),\label{eq:definition of j}
\end{equation}

\noindent where $\overrightarrow{X}\left(\tau,\sigma\right)\equiv\left(X^{z}\left(\tau,\sigma\right),X^{x}\left(\tau,\sigma\right)\right)$
denotes the embedding of the string in the $\left(z,x\right)$-plane,
and $x$ refers to the spatial components of the target spacetime.
Returning to our configuration, we evaluate $\overrightarrow{j}_{\pm}^{0}\left(x\right)$
for the two strings. Since the setup lies within a time slice of AdS$_{3}$,
the string charge density vectors are maximized along geodesics. As
a result, for each spacetime point $x$ on the D-brane, there exists
a unique $\overrightarrow{j}_{\pm}^{0}\left(x\right)$ vector flowing
into or out of the D-brane. Specifically, at every point $x_{n}$
on the D-brane, there is a pair of charge density vectors that converge.
The continuity conditions previously discussed ensure that the spatial
derivatives $\partial_{\sigma}\overrightarrow{X}$ are matched across
the junction. Consequently, the total string charge density vectors
can be smoothly connected as

\begin{eqnarray}
\overrightarrow{j}^{0}\left(x\right) & = & \overrightarrow{j}_{+}^{0}\left(x_{+}\right)\cup\overrightarrow{j}_{-}^{0}\left(x_{-}\right)\nonumber \\
 & = & \int d\sigma^{+}\delta\left(\overrightarrow{x}-\overrightarrow{X}_{+}\left(\tau,\sigma^{+}\right)\right)\partial_{\sigma^{+}}\overrightarrow{X}_{+}\left(\tau,\sigma^{+}\right)\nonumber \\
 &  & +\int d\sigma^{-}\delta\left(\overrightarrow{x}-\overrightarrow{X}_{-}\left(\tau,\sigma^{-}\right)\right)\partial_{\sigma^{-}}\overrightarrow{X}_{-}\left(\tau,\sigma^{-}\right)\nonumber \\
 & = & \int d\sigma\delta\left(\overrightarrow{x}-\overrightarrow{X}\left(\tau,\sigma\right)\right)\partial_{\sigma}\overrightarrow{X}\left(\tau,\sigma\right),
\end{eqnarray}

\noindent ensuring the global divergencelessness of $\overrightarrow{j}^{0}$
in the full D-brane--included configuration. In other words, the
endpoints of two open strings can join to form a new open string.
This mechanism has also been used to study open-string dynamics in
the closed-string vacuum and to explain how closed strings emerge
as weakly coupled excitations \cite{Kleban:2000pf}. In such a scenario,
if open strings are charged under a $U\left(1\right)$ gauge field
that becomes confined, the endpoints of an open string are connected
by an electric flux tube. The combination of the open string and this
flux tube then effectively forms a closed string. Moreover, this result
is also consistent with the edge-mode constraints on the two sides
of the horizon in the low-energy description, where the number and
positions of the string endpoints must match on both sides \cite{Ahmadain:2022eso}.
These endpoints are effectively frozen on the horizon due to the infinite
gravitational time dilation.

To relate this setup to the bit thread picture, we perform the coordinate
transformation introduced in \cite{Casini:2011kv,Espindola:2018ozt}
and further developed in \cite{Caggioli:2024uza}:

\begin{equation}
y_{\pm}=\frac{b\sinh\left(x_{\pm}/b\right)}{\cosh\left(x_{\pm}/b\right)\pm\sqrt{1-\left(z_{\pm}/b\right)^{2}}},\qquad w_{\pm}=\frac{z_{\pm}}{\cosh\left(x_{\pm}/b\right)\pm\sqrt{1-\left(z_{\pm}/b\right)^{2}}}.\label{eq:brangcor}
\end{equation}

\noindent This transforms the BTZ black brane metric (\ref{eq:BTZ brane})
into the familiar Poincaré patch of time slice of AdS$_{3}$:

\begin{equation}
dS_{\pm}^{2}=\frac{l_{\mathrm{AdS}}^{2}}{w_{\pm}^{2}}\left(dy_{\pm}^{2}+dw_{\pm}^{2}\right).
\end{equation}

\begin{figure}[h]
\begin{centering}
\includegraphics[scale=0.35]{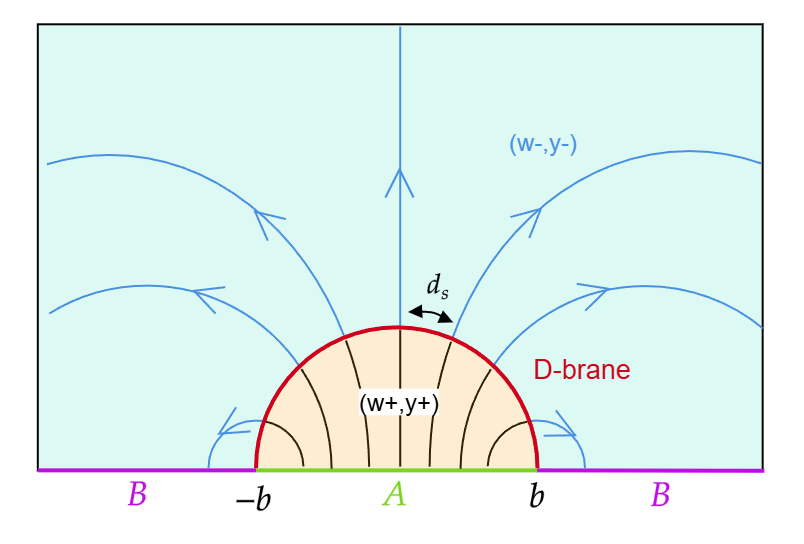}
\par\end{centering}
\caption{\label{fig:Dbrane2in1}Two open string worldsheets, originally illustrated
in figure (\ref{fig:2world}), are mapped into the Poincaré patch
of a time slice of AdS$_{3}$. The green, purple, and red lines denote
D-branes, with the red D-brane corresponding to the RT surface. Parallel
strings 1 (orange) is described by the region $y_{+}^{2}+w_{+}^{2}<b^{2}$,
while parallel strings 2 (cyan) occupies the region $y_{-}^{2}+w_{-}^{2}>b^{2}$,
where $b$ is the radius of the semicircle. The string charge density
vectors from both worldsheets are smoothly connected at the red D-brane,
ensuring the combined vector field remains divergenceless.}
\end{figure}

\noindent Under this map, the D-brane/RT surface is represented by
the semicircle $y_{\pm}^{2}+w_{\pm}^{2}=b^{2}$. Parallel strings
(set 1) occupies the interior orange region $y_{+}^{2}+w_{+}^{2}<b^{2}$,
and Parallel strings (set 2) covers the exterior cyan region $y_{-}^{2}+w_{-}^{2}>b^{2}$,
as illustrated in figure (\ref{fig:Dbrane2in1}).

\subsection{Deriving bit threads in string theory}

Building on the setup from the previous subsection, we now attempt
to show that bit threads can be exactly derived from multiple parallel
stationary strings carrying Kalb--Ramond charges. To this end, we
recall the solutions presented by Dabholkar et al. in the low-energy
analysis of macroscopic superstrings \cite{Dabholkar:1990yf,Sen:1992yt,Duff:1994an}.
A recent review is provided in \cite{Lu:2025awh}. These macroscopic
superstrings play a role analogous to solitons in supersymmetric field
theories. The relevant low-energy effective action, obtained as the
localization of our proposed action (\ref{eq:E theory action}) and
incorporating the macroscopic string as a source term, is given by

\begin{equation}
S=\frac{1}{2\kappa^{2}}\int d^{3}x\sqrt{-g}e^{-2\phi}\left(R+4\left(\nabla\phi\right)^{2}-\frac{1}{12}H^{2}+\frac{4}{k}\right)+S_{\sigma},
\end{equation}

\noindent with the string source described by the sigma model

\begin{equation}
S_{\sigma}=-\frac{1}{4\pi\alpha^{\prime}}\int d^{2}\sigma\left(\sqrt{\gamma}\gamma^{mn}\partial_{m}X^{\mathrm{a}}\partial_{n}X^{\mathrm{b}}g_{\mathrm{ab}}+\epsilon^{mn}\partial_{m}X^{\mathrm{a}}\partial_{n}X^{\mathrm{b}}B_{\mathrm{ab}}\right),
\end{equation}

\noindent This action describes a macroscopic string carrying Kalb--Ramond
charge, propagating in the background and interacting with background
fields. The equations of motion follow as

\begin{eqnarray}
R^{\mathrm{ab}}+2\nabla^{\mathrm{a}}\nabla^{\mathrm{b}}\phi-\frac{1}{4}H^{\mathrm{acd}}H_{\:\mathrm{cd}}^{\mathrm{b}} & = & \kappa^{2}T^{\mathrm{ab}},\nonumber \\
\nabla_{\mathrm{a}}\left(e^{-2\phi}H^{\mathrm{abc}}\right) & = & \frac{\kappa^{2}}{\pi\alpha^{\prime}}j^{\mathrm{bc}}\nonumber \\
4\nabla^{2}\phi-4\left(\nabla\phi\right)^{2}+R-\frac{1}{12}H^{2}+\frac{4}{k} & = & 0,\label{eq:EOM}
\end{eqnarray}

\noindent with source terms

\begin{eqnarray}
T^{\mathrm{ab}} & = & -\frac{e^{2\phi}}{2\pi\alpha^{\prime}\sqrt{-g}}\int d^{2}\sigma\sqrt{\gamma}\gamma^{mn}\partial_{m}X^{\mathrm{a}}\partial_{n}X^{\mathrm{b}}\delta^{\left(3\right)}\left(x-X\left(\tau,\sigma\right)\right),\nonumber \\
j^{\mathrm{ab}} & = & \frac{1}{\sqrt{-g}}\int d^{2}\sigma\left(\frac{\partial X^{\mathrm{a}}}{\partial\tau}\frac{\partial X^{\mathrm{b}}}{\partial\sigma}-\frac{\partial X^{\mathrm{b}}}{\partial\tau}\frac{\partial X^{\mathrm{a}}}{\partial\sigma}\right)\delta^{\left(3\right)}\left(x-X\left(\tau,\sigma\right)\right).\label{eq: source}
\end{eqnarray}

\noindent Note that we extract a factor of $1/\alpha^{\prime}$ from
$j^{\mathrm{ab}}$ in (\ref{eq: source}) and incorporate it into
the second equation of motion in (\ref{eq:EOM}). This adjustment
is necessary because the dimension of $j^{\mathrm{ab}}$ is $\left[L\right]^{-1}$
in the case of two spatial dimensions. The usual solutions of these
EOM (\ref{eq:EOM}) are known as string solitons \cite{Dabholkar:1990yf,Sen:1992yt,Duff:1994an}.
However, in our case, we require the solution to remain AdS$_{3}$,
which in turn demands that the string source does not significantly
modify the background geometry.

To approach this solution of the EOM (\ref{eq:EOM}), let us first
recall that the action originates from entanglement entropy via the
AdS$_{3}$/CFT$_{2}$ correspondence. This implies that bulk gravity
is weak, i.e. $\kappa^{2}=8\pi G_{N}^{\left(3\right)}\rightarrow0$,
while the central charge diverges, $c\rightarrow\infty$, in the CFT
at the conformal boundary. Parametrically, the three-dimensional Newton
constant scales as

\begin{equation}
16\pi G_{N}^{\left(D\right)}=\left(2\pi\sqrt{\alpha^{\prime}}\right)^{D-2}g_{s}^{2}\left(2\pi\right)^{-1},
\end{equation}

\noindent so that

\begin{equation}
\frac{\kappa^{2}}{\pi\alpha^{\prime}}=\frac{g_{s}^{2}}{2\pi\sqrt{\alpha^{\prime}}}.
\end{equation}

\noindent This factor multiplies the string-source terms in the EOM
(\ref{eq:EOM}). If the dilaton field is trivial, $\phi=0$, so that
$g_{s}=1$, then in the weak-gravity limit $\kappa^{2}\sim\sqrt{\alpha^{\prime}}\rightarrow0$
the prefactor diverges, making it difficult to find consistent AdS$_{3}$
solutions. To suppress the backreaction of $\mathcal{N}$ string sources
on the AdS$_{3}$ background, we therefore require

\begin{equation}
g_{s}^{2}=e^{2\phi_{0}}\ll\frac{\sqrt{\alpha^{\prime}}}{l_{\mathrm{AdS}}}\cdot\frac{1}{\mathcal{N}},
\end{equation}

\noindent where $l_{\mathrm{AdS}}$ is introduced to render the string
coupling dimensionless and $\phi_{0}$ is a constant dilaton. Thus,
the simplest choice for $g_{s}$ is

\begin{equation}
g_{s}^{2}=e^{2\phi_{0}}=\frac{\sqrt{\alpha^{\prime}}}{l_{\mathrm{AdS}}}\cdot\frac{\epsilon}{\mathcal{N}},\qquad\epsilon\ll1.\label{eq:string coupling}
\end{equation}

\noindent This necessity explains why the dilaton must be incorporated
into our proposed action (\ref{eq:E theory action}). 

This scaling for the string coupling is closely analogous to the D1--D5/F1-NS5
system \cite{Strominger:1996sh,Callan:1996dv,Maldacena:1998bw}. In
type IIB string theory on $\mathrm{R}^{4}\times\mathrm{R}^{2}\times\mathrm{T}^{4}$,
consider $Q_{1}$ fundamental strings on $\mathrm{R}^{2}$ bound to
$Q_{5}$ NS5-branes on $\mathrm{R}^{2}\times\mathrm{T}^{4}$. In the
near-horizon limit the geometry becomes AdS$_{3}$$\times$S$^{3}$$\times$T$^{4}$.
There the string coupling is \cite{deBoer:1998gyt}

\begin{equation}
g_{s}^{2}=\frac{1}{Q_{1}\sqrt{Q_{5}}},\qquad l_{\mathrm{AdS}}=\sqrt{Q_{5}}\sqrt{\alpha^{\prime}},
\end{equation}

\noindent which can be rewritten as

\begin{equation}
g_{s}^{2}=\frac{\sqrt{\alpha^{\prime}}}{l_{\mathrm{AdS}}}\cdot\frac{1}{Q_{1}},
\end{equation}

\noindent where the central charge of the dual CFT$_{2}$ is $c=6Q_{1}Q_{5}\gg1$.
Since $\mathcal{N}$ of (\ref{eq:string coupling}) also relates to
the charge of fundamental string (to be examined later), this agrees
with our condition above, showing that the same type of scaling naturally
appears in a well-understood brane system. Moreover, it is also shown
that when $Q_{1},Q_{5}\gg1$, the long string does not significantly
affect the background geometry in this system \cite{Seiberg:1999xz}.

Now, let us return to our setup. We begin by solving the EOM (\ref{eq:EOM})
in the limit $\frac{\kappa^{2}}{\pi\alpha^{\prime}}\rightarrow0$.
In this limit, we consider only the leading-order solution. The higher-order
corrections to the spacetime can then be interpreted as quantum corrections,
which complete the generalized entropy prescription \cite{Bekenstein:1973ur,Wall:2011hj}.
In this regime, the solution is precisely the BTZ black brane background
(\ref{eq:BTZ brane}) introduced earlier,

\begin{equation}
dS^{2}=\frac{l_{\mathrm{AdS}}^{2}}{z^{2}}\left(-\left(1-\left(\frac{z}{b}\right)^{2}\right)dt^{2}+\frac{dz^{2}}{1-\left(\frac{z}{b}\right)^{2}}+dx^{2}\right),\label{eq:black brane}
\end{equation}

\noindent where $x\in\mathbb{R}$, $z\in\left(0,b\right]$ and $z=b$
corresponds to the planar horizon. For simplicity, we compute only
one side ($+$ or $-$) of the brane, since the other side follows
analogously, and the corresponding charge density vectors $\overrightarrow{j}^{0}$
can be smoothly connected at $z=b$. Then, we place a stationary string
stretched along the $z$-direction, corresponding to the static gauge
choice

\begin{equation}
X^{t}\left(\tau,\sigma\right)=\tau,\qquad X^{z}\left(\tau,\sigma\right)=\sigma,\qquad X^{x}\left(\tau,\sigma\right)=0.
\end{equation}

\noindent In this gauge, the only non-vanishing current component
from the source term (\ref{eq: source}) is

\begin{equation}
j^{0z}=\frac{z^{3}}{l_{\mathrm{AdS}}^{3}}\delta\left(x\right),\qquad0\leq z\leq b.
\end{equation}

\noindent Now consider $\mathcal{N}$ multiple parallel open strings
uniformly distributed along the $x$-direction \cite{Dabholkar:1990yf,Sen:1992yt,Duff:1994an}.
This modifies the current to

\begin{equation}
j^{0z}=\frac{z^{3}}{l_{\mathrm{AdS}}^{3}}\sum_{n\in\mathbb{Z}}\delta\left(x-nd_{s}\right),
\end{equation}

\noindent where $d_{s}$ is the separation between adjacent strings.
The sum $\sum_{n}\delta\left(x-nd_{s}\right)$, which defines the
Dirac comb (or Sha function), indicates that the strings are located
at positions $\left(\ldots,-2d_{s},-d_{s},0,d_{s},2d_{s},\ldots\right)$.
To prepare for the comparison with bit threads in the next section,
let us recall that the threads originate from the boundary region
$A\in\left[-b,b\right]$ and end on the complementary region $B\in\left(-\infty,b\right)\cup\left(b,+\infty\right)$.
A consistent way to discretize this setup with open strings is to
place $\mathcal{N}$ parallel strings uniformly across the interval
of length $2b$. The separation between adjacent strings is then

\begin{equation}
d_{s}=\frac{2b}{\mathcal{N}}.
\end{equation}

\noindent With this choice, the charge density $j^{0z}$ can be expressed
as

\begin{equation}
j^{0z}=\frac{1}{d_{s}}d_{s}\frac{z^{3}}{l_{\mathrm{AdS}}^{3}}\sum_{n\in\mathbb{Z}}\delta\left(x-nd_{s}\right)=\frac{1}{2b}\frac{z^{3}}{l_{\mathrm{AdS}}^{3}}d_{s}\sum_{n\in\mathbb{Z}}\delta\left(x-nd_{s}\right),
\end{equation}

\noindent where in the second equality the factor of $\mathcal{N}$
is canceled against the $1/\mathcal{N}$ in the string coupling (\ref{eq:string coupling}),
so we suppress it here. In the continuum limit, $\mathcal{N}\rightarrow\infty$
(hence $d_{s}\rightarrow0$), the discrete comb of delta functions
converges weakly to the constant function 1, yielding the uniform
distribution

\begin{equation}
j^{0z}=\frac{1}{2b}\frac{z^{3}}{l_{\mathrm{AdS}}^{3}}\underset{d_{s}\rightarrow0}{\lim}d_{s}\sum_{n}\delta\left(x-nd_{s}\right)=\frac{1}{2b}\frac{z^{3}}{l_{\mathrm{AdS}}^{3}}.\label{eq:dirac comb}
\end{equation}

\noindent We emphasize that this equality holds distributionally in
the $x$-direction, not pointwise. The passage from the Dirac comb
to a constant function is analogous to techniques in string cosmology,
where a single fundamental string is generalized to a gas of $\mathcal{N}$
strings and the delta-function sources are averaged \cite{Brandenberger:1988aj,Tseytlin:1991xk,Battefeld:2005av}.
In that context, the total energy density of the string gas is obtained
by multiplying the number density of strings per spatial volume with
the corresponding single-string energy. Moreover, this corresponds
to the large-$\mathcal{N}$ decoupling limit in AdS/CFT \cite{Maldacena:1997re},
where $\mathcal{N}$ denotes the number of branes. Returning to our
case, the resulting current density vector is

\begin{equation}
j^{\mathrm{0a}}=\left(j^{0t},j^{0z},j^{0x}\right)=\frac{1}{2b}\frac{z^{3}}{l_{\mathrm{AdS}}^{3}}\left(0,1,0\right),\qquad\mathrm{a}=t,z,x,
\end{equation}

\noindent which satisfies the 3D divergenceless condition $\partial_{\mathrm{a}}\left(\sqrt{-g}j^{\mathrm{0a}}\right)=0$.
To compare with bit threads, we restrict to the 2D time slice $\left(z,x\right)$.
The projected divergenceless condition is

\begin{equation}
\partial_{i}\left(\sqrt{h^{\left(2\right)}}j^{0i}\right)=0,\qquad i=z,x,
\end{equation}

\noindent where the induced current on the time slice of the background
is

\begin{equation}
j^{0i}=\sqrt{-g_{tt}}j^{0i}=\frac{1}{2b}\frac{z^{2}}{l_{\mathrm{AdS}}^{2}}\sqrt{1-\left(\frac{z}{b}\right)^{2}}\left(1,0\right),\label{eq:open string charge}
\end{equation}

\noindent This describes the Kalb--Ramond charge density vector flowing
from $z=0$ to $z=b$, as depicted in figure (\ref{fig:2world}).
Mapping this configuration to figure (\ref{fig:Dbrane2in1}) via the
coordinate transformations (\ref{eq:brangcor}), we obtain

\begin{equation}
v^{i}=\frac{l_{\mathrm{AdS}}}{2G_{N}^{\left(3\right)}}j^{0i}=\frac{1}{4G_{N}^{\left(3\right)}}\frac{1}{l_{\mathrm{AdS}}}\left(\frac{2bw}{\sqrt{\left(b^{2}-y^{2}-w^{2}\right)^{2}+4b^{2}w^{2}}}\right)^{2}\left(\frac{b^{2}-y^{2}+w^{2}}{2b},\frac{yw}{b}\right),\label{eq:vandj}
\end{equation}

\noindent where the dimensionless prefactor $l_{\mathrm{AdS}}/2G_{N}^{\left(3\right)}$
is introduced to normalize the vector $\overrightarrow{j}^{0}$ such
that $v$ saturates the max-flow min-cut bound $\left|v\right|\leq1/4G_{N}^{\left(3\right)}$.
This expression exactly reproduces the bit thread vector field obtained
earlier \cite{Freedman:2016zud,Agon:2018lwq}. It is worth noting
that, when considering the full solutions for $+$ and $-$, the normal
flux matches on both sides of the sewing line at $z=b$. With boundary
conditions, this makes the stitched flow divergenceless. In other
words, the flow can not stop at the brane. 

Finally, a natural question is whether parallel strings would interact
and annihilate. To address this, we must include the missing equation
of motion from (\ref{eq:EOM}), obtained by varying the string embedding
field $X\left(\tau,\sigma\right)$:

\begin{equation}
\nabla_{m}\left(\gamma^{mn}\nabla_{n}X^{\mathrm{a}}\right)=-\Gamma_{\mathrm{bc}}^{\mathrm{a}}\partial_{m}X^{\mathrm{b}}\partial_{n}X^{\mathrm{\rho}}\gamma^{mn}+\frac{1}{2}H_{\;\mathrm{bc}}^{\mathrm{a}}\partial_{m}X^{\mathrm{b}}\partial_{n}X^{\mathrm{c}}\epsilon^{mn}.\label{eq:X EOM}
\end{equation}

\noindent Consider two adjacent strings: one taken as the source (at
the origin) and the other a stationary test string placed at some
finite separation, both stretched along the $z$-direction with the
same orientation. Using conformal gauge $X^{0}=\tau$, $X^{z}=\sigma$,
$\gamma^{mn}=\mathrm{diag}\left(-1,1\right)$, $\epsilon^{0z}=1$
for the test string and the EOM (\ref{eq:X EOM}), we will get the
acceleration for the test string in the transverse direction:

\begin{equation}
\frac{d^{2}}{d\tau^{2}}X^{x}\neq0.
\end{equation}

\noindent This result implies that the contribution of the graviton
to the force between two parallel strings cannot be cancelled by the
contribution of the Kalb--Ramond field. In other words, the parallel
strings in our case break the no-force condition and inevitably interact
with each other as time evolves. This can be understood intuitively:
at an initial moment, our setup of parallel strings corresponds to
a configuration of bit threads, from which the associated RT surface
can be simultaneously determined. Once time evolution is turned on,
the strings begin to interact, rearrange, and consequently redefine
both the bit threads and the RT surface, since the \emph{entangling
regions vary dynamically} with time. This key feature distinguishes
our setup from the conventional no-force condition and explains why
the latter is violated in our case. A familiar example is the phase
transition of the EWCS: when two entangled regions $A$ and $B$ evolve
to become sufficiently far apart, the EWCS between $A$ and $B$ vanishes,
splitting into two distinct systems, while a new EWCS between regions
$C$ and $D$ emerges. This observation is also consistent with our
earlier result in string field theory \cite{Jiang:2024noe}, where
open string scattering was shown to correspond to the evolution of
the entanglement wedge.

\subsection{Correspondence between string theory and entanglement entropy}

\begin{figure}[h]
\begin{centering}
\includegraphics[scale=0.35]{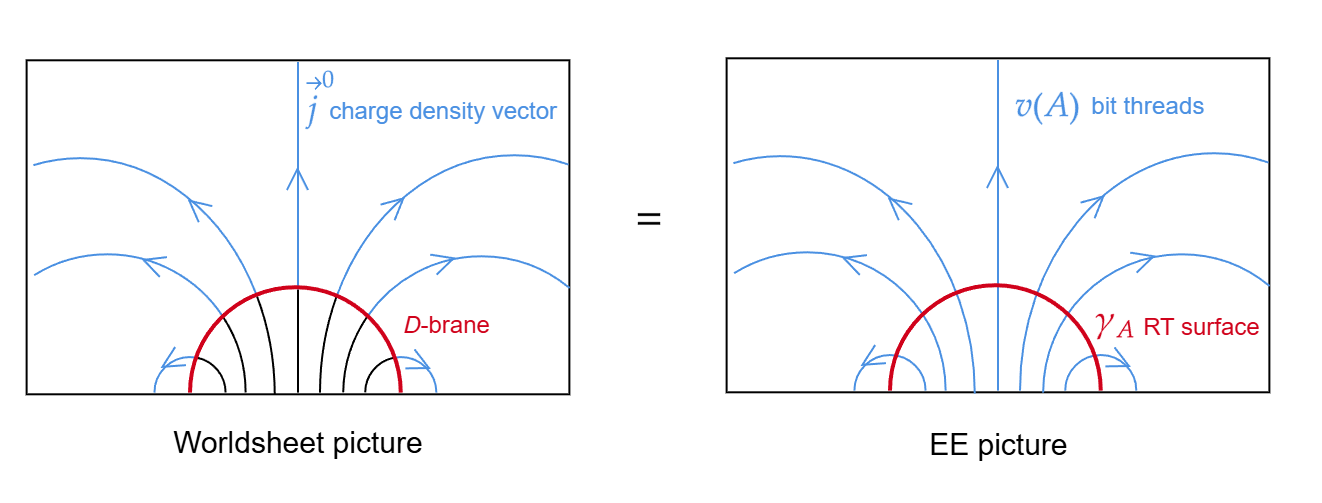}
\par\end{centering}
\caption{\label{fig:bit}This figure illustrates the correspondence between
the worldsheet formulation and the entanglement entropy framework.
In this picture, the $D$-brane is identified with the RT surface
$\gamma_{A}$ on a time slice of AdS$_{3}$. To realize the AdS$_{3}$
background from string theory, the inclusion of the Kalb--Ramond
field is essential. This field gives rise to the Kalb--Ramond charge
density vector $\protect\overrightarrow{j}^{0}$, which is tangent
to the open string parameterized from $\sigma=0$ to $\pi$. Notably,
this vector is orthogonal to the $D$-brane and corresponds to the
bit thread vector $v\left(A\right)$, which is orthogonal to the RT
surface $\gamma_{A}$. Both $\protect\overrightarrow{j}^{0}$ and
$v\left(A\right)$ satisfy a divergenceless condition, $\nabla\cdot\protect\overrightarrow{j}^{0}=\nabla\cdot v=0$,
reinforcing the parallel between the string worldsheet description
and the bit thread formulation of entanglement entropy.}
\end{figure}

Before proceeding to further analysis, let us pause briefly to summarize
the results obtained thus far. Based on the previous observations,
the correspondence between the string worldsheet and the RT surface
becomes manifest. The internal consistency of the worldsheet theory
requires the action to include not only the symmetric spacetime metric
$g_{\mathrm{ab}}$ but also the antisymmetric Kalb--Ramond field
$B_{\mathrm{ab}}$ and the constant dilaton $\phi$. All three fields
are indispensable:
\begin{enumerate}
\item Since our constructed action locally reduces to the Polyakov action
(\ref{eq:local action}) of string worldsheet theory, its equations
of motion admit the AdS$_{3}$ solution if and only if both $g_{\mathrm{ab}}$
and $B_{\mathrm{ab}}$ are included simultaneously.
\item The constant dilaton $\phi$ ensures the string coupling remains weak,
$g_{s}\ll1$. Consequently, multiple string sources do not backreact
on the AdS$_{3}$ geometry at leading order, allowing us to establish
a consistent correspondence between the string charge density and
the bit threads in AdS$_{3}$.
\end{enumerate}
Moreover, since entanglement entropy can be computed using the spacetime
metric $g_{\mathrm{ab}}$, it is natural to expect that there exists
a comparable entanglement-related quantity computable via the Kalb--Ramond
field $B_{\mathrm{ab}}$. This theoretical requirement provides compelling
evidence in support of the bit thread formulation of entanglement
entropy. Consequently, it enables us to establish an explicit correspondence
between quantities in the worldsheet theory and those in entanglement
entropy, as illustrated in figure (\ref{fig:bit}):
\begin{center}
\begin{tabular}{|c|c|}
\hline 
$\begin{array}{c}
\\
\\
\end{array}$\textbf{Worldsheet}$\begin{array}{c}
\\
\\
\end{array}$ & \textbf{Entanglement}\tabularnewline
\hline 
$\begin{array}{c}
\\
\\
\end{array}$$D$-brane/$E$-brane$\begin{array}{c}
\\
\\
\end{array}$ & RT surface\tabularnewline
\hline 
$\begin{array}{c}
\\
\\
\end{array}$Charge density vector $\overrightarrow{j}^{0}$$\begin{array}{c}
\\
\\
\end{array}$ & Bit threads $v$\tabularnewline
\hline 
\end{tabular}
\par\end{center}

\noindent Under this setup, the left-hand panel illustrates the oriented
blue lines in figure (\ref{fig:bit}), which represent the Kalb--Ramond
charge density vector $\overrightarrow{j}^{0}$, tangent to the string.
Specifically, this vector lies along the string in the direction of
increasing $\sigma$, reflecting the string\textquoteright s orientation.
The red curve in the same panel depicts the geodesic trajectory of
a string endpoint, corresponding to a $D$-brane. This $D$-brane
thus plays the role of the entangling surface, or more precisely,
the entangling brane ($E$-brane) \cite{Donnelly:2016jet}. This result
agrees with earlier studies, particularly when one considers extending
the open string beyond the D-brane, effectively slicing it into two
parts \cite{Susskind:1994sm,Donnelly:2016jet}. It is also consistent
with understanding from loop quantum gravity \cite{Donnelly:2008vx}.
In this configuration, entanglement entropy arises from tracing out
the degrees of freedom on one side of the D-brane, and its value is
proportional to the area (length, in this case) of the brane. This
correspondence is illustrated in the left and right panel of figure
(\ref{fig:bit}). Moreover, given that $g_{\mathrm{ab}}$ and $B_{\mathrm{ab}}$
appear on equal footing in the worldsheet action, their dual quantities
in the entanglement entropy framework---namely, the RT surface and
the bit threads---should likewise be treated on an equal footing.
Remarkably, this expectation is realized: entanglement entropy can
indeed be computed through both the geodesic length of the RT surface
and the flux of bit threads:
\begin{center}
\begin{tabular}{|c|c|c|}
\hline 
$\begin{array}{c}
\\
\\
\end{array}$$g_{\mathrm{ab}}$$\begin{array}{c}
\\
\\
\end{array}$ & RT surface & $S_{\mathrm{vN}}=\frac{\mathrm{area}\left(\gamma_{A}\right)}{4G_{N}}$\tabularnewline
\hline 
$\begin{array}{c}
\\
\\
\end{array}$\textbf{$B_{\mathrm{ab}}$}$\begin{array}{c}
\\
\\
\end{array}$ & Bit threads $v$ & $S_{\mathrm{vN}}=\underset{v}{\mathrm{max}}\int_{A}v$\tabularnewline
\hline 
\end{tabular}
\par\end{center}

\noindent The computational setup for evaluating entanglement entropy
using bit threads is visually illustrated in the right-hand panel
of figure (\ref{fig:bit}).

On the other hand, it is worth noting that this correspondence explicitly
demonstrates that spacetime geometry and dynamics are determined jointly
by entanglement entropy (via the RT surface) and bit threads, since
the equations of motion are those of $g_{\mathrm{ab}}$ and $B_{\mathrm{ab}}$.
Because the bit-thread flow $v$ necessarily incorporates the RT surface,
which connects the UV and IR regions, this provides a verification
of our earlier conjecture that bulk geometries can be fixed by two
distinct kinds of entanglement entropies: one defined by RT surfaces
whose endpoints both lie on the AdS boundary, and the other defined
by surfaces with one endpoint anchored on the boundary and the other
extending into the deep bulk \cite{Wang:2017ele,Wang:2018vbw,Wang:2018jva}.

Finally, our discussion has primarily focused on open strings. For
closed strings, the corresponding construction can be obtained through
a quotient procedure or the replica \textquotedbl copy-and-glue\textquotedbl{}
method, as in the case of reflected entropy on ordinary manifolds
\cite{Dutta:2019gen}. In such scenarios, the geodesic supporting
the D-brane corresponds to the initial and final configurations of
the closed string worldsheet. We will present a concrete example of
closed string case in the next section.

\section{Novel results}

In the previous section, we established an explicit correspondence
between the string worldsheet and the RT surface. One may naturally
ask whether this correspondence yields any novel or previously unknown
results. In this section, we will present some of them.

\subsection{Computing entanglement entropy from open string charge}

Remarkably, if the correspondence holds, it offers a new method to
compute the entanglement entropy of a CFT$_{2}$ using string-theoretic
quantities.

\begin{figure}[h]
\begin{centering}
\includegraphics[scale=0.35]{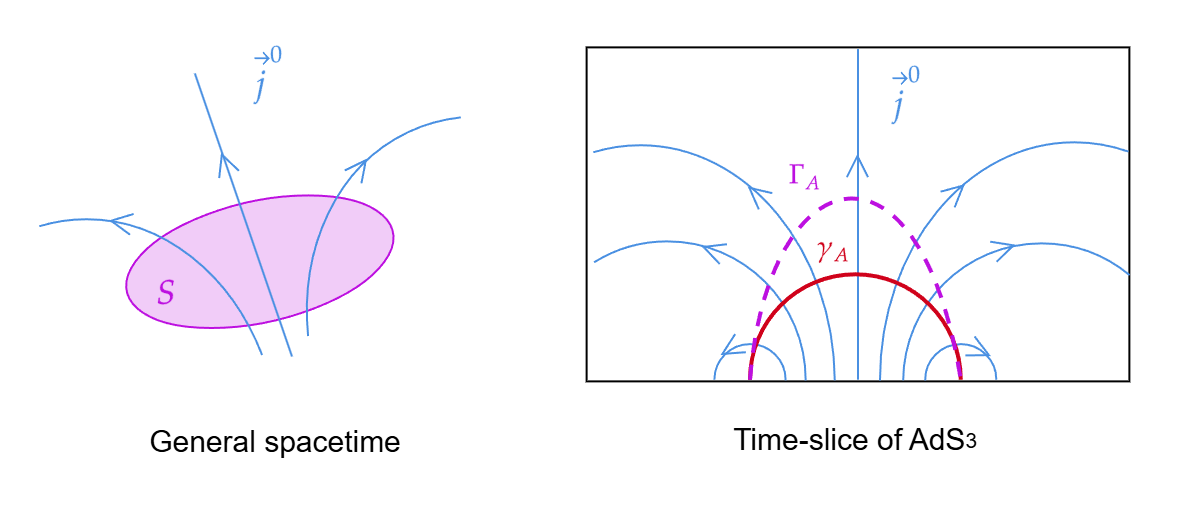}
\par\end{centering}
\caption{\label{fig:number}This figure illustrates the definition of the string
number $\mathcal{N}$. In the left-hand panel, for a general spacetime,
the two-dimensional surface $S$ is pierced by open strings, and the
string number is defined via the integration of the string charge
density vector $\protect\overrightarrow{j}^{0}$ across $S$. In the
right-hand panel, we depict the specific case relevant to our discussion:
the time slice of AdS$_{3}$, where the surface $S$ reduces to a
one-dimensional curve $\Gamma_{A}$. When $\Gamma_{A}$ coincides
with the geodesic $\gamma_{A}$ on which the $D$-brane resides, the
charge density vectors $\protect\overrightarrow{j}^{0}$ become normal
to $\gamma_{A}$.}
\end{figure}

To explore this, we recall a key observable in string theory--- the
\textbf{\emph{string number}} $\mathcal{N}$ \cite{Zwiebach:2004tj},
which can be expressed in terms of the open string charge density.
Concretely, it is defined via the Kalb--Ramond charge density vector
$\overrightarrow{j}^{0}$ as

\begin{equation}
\mathcal{N}\equiv\int_{S}\overrightarrow{j}^{0}\cdot d\overrightarrow{a}.
\end{equation}

\noindent where $S$ is a two-dimensional surface pierced by strings,
and the integrand counts the net flux of strings through $S$, as
illustrated in figure (\ref{fig:number}). Note that the string number
$\mathcal{N}$ defined here differs from its conventional definition.
Since $\overrightarrow{j}^{0}$ is obtained by summing over all parallel
stationary strings in (\ref{eq:dirac comb}), $\mathcal{N}$ is no
longer a discrete quantity but instead takes continuous values. On
a time slice of AdS$_{3}$, this reduces to an integral over a one-dimensional
curve $\Gamma_{A}$. Using the previous results (\ref{eq:vandj}),
the string number becomes

\begin{equation}
\mathcal{N}=\int\overrightarrow{j}^{0}\cdot dS_{\Gamma_{A}}=\frac{2G_{N}^{\left(3\right)}}{l_{\mathrm{AdS}}}\int\overrightarrow{v}\cdot dS_{\Gamma_{A}},
\end{equation}

\noindent where $\Gamma_{A}$ is any curve homologous to the entangling
region $A$ on the conformal boundary. When $\Gamma_{A}$ coincides
with the geodesic $\gamma_{A}$ on which the $D$-brane resides, the
vector $\overrightarrow{j}^{0}$ becomes the unit normal $\hat{n}$
to the $D$-brane. Then, the integral simplifies to

\begin{equation}
\mathcal{N}=\frac{2G_{N}^{\left(3\right)}}{l_{\mathrm{AdS}}}\int\overrightarrow{v}\cdot dS_{\Gamma_{A}}=\frac{2G_{N}^{\left(3\right)}}{l_{\mathrm{AdS}}}\int\overrightarrow{n}\cdot dS_{\gamma_{A}}=\frac{1}{2}\frac{L_{\gamma_{A}}}{l_{\mathrm{AdS}}},\label{eq:N}
\end{equation}

\noindent where $L_{\gamma_{A}}$ is the geodesic length. This allows
us to directly relate the entanglement entropy of region $A$ to the
string number by using the previous results (\ref{eq:vandj}) and
(\ref{eq:N}):

\begin{equation}
S_{\mathrm{vN}}=\frac{L_{\gamma_{A}}}{4G_{N}^{\left(3\right)}}=\frac{l_{\mathrm{AdS}}}{4G_{N}^{\left(3\right)}}\cdot2\mathcal{N}=\frac{c}{3}\ln\left(\frac{2b}{\epsilon}\right),\label{eq:open string EE}
\end{equation}

\noindent where $\epsilon$ is UV cut-off and central charge $c=3l_{\mathrm{AdS}}/2G_{N}^{\left(3\right)}$.
This result carries significant conceptual implications:
\begin{itemize}
\item The entanglement entropy arises from the number of open strings piercing
the entangling surface (i.e., the D-brane).
\item Each open string, endowed with a unit of Kalb--Ramond charge, contributes
a single bit of information. This bit encodes a microstate associated
with the entangling region $A$ and is stored holographically on the
entangling surface $\gamma_{A}$.
\end{itemize}

\subsection{Computing Bekenstein-Hawking entropy from closed string charge}

In the previous section, we derived the entanglement entropy using
open string charge. We now turn to the Bekenstein--Hawking entropy
of the BTZ black hole, deriving it through closed string charge. This
approach provides new insight into the conjecture proposed by Susskind
and Uglum.

To begin, let us recall the thermofield double (TFD) formalism and
its holographic interpretation \cite{Hartman:2015,Callebaut:2023fnf}.
The total Hilbert space is given as the tensor product of two identical
CFT Hilbert spaces,

\begin{equation}
\mathcal{H}_{total}=\mathcal{H}_{A}\otimes\mathcal{H}_{B},
\end{equation}

\noindent where each energy eigenstate satisfies $H\left|n\right\rangle =E_{n}\left|n\right\rangle $.
The TFD state, a pure entangled state in this doubled system, is defined
as

\begin{equation}
\left|TFD\right\rangle =\frac{1}{\sqrt{Z\left(\beta\right)}}\underset{n}{\sum}e^{-\beta E_{n}/2}\left|E_{n}\right\rangle _{A}\otimes\left|E_{n}\right\rangle _{B},\qquad Z\left(\beta\right)=\underset{n}{\sum}e^{-\beta E_{n}},
\end{equation}

\noindent where $T=1/\beta$ is the temperature. The corresponding
density matrix takes the form

\begin{equation}
\rho_{total}=\left|TFD\right\rangle \left\langle TFD\right|.
\end{equation}

\begin{figure}[h]
\begin{centering}
\includegraphics[scale=0.3]{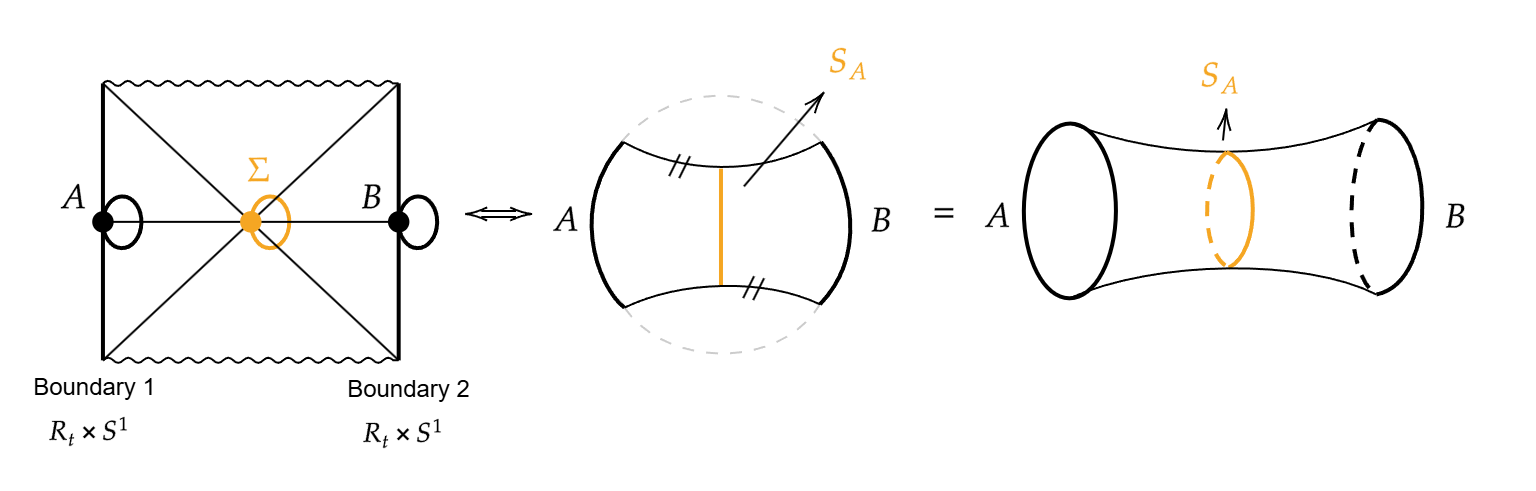}
\par\end{centering}
\caption{\label{fig:BTZ BH}This figure illustrates the holographic interpretation
of the TFD state. In the first panel, each point along the central
horizontal line represents an $S^{1}$ circle. The midpoint of this
line, marked by the orange circle, corresponds to the event horizon
$\Sigma$ at $t=0$ of the BTZ black hole. This horizontal line can
be reinterpreted as shown in the second and third panels, where the
orange geodesics denote the RT surface. The third panel is obtained
as a quotient of the geometry depicted in the second panel. Notably,
the event horizon and the RT surface coincide, capturing the same
surface.}
\end{figure}

\noindent The entropy of subsystem $A$ coincides with the entanglement
entropy between the two CFT copies,

\begin{equation}
S_{A}=-tr\rho_{A}\ln\rho_{A},\label{eq:TFD EE}
\end{equation}

\noindent where the reduced density matrix is $\rho_{A}=tr_{B}\rho_{total}=e^{-\beta H_{A}}$.
Once the TFD state is specified, the entangling regions $A$ and $B$
can be chosen on each respective boundary, and the entanglement entropy
$S_{A}$ can be computed using equation (\ref{eq:TFD EE}), as illustrated
in the two right panels of figure (\ref{fig:BTZ BH}). 

Holographically, the TFD state corresponds to the two-sided asymptotic
boundaries of the maximally extended Penrose diagram of the BTZ black
hole \cite{Maldacena:2001kr}. Each point in this diagram represents
a spatial $S^{1}$ circle, as shown in the left panel of figure (\ref{fig:BTZ BH}).
At fixed time, the entangling regions $A$ and $B$ are located on
the boundary endpoints of the horizontal line. The entanglement entropy
$S_{A}$, computed from the CFT equation (\ref{eq:TFD EE}), exactly
reproduces the area of the central orange point on this line, which
coincides with the area of the BTZ event horizon $\Sigma$. Consequently,
the entanglement entropy and the Bekenstein--Hawking entropy describe
the same surface \cite{Banados:1992wn},

\begin{equation}
S_{A}=S_{BH}=\frac{2\pi r}{4G_{N}^{\left(3\right)}}.\label{eq:BTZ entropy}
\end{equation}

\begin{figure}[h]
\begin{centering}
\includegraphics[scale=0.3]{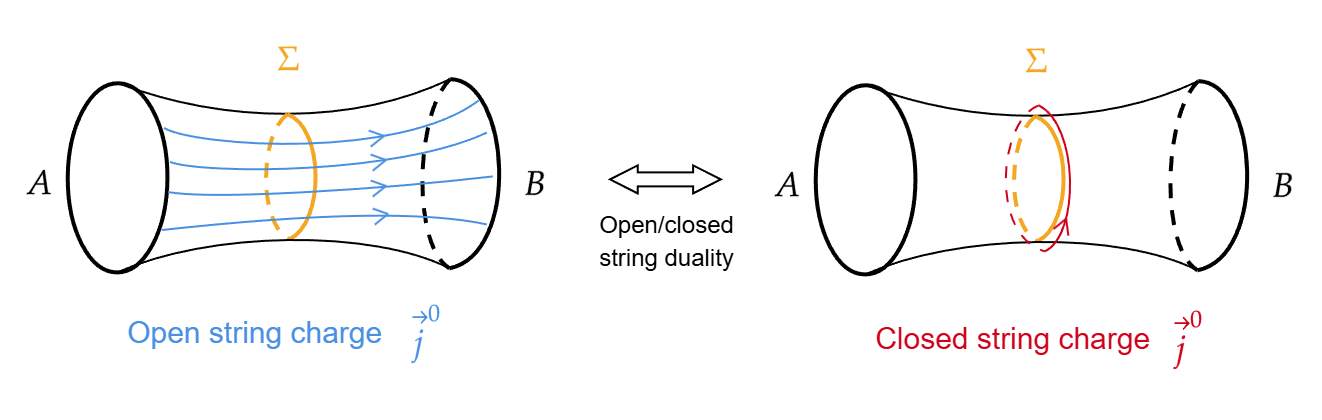}
\par\end{centering}
\caption{\label{fig:openclosed}This picture illustrates the open--closed
string duality in terms of the configuration of string charge density.
After taking the quotient of the geometry, the open string charge
can be interpreted as the current flowing from region $A$ to region
$B$. Its closed string dual corresponds to the current circulating
around the non-contractible circle, thereby producing a non-vanishing
total closed string charge. This closed string charge precisely encodes
the radius of the compactified circle, which is identified with the
horizon of the BTZ black hole.}
\end{figure}

Therefore, once we obtain the length of the orange circle, we can
directly compute the Bekenstein--Hawking entropy of the black hole.
Let us now return to the open string case, where the string carries
the Kalb--Ramond charge. When the background shown in figure (\ref{fig:openclosed})
is compactified by the quotient, we can invoke the open--closed string
duality. Under this duality, the open string charge maps to the closed
string charge. The corresponding charge flow precisely winds around
the orange circle, reproducing its area. If this interpretation is
correct, the Bekenstein--Hawking entropy of the BTZ black hole can
indeed be computed from the closed string charge.

To verify this idea, let us first recall the definition of the total
string charge $\overrightarrow{Q}$, given as the spatial integral
of the string charge density \cite{Zwiebach:2004tj}:

\begin{equation}
\overrightarrow{Q}=\int d^{d}x\sqrt{h^{\left(d\right)}}\overrightarrow{j}^{0}.
\end{equation}

\noindent Since $\overrightarrow{j}^{0}$ is defined by equations
(\ref{eq: source}) and (\ref{eq:open string charge}), in the background
(\ref{eq:black brane}) the factor $\sqrt{h^{\left(2\right)}}$ cancels
out, leaving the closed string charge

\begin{equation}
\overrightarrow{Q}=\int_{0}^{2\pi}d\sigma\partial_{\sigma}\overrightarrow{X}\left(t_{0},\sigma\right).
\end{equation}

\noindent For a contractible closed string, the total charge vanishes
due to $\overrightarrow{Q}=\overrightarrow{X}\left(t_{0},2\pi\right)-\overrightarrow{X}\left(t_{0},0\right)=0$.
In contrast, for a non-contractible closed string winding around the
compactified dimension $x$, the total charge is nonzero. Parameterizing
the embedding as $X=r\sigma$, where $r$ denotes the radius of the
compactified circle, we obtain

\begin{equation}
\overrightarrow{Q}=\left(0,Q\right),\qquad Q=2\pi r.
\end{equation}

\noindent The Bekenstein--Hawking entropy is then given by

\begin{equation}
S_{BH}=\frac{Q}{4G_{N}^{\left(3\right)}}=\frac{2\pi r}{4G_{N}^{\left(3\right)}},
\end{equation}

\noindent which exactly reproduces the Bekenstein--Hawking entropy
(\ref{eq:BTZ entropy}) of BTZ black hole. This result leads to two
key implications:
\begin{itemize}
\item The Bekenstein--Hawking entropy can be expressed in terms of the
closed string charge.
\item In higher dimensions, the corresponding generalization is provided
by D-brane charges, which naturally account for the entropy of higher-dimensional
black holes.
\end{itemize}
We believe this provides the most straightforward explanation of the
Bekenstein--Hawking entropy for the D1--D5 black hole system \cite{Strominger:1996sh,Callan:1996dv,Maldacena:1998bw}.
The corresponding Bekenstein--Hawking entropy is given by $S_{BH}=2\pi\sqrt{Q_{1}Q_{5}N}$,
where $Q_{1}$, $Q_{5}$ and $N$ are three different charges: the
wrapping number of the D1-branes, the wrapping number of the D5-branes,
and the momentum quantum number, respectively. This entropy can also
be captured by entanglement entropy. The near-horizon geometry of
the D1--D5 black hole is given by $\mathrm{BTZ}_{3}\times\mathrm{S}^{3}\times\mathrm{T}^{4}$,
and in the near-horizon limit of a near-extremal $\mathrm{BTZ}_{3}$
black hole, the geometry reduces to AdS$_{2}$$\times\mathrm{S}^{1}$.
In this case, the corresponding thermofield double (TFD) state in
the AdS$_{2}$/CFT$_{1}$ correspondence can also be used to compute
the entanglement entropy \cite{Azeyanagi:2007bj}. Therefore, the
equivalence between the Bekenstein--Hawking entropy and entanglement
entropy in this setting provides further evidence for our earlier
result that these two entropies represent the same physical quantity,
related by open--closed string duality. Related discussions can also
be found in \cite{Ying:2025rhb}.

Finally, the entanglement entropy and the Bekenstein--Hawking entropy,
computed from the open and closed string charges, provide a new avenue
to verify the conjecture of Susskind and Uglum. We will elaborate
on this explanation within the framework of the Susskind--Uglum conjecture
in Section 5.4. 

\subsection{ER=EPR from the string perspective}

\begin{figure}[h]
\begin{centering}
\includegraphics[scale=0.3]{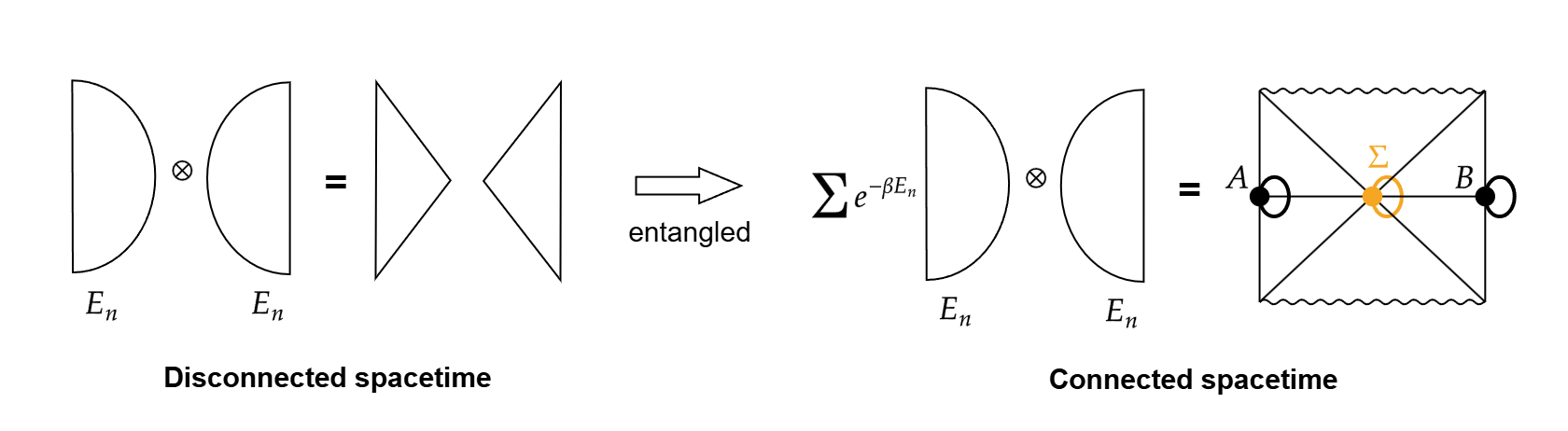}
\par\end{centering}
\caption{\label{fig:Raamsdonk}This picture illustrates Van Raamsdonk\textquoteright s
argument. The left panel shows that when the two subsystems are not
entangled, the bulk interpretation corresponds to a disconnected spacetime.
In contrast, the right panel shows that when the two subsystems are
entangled, a connected spacetime emerges through the quantum superposition
of disconnected geometries.}
\end{figure}

We can now revisit the ER = EPR conjecture from the perspective of
string theory, building on our previous results. We begin by revisiting
Van Raamsdonk\textquoteright s thought experiment. In the AdS/CFT
correspondence, if two subsystems are completely unentangled, their
joint state is simply a product state,

\begin{equation}
\left|\Psi\right\rangle =\left|\Psi_{A}\right\rangle \otimes\left|\Psi_{B}\right\rangle .
\end{equation}

\noindent which corresponds to two disconnected spacetimes, with each
$\Psi_{i}$ dual to the geometry of the respective subsystem. Once
the two subsystems are entangled, however, the product state is replaced
by the TFD state,

\begin{equation}
\left|\mathrm{TFD}\right\rangle =\underset{n}{\sum}e^{-\beta E_{n}/2}\left|E_{n}\right\rangle _{A}\otimes\left|E_{n}\right\rangle _{B},
\end{equation}

\noindent where $\beta$ is the inverse temperature and $E_{n}$ are
energy eigenvalues. The TFD state may be viewed as a quantum superposition
of disconnected geometries, which effectively yields a connected spacetime---such
as a wormhole---illustrated in figure (\ref{fig:Raamsdonk}). 

\begin{figure}[h]
\begin{centering}
\includegraphics[scale=0.3]{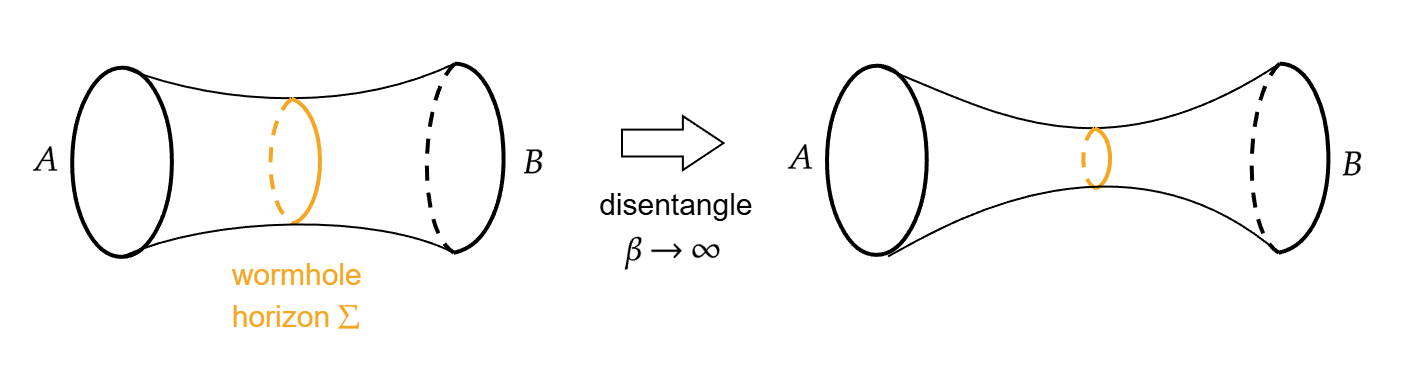}
\par\end{centering}
\caption{\label{fig:disentangle}Illustrating the disentanglement in the bulk
dual of the TFD. As the entanglement between subsystems $A$ and $B$
decreases to zero, the area of the minimal surface $\Sigma$ (the
wormhole horizon) separating the two regions also shrinks to zero.
Correspondingly, the proper distance between the associated bulk regions
diverges. In this limit, the spacetime regions are pulled apart and
eventually pinch off from one another, resulting in disconnected geometries.}
\end{figure}

This interplay between entanglement and geometry underlies the ER=EPR
conjecture, which posits that quantum entanglement (EPR pairs) is
associated with the formation of Einstein--Rosen (ER) bridges. In
the zero-temperature limit $\beta\rightarrow\infty$, the entangled
TFD state reduces to the unentangled product state, and the wormhole
disappears. In this regime, the wormhole horizon area shrinks to zero
while its proper length diverges, consistent with Raamsdonk\textquoteright s
argument and depicted in figure (\ref{fig:disentangle}). 

Since entanglement entropy can be realized in our proposed string
worldsheet picture, it follows naturally that there exists a string-theoretic
version of ER = EPR. In this picture, the closed string winds around
the wormhole horizon, with its winding charge contributing directly
to the Bekenstein--Hawking entropy. As the entanglement entropy between
two subsystems $A$ and $B$ decreases, the corresponding bulk horizon
shrinks. 

\begin{figure}[h]
\begin{centering}
\includegraphics[scale=0.3]{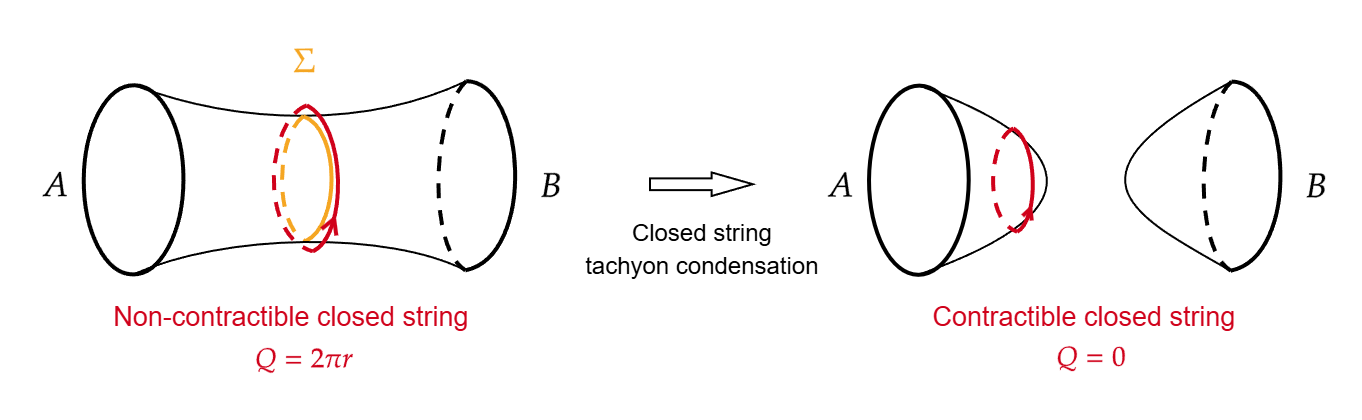}
\par\end{centering}
\caption{\label{fig:tachyon}We begin with a closed string winding around the
wormhole horizon in our proposed configuration. The topological change
of the wormhole proceeds in three steps: (1) Disentanglement between
the two subsystems, which reduces the horizon size; (2) once the horizon
shrinks below the string length scale, the winding closed string becomes
tachyonic; (3) finally, tachyon condensation drives the closed string
charge $Q=0$, signaling that the spacetime splits into two disconnected
components.}
\end{figure}

When the horizon radius falls below the string length scale, the winding
closed string develops tachyonic modes. This may trigger closed string
tachyon condensation, potentially reducing the total winding charge
$Q$ toward zero. It is important to emphasize that this process closely
parallels the mechanism of topology change via closed string winding
tachyons studied in \cite{Adams:2005rb}. The crucial difference is
that, in our setup, the shrinking of the radius is not driven by a
one-loop effective potential that destabilizes the modulus. Instead,
it arises dynamically through disentanglement in the dual CFT. A vanishing
or suppressed $Q$ would then suggest that the winding mode becomes
effectively contractible, indicating a possible transition where the
originally connected spacetime with a finite compact dimension could
fragment into two disconnected components, as illustrated in figure
(\ref{fig:tachyon}). While this scenario is consistent with known
results on closed string tachyon condensation, here we present it
as a conjectural realization of ER=EPR in the AdS$_{3}$/CFT$_{2}$
framework. Thus, closed string tachyon condensation provides a natural
microscopic mechanism for the disentanglement process in the ER =
EPR framework. This observation reinforces our earlier claim that
entanglement entropy can serve as a probe of the closed string tachyon
vacuum in closed string field theory \cite{Wang:2021aog,Jiang:2024noe}.

Recent studies on the relation between quantum extremal surfaces and
the island formula using replica wormholes can be found in \cite{Geng:2024xpj}.
It was further shown that the replica wormhole and the island formula
can be derived directly from CFT \cite{Geng:2025efs}. A comprehensive
review of recent developments on wormholes in holography is provided
in \cite{Kundu:2021nwp}.

\subsection{New realization of Susskind and Uglum's conjecture}

\begin{figure}[h]
\begin{centering}
\includegraphics[scale=0.35]{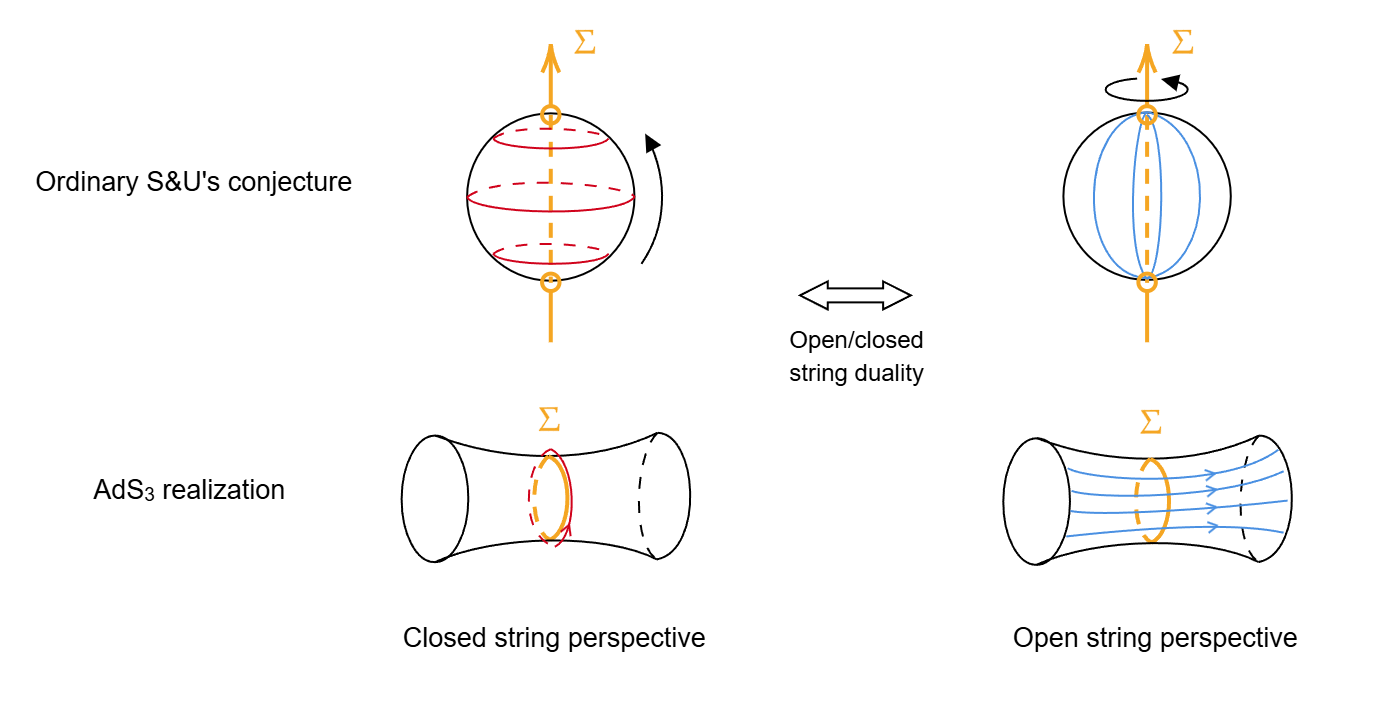}
\par\end{centering}
\caption{\label{fig:NewSandU}This figure illustrates how the Susskind--Uglum
conjecture can be generalized to the AdS$_{3}$/CFT$_{2}$ framework.
In our proposed configuration, the two-punctured spheres of the original
Susskind--Uglum setup (shown in the first row) are replaced by hyperbolic
cylinders in AdS$_{3}$ (shown in the second row). The open-- and
closed--string descriptions extend naturally to the cylinder geometry.
In this context, the entangling surface or horizon, which punctures
the two-sphere in the original picture, is realized as the waist of
the hyperbolic cylinder in AdS$_{3}$ , corresponding to the throat
of the wormhole.}
\end{figure}

In this subsection, we aim to extend the understanding of Susskind
and Uglum\textquoteright s conjecture in light of our previous results.
In Susskind and Uglum\textquoteright s framework, black hole entropy
naturally arises in string theory. The sphere diagram accounts for
the classical Bekenstein--Hawking entropy of a black hole: it may
be interpreted as a closed string emitted from one point on the horizon
and reabsorbed at another. Equivalently, it corresponds to the one-loop
diagram of open string theory, with both endpoints anchored on the
horizon, as illustrated in figure (\ref{fig:SU}). This dual interpretation
reflects the open--closed string duality, where the one-loop open
string diagram and the two-punctured closed string sphere share the
same underlying Riemann surface. Our results refine this picture in
the context of AdS$_{3}$/CFT$_{2}$, as illustrated in figure (\ref{fig:NewSandU}).

\begin{figure}[h]
\begin{centering}
\includegraphics[scale=0.3]{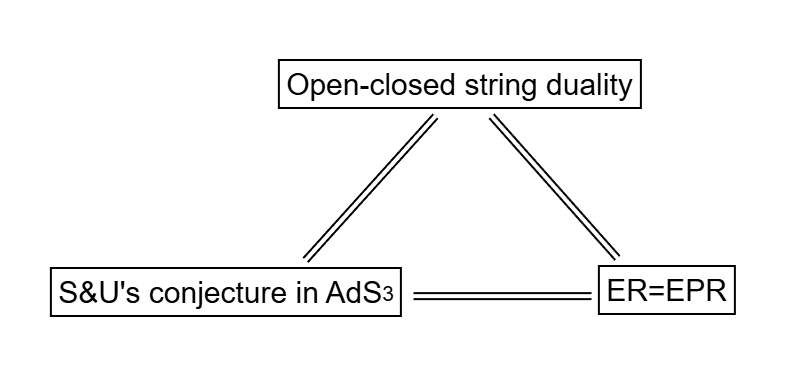}
\par\end{centering}
\caption{\label{fig:tri}This figure illustrates the equivalence between open--closed
string duality, ER=EPR, and the Susskind-- Uglum conjecture.}
\end{figure}

In this setting, the two-punctured sphere is naturally replaced by
the hyperbolic cylinder. Consequently, the entangling surface is no
longer represented by two punctures but instead sits at the waist
of the hyperbolic cylinder: in the open string picture, strings pass
through the horizon, while in the closed string picture, they wind
around it. This modification both confirms and sharpens the Susskind--Uglum
conjecture, as we demonstrated that open string charge contributes
to entanglement entropy, whereas closed string charge contributes
to black hole entropy. Furthermore, since our construction can be
naturally extended to the ER=EPR framework, we arrive at an equivalent
interpretation in which entanglement and geometry are unified through
string-theoretic charges, as illustrated in figure (\ref{fig:tri}).

Obviously, this equivalence naturally shares the same underlying property.
To see this, let us first recall the open--closed string duality.
In string theory, the appropriate description of the string cylinder---either
as a one-loop open string stretched between D-branes or as a tree-level
closed string exchanged in the bulk---depends on the brane separation.
When the separation is large, the propagation of closed strings dominates
at long distances: the massive modes decouple, leaving only the massless
sector. However, the corresponding open strings are long, so their
massive excitations significantly contribute, leading to strong gauge
coupling. Conversely, when the D-branes are close together, the supergravity
approximation breaks down, while the massless sector of open strings
dominates. Thus, as the D-brane separation increases, the open string
description is gradually replaced by the closed string description,
and the two descriptions ultimately become causally disconnected as
the closed-string propagation time tends to infinity.

A similar mechanism emerges in the other two frameworks we have discussed.
In the ER=EPR context, the degree of quantum entanglement plays the
role of the brane separation: strongly entangled states correspond
to a connected wormhole geometry, while weak entanglement invalidates
the usual von Neumann entropy measure, requiring instead a description
in terms of black hole entropy. When the systems become completely
disentangled, the wormhole throat shrinks to zero and the spacetimes
disconnect, reflecting a transition in the dominant description. Likewise,
in the EWCS, phase transitions occur when tuning the size or separation
of subsystems, with different extremal surfaces controlling the entropy
in different regimes. Finally, the Susskind--Uglum conjecture can
also be understood within this duality structure: black hole entropy
admits both a microscopic entanglement interpretation and a macroscopic
geometric area law, with the relevant description depending on the
physical regime.

Taken together, these parallels suggest that open--closed string
duality, ER=EPR, and the Susskind--Uglum conjecture are not merely
analogous but manifestations of a deeper equivalence. In each case,
the transition between dual descriptions---whether in string theory,
spacetime geometry, or entanglement entropy---reflects the same underlying
principle.

Finally, we also wish to emphasize the distinction between the method
used in the ordinary Susskind--Uglum conjecture and our picture in
AdS$_{3}$. In the original Susskind--Uglum setup, the compact spherical
worldsheet intersects the horizon infinitely many times. The reason
is that the worldsheet field $X^{\mu}$ exhibits short-distance quantum
fluctuations \cite{Susskind:1993aa,Mousatov:2020ics}. Consequently,
an unregulated ($\epsilon=0$) spherical worldsheet cannot intersect
the horizon at only two points. This leads directly to the problem
that the punctured-sphere partition function cannot be defined. To
address this issue, off-shell string theory introduces a UV regulator
$\epsilon$; see \cite{Ahmadain:2022eso} for details. In contrast,
in our work the entropy is computed from the string number, which
is determined by the number of strings---each carrying Kalb--Ramond
charge---that cross the horizon at any fixed time slice. From this
viewpoint, although a single macroscopic string may intersect the
horizon infinitely many times due to short-distance fluctuations,
the total number of strings remains well-defined and unique as long
as the parallel strings do not interact. This is precisely why our
approach differs conceptually from the off-shell sphere-partition-function
method and avoids the necessity of introducing a UV regulator.

\subsection{Are the entanglement entropy and RT surface quantized?}

In the previous discussion, we considered multiple parallel open strings
carrying Kalb--Ramond charge, each contributing to the entanglement
entropy. A natural question arises: if we consider only a single string,
does it correspond to one quantum of entanglement entropy---i.e.
to the minimal \textquotedblleft bit\textquotedblright{} of entanglement?
Since the entanglement entropy is proportional to the area of the
RT surface, this would suggest that the RT surface itself possesses
a minimal value and is quantized.

To examine this, let us recall the charge density vector for a single
string source,

\begin{equation}
j^{\mathrm{0a}}=\left(j^{0t},j^{0z},j^{0x}\right)=\frac{z^{3}}{l_{AdS}^{3}}\left(0,\delta\left(x\right),0\right),\qquad0\leq z\leq b.
\end{equation}

\noindent As in our earlier argument, we consider the induced current
on the time slice of the background,

\begin{equation}
j^{0i}=\sqrt{-g_{tt}}j^{0i}=\frac{z^{2}}{l_{AdS}^{2}}\sqrt{1-\left(\frac{z}{b}\right)^{2}}\left(\delta\left(x\right),0\right),\qquad i=z,x.
\end{equation}

\noindent Performing the coordinate transformation (we restrict attention
to the $+$ branch)

\begin{equation}
y=\frac{b\sinh\left(x/b\right)}{\cosh\left(x/b\right)+\sqrt{1-\left(z/b\right)^{2}}},\qquad w=\frac{z}{\cosh\left(x/b\right)+\sqrt{1-\left(z/b\right)^{2}}},
\end{equation}

\noindent with inverse
\begin{equation}
z=\frac{2b^{2}w}{\sqrt{\left(b^{2}-y^{2}-w^{2}\right)+4b^{2}w^{2}}},\qquad x=\frac{b}{2}\log\left(\frac{\left(b+y\right)^{2}+w^{2}}{\left(b-y\right)^{2}+w^{2}}\right).\label{eq:inverse tran}
\end{equation}

\noindent the charge density vector becomes

\begin{eqnarray}
j^{0y} & = & \frac{2yw^{2}\left(b^{2}+w^{2}\right)}{l_{AdS}^{2}\left[\left(b^{2}-y^{2}-w^{2}\right)^{2}+4b^{2}w^{2}\right]}\delta\left(y\right),\nonumber \\
j^{0w} & = & \frac{w^{2}\left(b^{2}-y^{2}+w^{2}\right)\left(b^{2}+w^{2}\right)}{l_{AdS}^{2}\left[\left(b^{2}-y^{2}-w^{2}\right)^{2}+4b^{2}w^{2}\right]}\delta\left(y\right).
\end{eqnarray}

\noindent Since the delta function enforces $y=0$, this simplifies
to

\begin{equation}
\left(j^{0w},j^{0y}\right)=\left(\frac{w^{2}}{l_{AdS}^{2}}\delta\left(y\right),0\right),
\end{equation}

\noindent which lives on the Poincaré patch of a time slice of AdS$_{3}$:

\begin{equation}
dS^{2}=\frac{l_{\mathrm{AdS}}^{2}}{w^{2}}\left(dy^{2}+dw^{2}\right).
\end{equation}

\noindent Using the correspondence (\ref{eq:vandj}),

\begin{equation}
v=\frac{l_{\mathrm{AdS}}}{2G_{N}^{\left(3\right)}}\left(j^{0w},j^{0y}\right).
\end{equation}

\noindent we compute the entanglement entropy. On a constant $w$
surface, the induced metric is $h=\frac{l_{\mathrm{AdS}}^{2}}{w^{2}}$,
and the unit normal vector is $n_{\mu}=\left(l_{\mathrm{AdS}}/w,0\right)$.
Thus, 

\begin{equation}
\sqrt{h}n_{\mu}v^{\mu}=\frac{l_{\mathrm{AdS}}}{2G_{N}^{\left(3\right)}}\delta\left(y\right)
\end{equation}

\noindent Integrating over the boundary interval,

\begin{equation}
S_{\mathrm{vN}}=\int_{A}\sqrt{h}n_{\mu}v^{\mu}=\int_{\epsilon-b}^{b-\epsilon}\sqrt{h}n_{\mu}v^{\mu}dy=\frac{l_{\mathrm{AdS}}}{4G_{N}^{\left(3\right)}}\cdot2,
\end{equation}

\noindent This exactly matches our earlier result (\ref{eq:open string EE})
for the entanglement entropy of open strings when $\mathcal{N}=1$.
It therefore represents the minimal value of the entanglement entropy
and the corresponding minimal length of the RT surface. For a general
number $\mathcal{N}$ of strings, the result extends straightforwardly
to

\begin{equation}
S_{\mathrm{vN}}=\frac{l_{\mathrm{AdS}}}{4G_{N}^{\left(3\right)}}\cdot2\mathcal{N}.\label{eq:N EE}
\end{equation}

\noindent Hence, the entanglement entropy is discretized according
to the number of strings, providing strong evidence that the RT surface
itself should be quantized. In the limit $\mathcal{N}\rightarrow\infty$
with inter-string spacing $d_{s}\rightarrow0$, $\mathcal{N}$ becomes
effectively continuous, and the result reduces to the well-known CFT$_{2}$
entanglement entropy (\ref{eq:open string EE}).

Based on this observation, the pure state $\left|\psi\right\rangle $
of the two segments of open string, string$_{1}$ and string$_{2}$
on two sides of RT-surface shown in figure (\ref{fig:Dbrane2in1}),
can be decomposed as $\left|\psi\right\rangle \in\mathcal{H}\subseteq\mathcal{H}_{\mathrm{string_{1}}}\otimes\mathcal{H}_{\mathrm{string_{2}}}$.
Consequently, we obtain

\begin{equation}
\frac{l_{\mathrm{AdS}}}{4G_{N}^{\left(3\right)}}\cdot2\widehat{\int_{\gamma_{A}}\overrightarrow{j}^{0}}\left|\psi\right\rangle =\frac{l_{\mathrm{AdS}}}{4G_{N}^{\left(3\right)}}\cdot2\mathcal{N}\left|\psi\right\rangle ,\label{eq:quanti rt}
\end{equation}

\noindent where the current $\overrightarrow{j}^{0}$ is promoted
to an operator.

\begin{figure}[h]
\begin{centering}
\includegraphics[scale=0.4]{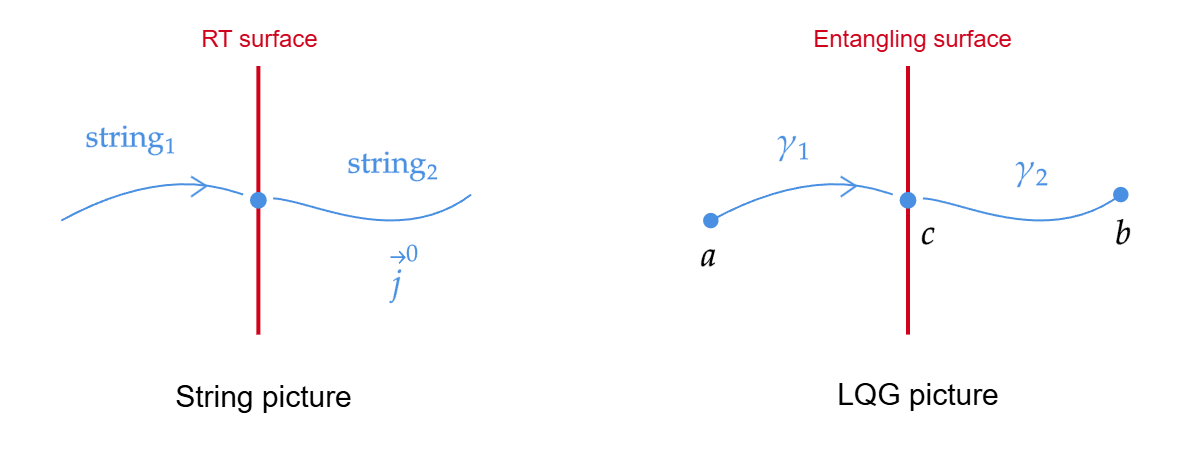}
\par\end{centering}
\caption{\label{fig:LQG}This picture illustrates the similarities between
the entanglement entropies obtained from string theory and from LQG.
In the string theory picture, the oriented open string charge density
$\protect\overrightarrow{j}^{0}$ crossing the RT surface (or D-brane)
contributes to the minimal value of entanglement entropy. As the number
of open strings increases, the entanglement entropy and the area of
the RT surface both increase in discrete steps. In the LQG picture,
the entangling surface cuts the oriented Wilson line into two parts,
to which the Schmidt decomposition can be applied. This again yields
minimal values of entanglement entropy. As more Wilson lines cross
the entangling surface, introducing additional nodes on it, the entanglement
entropy and the area of the entangling surface also increase in discrete
steps.}
\end{figure}

To further substantiate the evidence for the quantization of entanglement
entropy and the RT surface, let us revisit our setup, in which an
oriented open string is intersected by the RT surface and thereby
divided into two segments, as shown in the left panel of figure (\ref{fig:LQG}).
In this picture, the entanglement entropy (\ref{eq:N EE}) naturally
measures the quantum correlations between the two string segments.
Although the explicit Schmidt decomposition of the open string Hilbert
space is technically challenging, the bit-thread formalism provides
a well-defined method to compute this entanglement entropy. Interestingly,
a closely analogous situation arises in loop quantum gravity (LQG).
There, the entangling surface cuts an oriented Wilson line state $\gamma$
into two parts, as illustrated in the right panel of figure (\ref{fig:LQG}).
The entanglement entropy between these two parts is known to be quantized,
yielding discrete values. This suggests that quantization of entanglement
entropy and its dual RT surface may be natural both in string theory
and LQG. It therefore provides a potential bridge between string theory
and LQG, realized through the discrete nature of entanglement entropy.

To be specific, let us recall the calculation of entanglement entropy
in LQG \cite{Donnelly:2008vx}. The first step is to perform a Schmidt
decomposition. Our previous study of the TFD state already provides
a canonical physical realization of the Schmidt decomposition for
thermal states. Consider a region divided into two subregions, $A$
and $B$ , by entangling surface. Any pure state $\left|\psi\right\rangle \in\mathcal{H}=\mathcal{H}_{A}\otimes\mathcal{H}_{B}$
can be decomposed into a product basis as

\begin{equation}
\left|\psi\right\rangle =\underset{i}{\sum}\lambda_{i}\left|\varphi_{i}^{A}\right\rangle \otimes\left|\varphi_{i}^{B}\right\rangle .
\end{equation}

\noindent which is the Schmidt decomposition. Here, $\lambda_{i}$
are the Schmidt coefficients and the number of non-zero terms defines
the Schmidt rank. Tracing over one subregion gives the reduced density
matrix for region $A$:

\begin{equation}
\rho\left(A\right)=\underset{i}{\sum}\lambda_{i}^{2}\left|\varphi_{i}^{A}\right\rangle \left\langle \varphi_{i}^{A}\right|,
\end{equation}

\noindent and similarly for region $B$:

\begin{equation}
\rho\left(B\right)=\underset{i}{\sum}\lambda_{i}^{2}\left|\varphi_{i}^{B}\right\rangle \left\langle \varphi_{i}^{B}\right|.
\end{equation}

\noindent The von Neumann entanglement entropy of either subsystem
is then

\begin{equation}
S_{\mathrm{vN}}\left(A\right)=S_{\mathrm{vN}}\left(B\right)=-\underset{i}{\sum}\lambda_{i}^{2}\log\lambda_{i}^{2}.
\end{equation}

Now, let us see how this decomposition can be used to calculate the
entanglement entropy of the simplest spin-network states. We begin
by recalling the Wilson line state $\left|\gamma,j,a,b\right\rangle $.
In LQG, such states are functionals that map the $\mathfrak{su}\left(2\right)$
connection $A$ on a path $\gamma$ to $\mathbb{C}$. Explicitly,
the Wilson line state can be written as

\begin{equation}
\psi_{\gamma,j,a,b}\left[h\left(A\right)\right]=\left.\left\langle h\left(A\right)\right|\gamma,j,a,b\right\rangle =\sqrt{2j+1}D_{ab}^{\left(j\right)}\left(h_{\gamma}\left(A\right)\right),
\end{equation}

\noindent where $h\left(A\right)=\mathcal{P}\exp\left(\int A\right)$
denotes the holonomy of the connection along $\gamma$, and $D_{ab}^{\left(j\right)}\left(g\right)$
are Wigner matrices giving the spin-$j$ irreducible representation
of $g\in SU\left(2\right)$, with matrix indices $a$, $b$. The prefactor
$\sqrt{2j+1}$ ensures orthonormality and proper normalization of
the link states. Now suppose the path $\gamma$ is split into two
segments, $\gamma=\gamma_{1}\circ\gamma_{2}$, as illustrated in the
right panel of figure (\ref{fig:LQG}). The corresponding Hilbert
space factorizes as

\begin{equation}
\mathcal{H}_{\gamma}\subseteq\mathcal{H}_{\gamma_{1}}\otimes\mathcal{H}_{\gamma_{2}}.
\end{equation}

\noindent Accordingly, the Wilson line state admits the decomposition

\begin{eqnarray}
\psi_{\gamma_{1}\circ\gamma_{2},j,a,b}\left[h\left(A\right)\right] & = & \frac{1}{\sqrt{2j+1}}\stackrel[c=1]{2j+1}{\sum}\psi_{\gamma_{1},j,a,c}\left[h_{\gamma_{1}}\left(A\right)\right]\psi_{\gamma_{2},j,c,b}\left[h_{\gamma_{2}}\left(A\right)\right]\nonumber \\
 & = & \frac{1}{\sqrt{2j+1}}\stackrel[c=1]{2j+1}{\sum}\left.\left\langle h\right|\gamma_{1},j,a,c\right\rangle \left.\left\langle h\right|\gamma_{2},j,c,b\right\rangle .
\end{eqnarray}

\noindent Equivalently, in Hilbert space notation,

\begin{equation}
\left|\gamma,j,a,b\right\rangle =\frac{1}{\sqrt{2j+1}}\stackrel[c=1]{2j+1}{\sum}\left|\gamma_{1},j,a,c\right\rangle \otimes\left|\gamma_{2},j,c,b\right\rangle ,
\end{equation}

\noindent This expression is precisely a Schmidt decomposition for
the state $\left|\gamma,j,a,b\right\rangle $. From it, the Schmidt
coefficients are immediately read off as

\begin{equation}
\lambda_{i}=\frac{1}{\sqrt{2j+1}},\qquad i=1,2,\ldots,2j+1.
\end{equation}

\noindent The entanglement entropy is then

\begin{equation}
S_{A}=S_{B}=-\underset{i}{\sum}\lambda_{i}^{2}\log\lambda_{i}^{2}=\log\left(2j+1\right).
\end{equation}

Similarly, in LQG, one may ask whether there exists an operator whose
eigenvalues directly correspond to entanglement entropy. More precisely,
consider a state $\left|\psi\right\rangle \in\mathcal{H}\subseteq\mathcal{H}_{\gamma_{1}}\otimes\mathcal{H}_{\gamma_{2}}$.
In this framework, there is a Noether charge $Q$ originating from
the Lagrangian such that \cite{Donnelly:2008vx}:

\begin{equation}
\widehat{\int_{\gamma_{A}}Q}\left|\psi\right\rangle =\log\left(2j+1\right)\left|\psi\right\rangle ,
\end{equation}

\noindent which yields a result analogous to that obtained in the
string case (\ref{eq:quanti rt}).

Since the computation of entanglement entropy in both the open string
and the LQG pictures requires performing the Schmidt decomposition
in an analogous configuration, and given that both the open string
charge density and the Wilson line are oriented objects, it becomes
natural to investigate deeper relations between string theory and
LQG. To approach this issue, let us recall Wall\textquoteright s conjecture
\cite{Wall:2023myf}. Consider two theories: the first is string theory
in AdS, which is dual to $N=4$ super Yang--Mills theory; the second
is LQG. If the LQG description is obtained through the quantization
of AdS gravity, then it also admits a CFT dual. These two CFTs---the
\textquotedblleft stringy CFT\textquotedblright{} and the \textquotedblleft LQG
CFT\textquotedblright ---can in principle become strongly entangled,
thereby giving rise to a wormhole geometry. The wormhole throat thus
unifies string theory and LQG, suggesting that the two frameworks
are different descriptions of the same underlying theory.

\begin{figure}[h]
\begin{centering}
\includegraphics[scale=0.35]{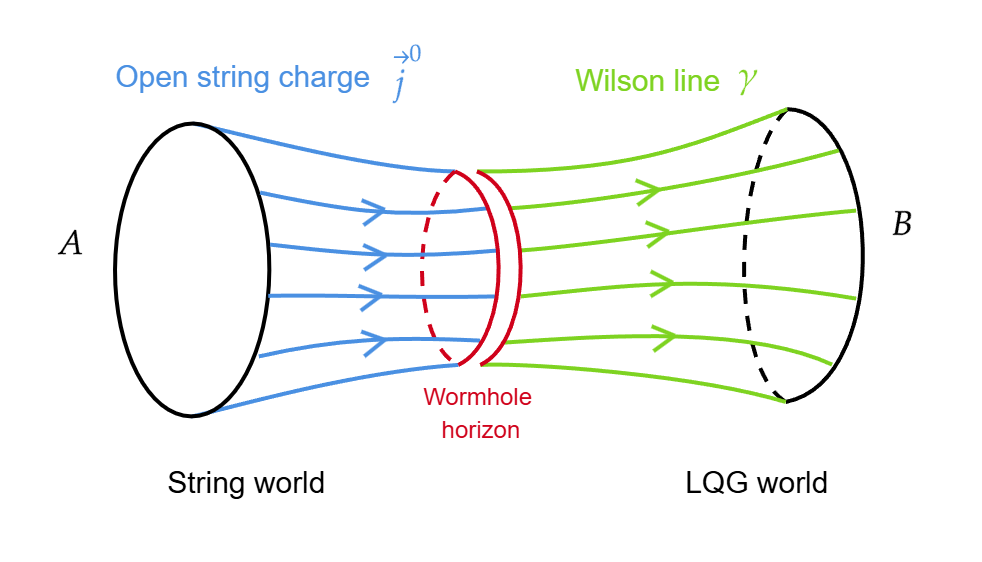}
\par\end{centering}
\caption{\label{fig:LQGworm}This picture illustrates the holographic dual
of the Hilbert space $\mathcal{H}\subseteq\mathcal{H}_{string}\otimes\mathcal{H}_{LQG}$,
under the assumption that one day AdS gravity can be quantized within
the framework of LQG. The left part represents half of the BTZ wormhole,
characterized by the open string charge density discussed previously
in figure (\ref{fig:openclosed}). The right part arises from the
LQG construction, which can be obtained in a manner analogous to string
theory, as suggested by the correspondence shown in figure (\ref{fig:LQG}).
In this way, half of the Wilson line states can also be interpreted
as constituting half of the wormhole. It is then natural to glue these
two halves together along the horizon to form a new wormhole, with
open strings and Wilson lines meeting at the same points on the horizon.
The resulting theory therefore admits two equivalent descriptions:
one in terms of string theory, and the other in terms of LQG. }
\end{figure}

Building on our earlier observations, we can refine this conjecture.
Specifically, by considering the Schmidt decomposition of the open
string charge density and the LQG Wilson line, one can split each
system into two parts and then regroup half of them to form a new
configuration, namely $\left|\psi\right\rangle \in\mathcal{H}\subseteq\mathcal{H}_{\mathrm{string_{1}}}\otimes\mathcal{H}_{\mathrm{\gamma_{2}}}$.
Since both descriptions share the same quantized wormhole horizon,
and the oriented open strings can connect to the oriented Wilson lines
at the same points on the horizon, the two theories must coincide
on the wormhole throat. This new perspective provides a concrete bridge
between the string and LQG pictures through the language of entanglement,
as illustrated in figure (\ref{fig:LQGworm}).

\section{Conclusion and discussion}

In this paper, we began with the entanglement entropy $S_{\mathrm{vN}}$
of a CFT$_{2}$. Utilizing this quantity and the RT prescription,
we constructed Synge\textquoteright s world function $\Omega\left(X,X^{\prime}\right)=8G_{N}^{2}S_{\mathrm{vN}}^{2}$
in the bulk of AdS$_{3}$, corresponding to the square of the geodesic
length between bulk points $X$ and $X^{\prime}$. Since both $X$
and $X^{\prime}$ were boundary-determined bulk points fixed by the
entangling regions in the CFT, they could be parameterized by the
coordinates $\left(\tau,\sigma\right)$ of the entangling regions
of CFT$_{2}$. When the entangling regions varied, the entanglement
entropy and its dual geodesic effectively generated a two-dimensional
surface in the bulk, described by an embedding function $X^{\mu}\left(\tau,\sigma\right)$.
Based on this observation, we proposed an action for entanglement
entropy constructed from $\Omega\left(X,X^{\prime}\right)$ and parameterized
by $\left(\tau,\sigma\right)$. This action described the dynamics
of this two-parameter family of geodesics. In the near-coincidence
limit and employing RNC, the action reduced to a Polyakov-type string
worldsheet action for the small separation field $\hat{X}^{\mathrm{a}}$,
with higher-order corrections. Requiring conformal symmetry of the
worldsheet theory led to consistency conditions on the background
fields, interpreted as the equations of motion for massless closed
string modes---namely, the gravitational field equations. Since $\Omega\left(X,X^{\prime}\right)$
was constructed in AdS$_{3}$, the resulting equations admitted AdS$_{3}$
as a solution. This consistency required the inclusion of an additional
antisymmetric background field---the Kalb--Ramond field---and the
dilaton field. This insight suggested that a corresponding quantity
must also exist in the entanglement entropy picture.

Because entanglement entropy could be calculated from the symmetric
spacetime metric, i.e., by computing the area of the RT surface, and
since the antisymmetric Kalb--Ramond field entered the action on
equal footing with the metric, there had to exist a way to compute
entanglement entropy using the Kalb--Ramond field. The breakthrough
came from bit threads, which are divergenceless and bounded vector
fields perpendicular to the RT surface and provide an alternative
formulation of entanglement entropy. In our case, the Kalb--Ramond
field introduced the charge density vector $\overrightarrow{j}^{0}$,
which is also divergenceless. To compare this charge density with
bit threads, we embedded $\overrightarrow{j}^{0}$ into the AdS background.
This required employing methods from the study of string solitons,
where $\overrightarrow{j}^{0}$ played the role of a string source
interacting with background fields. By considering multiple parallel
string sources and taking the limits $\kappa^{2}=8\pi G_{N}^{\left(3\right)}\rightarrow0$
and $g_{s}\ll1$, we successfully derived bit threads from the Kalb--Ramond
charge density $\overrightarrow{j}^{0}$. This confirmed our conjecture
that the antisymmetric Kalb--Ramond field could be used to compute
entanglement entropy. Furthermore, this result allowed us to establish
a concrete correspondence between the string worldsheet and the RT
surface.

Exploiting this correspondence, we demonstrated that entanglement
entropy could be computed within string theory: specifically, it was
proportional to the number of open strings---each carrying Kalb--Ramond
charge---that pierced the entangling surface. This result aligned
with and extended previous studies connecting entanglement entropy
to string-theoretic descriptions. Moreover, by applying open--closed
string duality, the oriented open strings became closed strings winding
around the horizon of the BTZ black hole. The corresponding closed
string charge then captured the Bekenstein--Hawking entropy of the
black hole. Based on this result, the ER=EPR correspondence could
be reinterpreted from the perspective of string charge. The holographic
disentanglement procedure, dual to the breaking of the wormhole horizon,
could be interpreted as tachyon condensation driven by the decrease
of the closed string charge $Q$. When $Q=0$, the closed string became
contractible in the compactified background, signifying that spacetime
split into two disconnected components. This picture also provided
a new realization of the Susskind--Uglum conjecture, with open--closed
string duality, ER=EPR, and the Susskind--Uglum conjecture manifesting
within the same configuration.

Finally, we discussed the possibility of quantizing entanglement entropy
or the RT surface, since entanglement entropy could be realized as
the number of strings crossing the RT surface. We also compared this
result with entanglement entropy calculations in LQG. In both frameworks,
the Schmidt decomposition could be employed to obtain entanglement
entropy, implying that each system could be split into two parts and
then regrouped to form a new configuration. This implication further
refined Wall\textquoteright s conjecture.

We conclude with the following remarks:
\begin{itemize}
\item In our work, after obtaining the action for the entanglement entropy,
we reproduced the result using the bit thread method. This approach
is well-defined and does not rely on the replica trick. Nevertheless,
it remains important to understand how the replica trick could be
implemented in our framework, and more specifically, how it can be
realized directly on the string worldsheet. We believe our work provides
a new perspective that may open the way toward such an approach.
\item Based on recent developments of hyperbolic string vertices in closed
string field theory (CSFT) \cite{Costello:2019fuh,Cho:2019anu,Firat:2021ukc,Erbin:2021smf,Erbin:2022rgx,Firat:2023glo,Firat:2023suh,Firat:2023gfn,Firat:2024ajp},
we have established a close relation between the entanglement wedge
and hyperbolic string vertices \cite{Wang:2021aog,Jiang:2024noe}.
Specifically, the entanglement wedge cross-section (EWCS) corresponds
to the boundary length $L$ of the hyperbolic vertices $\mathcal{V}_{g,n}\left(L\right)$,
where $g$ denotes the genus and $n$ the marked points of compact
Riemann surfaces. Since $\mathcal{V}_{g,n}\left(L\right)$ exactly
solves the geometric master equation, complete knowledge of all string
vertices $\mathcal{V}_{g,n}\left(L\right)$ would in principle provide
a full specification of string field theory. Moreover, $\mathcal{V}_{g,n}\left(L\right)$
is uniquely determined by $L$. In our framework, there exists a parallel
description for EWCS in terms of the string charge density. This observation
suggests the possible existence of a string-charge formulation of
string field theory. 
\end{itemize}
\vspace{5mm}

\noindent {\bf Acknowledgements} 
We would especially like to thank Aron Wall for reading the draft and providing valuable suggestions. We are deeply indebted to Xin Gao, Yongge Ma, Jun Nian, and Jia-Rui Sun for helpful discussions. HW is grateful for the collaboration with Amr Ahmadain and Zihan Yan on a related project, which helped motivate this work. We are also grateful for the hospitality of Strings 2025 at New York University Abu Dhabi and of ICTP-AP, during which part of this work was completed.

HW is supported by NSFC Grant No.12105191. SY is supported by NSFC Grant No.12105031 and cstc2021jcyj-bshX0227.

\end{document}